\documentclass[10pt, a4size, journal]{IEEEtran}

\renewcommand{\baselinestretch}{0.9}

\IEEEoverridecommandlockouts

\usepackage{cite}
\usepackage{amsmath,amssymb,amsfonts}
\usepackage{graphicx}
\usepackage{textcomp}
\usepackage[dvipsnames]{xcolor}

\usepackage{stackengine}
\def\delequal{\mathrel{\ensurestackMath{\stackon[1pt]{=}{\scriptstyle\Delta}}}}
\usepackage{float}
\usepackage{bbm}
\usepackage{multirow}
\usepackage{fancyhdr, epsfig, epsf, amsthm, amsmath, amssymb, amsfonts, subfigure,color,dsfont}

\usepackage{dblfloatfix} 
\usepackage[most]{tcolorbox}
\tcbuselibrary{breakable} 
\usepackage{enumitem}
\usepackage{url}
\usepackage[keeplastbox]{flushend}
\usepackage{algorithm,algorithmic,xpatch}

\usepackage{indentfirst}
\usepackage{wrapfig,lipsum,booktabs}

\def\uav{u}

\def\l{\left}
\def\r{\right}
\def\({\l(}
\def\){\r)}
\def\[{\l[}
\def\]{\r]}
\def\T{\intercal}

\begin{document}
	
\title{\fontsize{19}{24}\selectfont {
	Communication-Efficient Massive UAV Online Path Control: Federated Learning Meets Mean-Field Game Theory}  \thanks{This work was supported in part by Academy of Finland under Grant 294128, in part by the 6Genesis Flagship under Grant 318927, in part by the Kvantum	Institute Strategic Project NOOR, in part by the EU-CHISTERA projects LeadingEdge and CONNECT, and in part by the Academy of Finland through the MISSION Project under Grant 319759.}}

	
	
	\author{\IEEEauthorblockN{Hamid Shiri, Jihong Park, and Mehdi Bennis \\
			\textit{(Invited Paper)}}
		\thanks{Hamid Shiri and Mehdi Bennis are with the Faculty of Information
			Technology and Electrical Engineering, University of Oulu, 90570 Oulu,
			Finland (email: $\{$hamid.shiri, mehdi.bennis$\}$@oulu.fi). }
		\thanks{Jihong Park is with the School of Information Technology, Deakin University, Geelong, VIC 3220, Australia (email: $\{$jihong.park$\}$@deakin.edu.au).}
	}
	
	\maketitle
	
	\begin{abstract}
		This paper investigates the control of a massive population of UAVs such as drones. The straightforward method of control of UAVs by considering the interactions among them to make a flock requires a huge inter-UAV communication which is impossible to implement in real-time applications. One method of control is to apply the mean field game  (MFG) framework which substantially reduces communications among the UAVs. However, to realize this framework, powerful processors are required to obtain the control laws at different UAVs. This requirement limits the usage of the MFG framework for real-time applications such as massive UAV control. Thus, a function approximator based on neural networks (NN) is utilized to approximate the solutions of  Hamilton-Jacobi-Bellman (HJB) and Fokker-Planck-Kolmogorov (FPK) equations. Nevertheless, using an approximate solution can violate the conditions for convergence of the MFG framework. Therefore, the federated learning (FL) approach which can share the model parameters of NNs at drones, is proposed with NN based MFG to satisfy the required conditions. The stability analysis of the NN based MFG approach is presented and the performance of the proposed FL-MFG  is elaborated by the simulations.
	\end{abstract}
	

\section{Introduction}

Real-time control of a large number of unmanned aerial vehicles (UAVs) is instrumental in enabling mission-critical applications, such as covering wide disaster sites in emergency cell networks \cite{b3}, search-and-rescue missions to deliver first-aid packets, and firefighting scenarios~\cite{Ackerman:18, b45}. One key challenge is inter-UAV collision, notably under random wind dynamics~\cite{b3, shiri2019massive}. A straightforward solution is to exchange instantaneous UAV locations, incurring huge communication overhead, which is thus unfit for real-time operations. Alternatively, in this article we propose a novel real-time massive UAV control framework leveraging mean-field game (MFG) theory \cite{b10, b11, b12} and federated learning (FL) \cite{mcmahan2016communication, shokri2015privacy}.
			
		In our proposed \emph{FL-MFG control} method, each UAV determines its optimal control decision (e.g., acceleration) not by exchanging UAV states (e.g., position and velocity), but by locally estimating the entire UAV population's state distribution, hereafter referred to as \emph{MF distribution}. 
		According to MFG~\cite{b11}, such a distributed control decision asymptotically achieves the epsilon-Nash equilibrium as the number of UAVs goes to infinity. To implement this, the UAV needs to solve a pair of coupled stochastic differential equations (SDEs), namely, the Fokker-Plank-Kolmogorov (FPK) and Hamilton–Jacobi–Bellman (HJB) equations, for the population distribution estimation and optimal control decision, respectively. The complexity of solving FPK and HJB increases with the state dimension, creating another bottleneck in real-time applications.

		To resolve this complexity issue, FL-MFG control utilizes neural-network (NN) based approximations~\cite{b14, shiri2019massive, shiri2019remote} and FL \cite{mcmahan2016communication}. Specifically, instead of solving HJB and FPK equations, every UAV runs a pair of two NNs, HJB NN and FPK NN whose outputs approximate the solutions of HJB and FPK equations, respectively. The approximation accuracy increases with the number of UAV state observations, i.e., NN training samples. To accelerate the NN training speed, by leveraging FL, each UAV periodically exchanges the HJB NN and FPK NN model parameters with other UAVs, thereby reflecting the locally non-observable training samples. In a source-destination UAV dispatching scenario shown in Fig.~1, simulation results corroborate that FL-MFG control achieves up to $ 50\% $ shorter travel time, $ 25\% $ less motion energy, $ 75\% $ less total transmitted bits, and better collision avoidance measured by $ 50\% $ lower collision probability number, compared to baseline schemes: FL-MFG exchanging only either HJB NN or FPK NN model parameters, and a control scheme running only HJB NN while exchanging state observations.

\begin{figure}[t]
		\centering
		\includegraphics[width = .8\columnwidth]{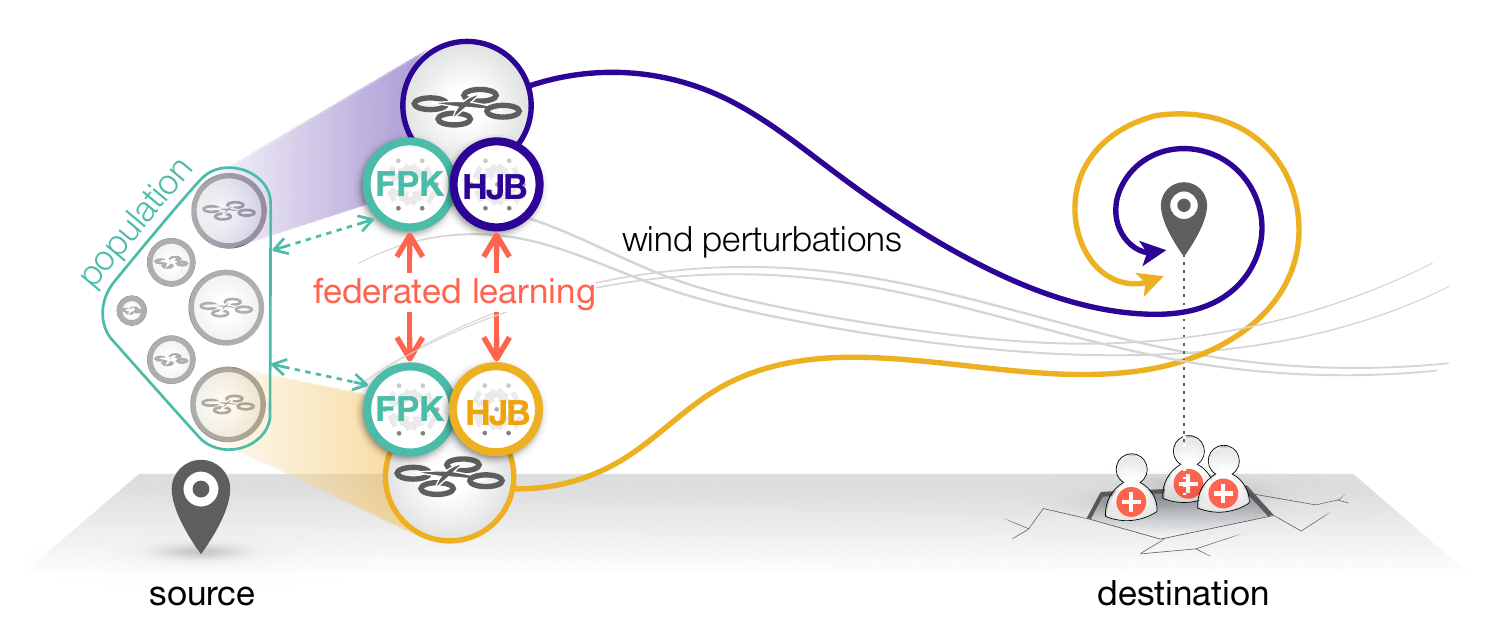}\vskip -5pt
		\caption{ {An illustration of dispatching massive UAVs from a source point to a destination site. Each UAV communicates with neighboring UAVs for achieving: 1) the fastest travel, while jointly minimizing 2) motion energy and 3) inter-UAV collision, under wind perturbations.}} \vskip -5pt
		\label{Fig:Overview} 
\end{figure}
	
\subsection{Background and Related Works}

\noindent\textbf{1. UAV Path Planning:}\quad
Path planning control is about controlling the movement of UAVs to accomplish a target mission. The mission can be broadly categorized into two scenarios. One is to control the UAVs to provision a service, such as ground surveillance \cite{valavanis2015handbook}, emergency networks \cite{b3}, and hotspot aerial cellular networks \cite{8329013}. In this case, the mission can be achieved by maximizing the provisioning network coverage \cite{Feng19Coverage}, surveillance range \cite{Lim10Surveillance}, and traffic offloading rate \cite{lee2020integrating}, while minimizing a cost such as the motion and communication energy consumption. 
The energy consumption is highly dependent on the altitude of UAVs and the types of UAVs (e.g., fixed-wing or multi-copter) \cite{7888557}, as well as communication channels as we shall briefly review in the next part of this subsection.

The other mission is to reach a target destination for the purpose of disaster aid delivery, firefighting, or search and rescue \cite{Ackerman:18, b45}. To this end, the objective is to minimize the travel time/path, while also minimizing a cost function of their motion and communication energy consumption \cite{7888557}, the risk of collision \cite{shiri2019massive}, base-station-UAV disconnection \cite{CellUAV2019Zhang} and interference of UAVs on the ground base-stations (BS) \cite{Walid2019InterferenceUAVrl}. Particularly, the collision avoidance is one key issue in such a mission wherein inter-UAV distances can be very short when collectively lifting a heavy load and/or sharing the same shortest path to the destination. Furthermore, the mission should be executed in real time under a harsh environment \cite{Yang18, oatao19443}, e.g., disturbed by wind due to the low altitude in a rescue mission, aggravating the mission completion difficulty, compared to the aforementioned UAV-as-an-infrastructure missions wherein the optimal UAV paths can be pre-programmed for their long-term operations under stationary environments. The latter type mission is of prime interest throughout this article. 

For a given mission, the UAV path planning can be implemented in an offline or online way. In an offline method, the optimal path is pre-programmed \cite{Comm19UAVmardani,b3}, so is vulnerable to environmental dynamics such as random wind perturbations and moving obstacles. Online methods \cite{zeng2019accessing,shiri2019remote} resolve such a problem by continuously updating their models during exploitation, at the cost of the difficulty in training, in terms of accuracy and convergence speed due to the lack of training samples. To accelerate the training speed to achieve higher accuracy in real time, in this work we apply federated learning across multiple UAVs through UAV-to-UAV communications. The communication models will be briefly reviewed in the next part of this subsection.

\noindent\textbf{2. UAV Communication:}\quad
In the existing literature, UAV communications are grouped into UAV-to-ground communication and inter-UAV communication. For the UAV-to-ground communication, the channel characteristics are sensitive to the blockages and fading near the ground, as well as to the interference from UAVs, ground base stations (BSs), and ground users.
This communication scenario includes a remotely controlled UAV path planning wherein UAVs' states are downloaded by a ground station, and the control commands are uploaded to UAVs \cite{shiri2019remote}.
Since direct wireless communications between the UAV and its ground controller are limited by the signal attenuation in long-range UAV operations, the 3GPP Release 15 has enabled long-range UAV-to-ground communications through cellular networks \cite{zeng2019accessing, austin2011unmanned, 3gpp15}. In this context, interference management is one major issue in UAV-to-ground communication. Recent studies have addressed this issue by adjusting the UAV altitude \cite{Walid2019InterferenceUAVrl}, ground BS height and antenna tilting angle \cite{azari2017coexistence}, while exploiting multi-antenna UAVs \cite{liu2018multibeam}, transmit power control \cite{Taleb2018GC}, and optimal sub-carrier assignment \cite{Taleb2019WCNC}. For given channel conditions, the UAV-to-ground connectivity should be ensured by limiting the range of UAV operations \cite{CellUAV2019Zhang, RZHANG19}. Alternatively, one can partly offload the controlling operations to the UAVs so that they can locally carry out decisions even when the connectivity is temporarily lost \cite{shiri2019remote}.

On the other hand, inter-UAV communications have different channel characteristics compared to UAV-to-ground communications. For instance, in a high altitude, signal scattering becomes sparse due to the thin atmosphere and lack of obstacles. In this case, only path loss may dominate the channel quality without fading \cite{Wang19A2Achannel}. An example of this case is autonomously controlled UAVs that communicate with each other for collective path planning while avoiding inter-UAV collision \cite{b3, shiri2019massive}. Some of these UAVs may communicate with the ground BS or satellite system \cite{8579209}. In this case, by balancing the UAV-to-UAV communications, UAV-to-ground communications, and UAV action optimization, in terms of the energy budget and spectrum bandwidth, the UAVs can achieve efficient data transmission rate \cite{8624565}. In this study, we focus only on the case where UAVs communicate only with neighbors for avoiding collision while accelerating their online training speed.

\noindent\textbf{3. (Windy) Environment:}\quad
Most works in UAV control rely on the knowledge of the environment and the quality of measurements \cite{b20, b21, b26}. Wind profile information is one of the most significant challenging requirements for the UAV control, which can be obtained from various sources such as land-based systems (e.g., weather station, costly meteorological mast, or portable sonar/lidar sensors), airborne platforms (e.g., tethered balloon, or kite), and UAVs (e.g., by installing anemometers on them or by calculations on the state of the UAV) \cite{cano2013quadrotor, abichandani2020wind}.
Still, by utilizing the most accurate tools, obtaining the exact information about the environment in realistic real-time control applications is not possible. 

In addition to wind, presence of static and moving obstacles, threats, other UAVs, and the uncertainties of state observation and the action generators make the real-time path planning of UAV a complex and challenging problem \cite{Agate13Env, cui2018uav, wang20Prediction}. 
For uncertain and dynamic environments, the learning tools such as reinforcement learning models \cite{b22, b23}, Bayesian methods \cite{b27}, and models based on MFG \cite{b3, b6} can be promising control alternatives since they can adapt themselves within the environment.
To the best of our knowledge, there are not many works considering the massive number of UAVs scenario in a windy environment in an online manner all together for real-time applications, and the existing works usually suffer from complex computations or communications cost. 

\noindent\textbf{4. UAV Control:}\quad
In general, two models are used to handle the massive UAV flocking problem. First, the direct control model where a stochastic differential equation (SDE) at each UAV should be solved to obtain the control inputs. Second, the mean-field approximation approach, where the global behavior of the agents is obtained to control the UAVs. 

Following \cite{b6}, in the direct control model,  agents can obtain the optimal solution by exchanging the exact information of their states with each other and solving their SDEs. Nevertheless, due to high complexity and cost of communications for high number of agents, the direct SDE method is only limited to a small number of agents.

Two mean-field based approaches, i.e., the mean-field game (MFG) theory and mean-field control (MFC) theory, are developed to study the behavior of a large population by approximating the behavior with mean-field under exchangeability assumptions, and describing it by a pair of partial differential equations \cite{b10, b11, b12}.

In MFC, an optimal action rule for the whole population is obtained by solving the optimization problem in a collaborative way. 
However, in MFG theory  which  is developed mainly in \cite{b10, b11, b12}, the agents are competing in the $ N $-player non-cooperative game, where the population behavior is approximated by a mean-field and as a result, the game problem becomes tractable. 

In MFG, the optimal action at each agent is obtained by solving a pair of coupled partial differential equations (PDEs): one, called HJB equation which depends on the agent's own state and the interaction term which depends on the distribution of the states of all the agents; and the other one, called FPK equation, describes the evolution of the distribution of the agents depending on the general control rule obtained from HJB equation.

\noindent\textbf{5. ML Aided Control:}\quad
Many of the numerical methods to solve the HJB-FPK equations are computationally expensive especially because of curse of dimensionality, and are ill-suited for real-time applications  \cite{b15, XU2011279, kharroubi2014numerical, b19, b8, b9, b25}. 
In \cite{b15} a numerical approximation method is proposed to approximate the Kolmogorov PDE considering the curse of dimensionality issue. In \cite{XU2011279} a type of iterative method for discrete HJB is obtained and its convergence is investigated. In \cite{kharroubi2014numerical} a probabilistic numerical method based on least-square regressions to solve nonlinear HJB equation together with its analysis is obtained.
Besides, there are many more numerical methods to solve PDEs as in \cite{b19, b8, b9, b25} with high processing requirements.
Therefore, new methods based on machine learning (ML), e.g., (deep) reinforcement learning and neural-network-based online methods, are important to obtain the solution for PDEs with more accuracy or speed \cite{b16, b17, b18, b14}. 
The (deep) reinforcement learning methods are developed to learn the solution of the HJB equation and control rule in \cite{b16, b17, b18}.  
In \cite{b14} a neural-network-based online solution of the HJB equation is used to explore the infinite horizon optimal robust guaranteed cost control of uncertain nonlinear systems. 

Classical centralized multi-agent learning requires communication resources, such as bandwidth and energy, in order to gather all data samples from the agents in  the central unit or server.
Federated learning is proposed in \cite{mcmahan2016communication, shokri2015privacy} to enable model training without sharing data. Instead, every agent trains its local model with its local data samples, and the global training model is obtained by sharing and averaging the local models (rather than the local samples). FL has several benefits such as communication cost reduction and privacy preservation. FL was also shown to be useful in enabling URLLC \cite{samarakoon2018federated}. In addition, there are several research works considering communication cost in control, such as the work in \cite{zeng2019joint} that investigates channel delays in swarm stability for autonomous vehicular platoon systems, \cite{8593118} that improves communication efficiency of a control system, and \cite{ShiriSparseIET2018} which improves communication cost and adaptive estimation error by leveraging  sparsity.
In summary, there is a need for communication-efficient methods to enable ML and control in autonomous applications such as UAVs.

\noindent\textbf{6. Major Challenges:}\quad
Based on the research works mentioned above, still there are some important challenges of implementing MFG framework for real-time control of UAVs in a dynamic environment including: \textbf{1)} except for few cases obtaining an analytical solution for HJB-FPK differential equations is impossible due to its untractability, and the available numerical methods incur high processing power which is not suitable for real-time applications; \textbf{2)} the approximation methods depending on each agents solution might result in different control rules at the agents which violates the interchangeability condition of MFG; \textbf{3)} for distributed approximation methods, the effect of communication channel and the payload size is an important issue in multi agents systems; \textbf{4)} ML based methods require  enough samples for training, and have convergence concerns, which are not considered in MFG frameworks.

\subsection{Contributions and Organization}
In this paper, we study the real-time control of a large population of UAVs in a windy environment. We start this by explaining the scenario of multiple UAVs to be moved from a starting region to a destination region as shown in \figurename{\ref{Fig:Overview}}. 
Then,  two control methods are described in detail, i.e. the HJB control method and the MFG control method to dispatch multiple UAVs quickly, safely, and with low energy consumption. 
The main contribution of this paper are summarized as follows:
\begin{itemize}[leftmargin=*]
	\item To reduce the computational cost of HJB control and MFG control, an NN-based function approximator is utilized to approximate the solution of HJB and FPK equations adaptively. The method used here is a variant of our previous work in \cite{shiri2019massive}, where one single-layer network with two outputs is utilized to estimate the solution of each HJB and FPK equations. This method gives an approximate solution to the HJB and MFG framework. It is shown in \cite{shiri2019massive} that MFG-based learning method requires less communications cost than HJB-based control.
	
	\item To validate the feasibility of the proposed method, the Lyapunov stability analysis is used for HJB and FPK approximation error. These analyses show that the error of approximate solutions for HJB and FPK is bounded, which means that the obtained approximate control actions from NNs are an approximation of the optimal control actions. However, one main requirement for the stability of the approximate solution of MFG is that enough data samples should be provided to update NN's weights, which is challenging in real-time applications.
	
	\item To make sure the UAVs do not lack data samples and to mitigate the communication costs and stability concerns of MFG, an FL based MFG strategy is proposed, which will be named as {\textsf{MfgFL-HF} control} in this paper. This method benefits from the communication reduction property of the FL method for better training of the NN models.
	The performance and stability of \textsf{MfgFL-HF} are verified by the simulations. It is shown that adopting FL can yield faster, safer, energy-efficient, and communication-efficient control over the baseline methods. 
\end{itemize}

The remainder of this paper is organized as follows. Section \ref{SE:02} describes the system model for controlling the population of  UAVs. Section \ref{SE:03} explains the HJB and MFG control methods and the necessity to propose an alternative method. Section \ref{SE:04} proposes the online NN-base method to obtain an approximate solution for HJB and FPK equations, with their stability analysis brought in the appendices. Section \ref{SE:05} proposes FL-based MFG methods in detail. Section \ref{SE:06} validates the performance of the proposed method by simulations, followed by our conclusions in Section \ref{SE:07}.

\section{System Model} \label{SE:02}
Consider the scenario of \figurename{\ref{Fig:Overview}}, where a set $ \mathcal{N} $ of $ N $  UAVs are set to go from a starting position to a specified destination in a windy environment. There are three major issues in this problem as: \textbf{A) Dynamics of the control system}, which reflects the relationship between the parameters of the system, and also the effect of the environment in the system. The more information we have about the environment, the better model we can utilize for control. Here we will assume the wind perturbations as the main source of randomness in the system.
\textbf{B) Control problem}, which will consider the costs and interests to formulate a problem where its solution can control the UAVs to the destinations. Here, one major assumption for control is that the number of UAVs, i.e. $ N $, is large. When the number of UAVs is getting larger, the complexity and the risk of the problem increases consequently, especially in the real-time application with expensive UAVs such as UAVs. 
\textbf{C) Channel}, in a multi-UAV control, the communications among the  UAVs are of critical importance to achieving the control objectives. Here, following the explanations in the Introduction, we will only consider inter-UAV channels, which are modeled as Rician in \cite{Goddemeier}.
In the following, we will consider these three challenges to address the objective of the paper. However, the more focus will be on control with more details on the next sections.

\subsection{Dynamics of Control System}
In order to solve the UAV control problem, we should obtain the relationships among its location, velocity, acceleration, and effect of wind on them at the coordinate system.
Then let us use a Cartesian coordinate system with the origin at the target position as the global reference coordinate. 
We define \small$ r_{i}(t)\in\mathbb{R}^2 $\normalsize\, as the vector from the target destination to the current position of \small$i$\normalsize-th UAV \small$\uav_i$\normalsize\, at time \small$ t \geq 0 $\normalsize. Therefore, the objective of each \small$\uav_i$, $ 1 \leq i \leq N $\normalsize, is to gradually reduce the distance between destination point and the \small$\uav_i$\normalsize's current position, by tuning its velocity \small$ v_{i}(t) \in\mathbb{R}^2  $\normalsize\, by controlling the acceleration \small$ a_{i}(t) \in\mathbb{R}^2 $\normalsize\, under random wind dynamics. 
Following \cite{b49}, the wind dynamics are assumed to follow an Ornstein-Uhlenbeck process with an average wind velocity~\small$v_o$\normalsize. The temporal state dynamics are thereby given~as:

\vspace{-0pt}
\fontsize{8}{0}\selectfont
\begin{subequations}
	\begin{align}
	\text{d}v_i(t) &= a_i(t) \text{d}t - c_0 \left ( v_i(t) - v_o \right ) \text{d}t + V_o \text{d}W_i(t)\label{Eq:sd_a} \end{align}  \vspace{-10pt}
	\begin{align}
	\text{d}r_i(t) &= v_i(t)\text{d}t, \label{Eq:sd_b} 
	\end{align}
\end{subequations}\normalsize
where \small$ c_0 $\normalsize\, is a positive constant, \small$ V_o\in\mathbb{R}^{2\times 2}$\normalsize\, is the covariance matrix of the wind velocity, and \small$W_i(t)\in\mathbb{R}^2 $\normalsize\, is the standard Wiener process independently and identically distributed (i.i.d.) across UAVs.

Now in order to write the dynamics of the controlled system (\ref{Eq:sd_a}-\ref{Eq:sd_b}) in a compact form, let us define the state of each \small$\uav_i$\normalsize\, as \small$s_i(t) \delequal [r_i(t)^{\T},v_i(t)^{\T}]^{\T}\in\mathbb{R}^4$\normalsize; so the SDEs (\ref{Eq:sd_a}-\ref{Eq:sd_b}) can be rewritten as 
\vspace{-0pt}
\fontsize{8}{0}\selectfont
\begin{equation} \label{Eq:StateDy_01}
\text{d} s_i(t)= \left( A s_i(t) + B (a_i(t) + c_0 v_o ) \right) \textup{d}t + G \text{d} W_i(t)\text{,}
\end{equation} \normalsize
where \small$A\!\delequal\!\(\begin{smallmatrix}0 & I\\0 & -c_0 I\end{smallmatrix}\)$\normalsize, \small$B\!\delequal\!\(\begin{smallmatrix}0 \\ 
I\end{smallmatrix}\)$, $G\!\delequal\!\(\begin{smallmatrix}0 \\ V_o\end{smallmatrix}\)$\normalsize, and \small$I$\normalsize\, denotes the two-dimensional identity matrix.
Furthermore, by defining \small$ f(s_i(t)) =  A s_i(t) + c_0 B v_o $\normalsize, we rewrite the equation (\ref{Eq:StateDy_01}), in a compact form as

\vspace{-0pt}
\fontsize{8}{0}\selectfont
\begin{equation} \label{Eq:StateDy_02}
\text{d} s_i(t)= \left( f(s_i(t))+ B a_i(t)  \right) \textup{d}t + G \text{d} W_i(t)\text{.}
\end{equation} 
\normalsize

\subsection{Control Problem}
In general, a model to solve the mentioned control problem should consider three high-level interests for the UAVs. \textbf{First, travel time minimization:} each UAV \small$ i $\normalsize\, should increase speed in the direction to the destination point, to reduce the remaining distance to the destination point while considering to limit the total speed of the UAV. \textbf{Second, motion energy}: each UAV \small$ i $\normalsize\, should reduce the (motion) energy consumption since the UAVs flight time depends on its battery capacity. \textbf{Lastly, collision avoidance}: the collective interest of the whole population is to make a flock of the UAVs traveling together to avoid UAVs colliding each other and also to complete the mission quickly. Nevertheless, there is a trade-off among the interests which should be considered in the control problem. 

To achieve the aforementioned points, UAV~$\uav_i$ at time \small$t\!<\!T_f$\normalsize\, aims to minimize its average cost~\small$ \psi_{i}^{a_i}(s_i,t;s_{-i} ) $\normalsize, where  \small$ {T_f} $\normalsize\, is the terminal control time, and \small$ s_{i}(t) = s_{i}, {s}_{-i}(t) = s_{-i}  $\normalsize\, are the state of UAV $ \uav_i $ and the set of states of all UAVs excluding UAV $ \uav_i $ at time \small$ t $\normalsize, respectively. The average is taken with respect to a measure (of the integral inside the expectation) with a probability distribution depending on \small$ (s_i,t) $\normalsize, and it is calculated for the trajectory \small$ \{s_i(\tau)\}_{[t,T]} $\normalsize\, obtained by the control law \small$ a_i $\normalsize. The cost \small$\psi_{i}^{a_i}(s_i,t;s_{-i} )$\normalsize\, consists of the term \small$ g( a_{i}(\tau), s_{i}(\tau); s_{-i}(\tau))$\normalsize\, depending on the local state \small$s_i(\tau)$\normalsize\, and the control action \small$a_i(\tau)$\normalsize\, with given states of other UAV's as \small$ {s}_{-i}(\tau) $\normalsize, i.e.,

\vspace{-0pt}
\fontsize{8}{0}\selectfont
\begin{align} \label{Eq:LRA}
\hspace{-5pt} \psi^{a_i}(s_i,t;s_{-i} ) \!=\!  \mathbb{E} \[\; \int_{t}^{T_f}  g( a_{i}(\tau), s_{i}(\tau); s_{-i}(\tau)) \textup{d}\tau \] \text{,}
\end{align}\normalsize
where \small$ \mathbb{E} $\normalsize\, is the expectation operator, and \small$  g( a_{i}(\tau), s_{i}(\tau); s_{-i}(\tau)) $\normalsize\, is
\vspace{-5pt} \fontsize{8}{0}\selectfont
\begin{align} \label{Eq:utility}
\hspace{-5pt}  g( a_{i}(\tau), s_{i}(\tau); s_{-i}(\tau)) &\!=\!  \phi_L(s_i(\tau)) \! +  c_3 {\left\| a_{i}(\tau) \right\|^{2}  } 
 \! + \! c_2 \phi_G(s_{i}(\tau);s_{-i}(\tau))
\end{align}\normalsize
in which, the term \small$\phi_L(s_i(\tau))$\normalsize\, depends only on the local state \small$s_i(t)$\normalsize\, and the term \small$ \phi_G(s_{i}(\tau);s_{-i}(\tau))$\normalsize\, relies on the global state \small$ \{ s_i(t), s_{-i}(t) \}  $\normalsize, given~as:

\vspace{-0pt}
\fontsize{8}{0}\selectfont
\begin{align}
&\;\;\phi_L(s_i(\tau)) = {\frac{v_i(\tau) \cdot r_i(\tau)}{\left \| r_i(\tau) \right \|}  }  + {c_1 {\left\| v_i(\tau) \right\|^{2}} }, \\
&\;\;\phi_G(s_{i}(\tau);s_{-i}(\tau))  =  {\frac{1}{N} \sum_{\uav_j\in \mathcal{N}}   \frac{ \left\| v_{j}(\tau) - v_i(\tau) \right\|^{2} }{\left( \varepsilon + \left \| r_j(\tau) - r_i(\tau) \right \|^2  \right)^{\beta}}}, \label{Eq:InterAction01}
\end{align}\normalsize
and the terms $ c_1 $, $ c_2 $,  $ c_3 $, $\beta$, and $\varepsilon$ are positive constants. 

The local term $\phi_L(s_i(\tau))$ and the second term in \eqref{Eq:utility} focus on the the two objectives, i.e. \textbf{travel time} and \textbf{motion energy} minimization. It is intended to minimize the remaining travel distance $\|r_i(\tau)\|$ by maximizing the speed towards the destination, i.e., minimizing the projected speed $v_i(\tau)\cdot r_i(\tau)/\|r_i(\tau)\|$ towards the opposite direction to the destination. 
Also, we minimize the kinetic energy and the acceleration control energy by minimizing proxy  terms \small$\|v_i(\tau)\|^2$\normalsize\, (speed) and  \small$\| a_i(\tau) \|^2$\normalsize\, (acceleration), respectively~\cite{b35, b6}.
The actual instantaneous motion power consumption \small$ P(\tau) $\normalsize\, of a UAV in the environment, knowing the UAV's speed {\small$ \| {v}(\tau) \|  $\normalsize},  and characteristics of the UAV and air, is calculated by

\fontsize{7.5}{0}\selectfont 
	\begin{align}
	P(\tau)  \!=\!  \lambda_0 (1\!+\! \frac{3 \| {v}(\tau) \|^2 }{\omega_\text{tip}^2})+\lambda_1( \sqrt{1\!+\! \frac{ \| {v}(\tau) \|^4 }{4 \chi_{o}^4}} \!-\! \frac{ \| {v}(\tau) \|^2 }{2 \chi_{o}^2} )^\frac{1}{2} \!+\! \lambda_2  \frac{\| {v}(\tau) \|^3 }{2}    \text{,}  \label{Eq:Power1}
	\end{align}\normalsize
{\noindent}where  \small$ \lambda_0  $\normalsize, \small$ \lambda_1  $\normalsize, and \small$ \lambda_2 $\normalsize, rotor blade tip speed as \small$ \omega_\text{tip}$\normalsize, and mean rotor induced velocity in hovering \small$ \chi_{o} $\normalsize\, are the physical characteristics of UAV in the environment \cite{ZengeRotary}.
Then, the motion energy \small$ E(t) $\normalsize\, for each UAV \small$ i $\normalsize\, at time \small$ t $\normalsize, is defined as 

\vspace{-0pt}
\fontsize{8}{0}\selectfont
\begin{align}
E(t) = \int_{\tau = 0}^{t} P(\tau)\text{d} \tau,
\end{align}
\normalsize
which is used as the energy metric to compare different algorithms in our work.

In addition, for comparison purposes, the mission completion metric for each agent \small$ i $\normalsize\, is defined as the time \small$ t = T_i $\normalsize, when the state \small$ s_i(t) $\normalsize\, of the agent  enters the area defined as  \small$ \{ s_{\text{dest}} : \| s_{\text{dest}} \| = \text{const.}  \}  $\normalsize\, for the first time, i.e., \small$\| s_i(T_i) \|   \! \leq \! \| s_{\text{dest}} \| $\normalsize. Then, the average travel time is defined as \small$  T_\text{avg} = \sum_{\uav_i\in \mathcal{N}} T_i $\normalsize, and the mission completion time is defined as  \small$  T_\text{max} = \arg \max_{T \in \{T_i\}} T  $\normalsize.

The global term \small$\phi_G(s_{i}(\tau);s_{-i}(\tau))$\normalsize\, in \eqref{Eq:utility} refers to \textbf{collision avoidance} and is intended to form a flock of UAVs moving together~\cite{b1}. The flocking leads to small relative inter-UAV velocities for avoiding collision even when their controlled velocities are slightly perturbed by wind dynamics. 
A collision happens when the inter-UAV distance is less than a defined distance $ r_\text{coll}$. 
Furthermore, the flocking yields closer inter-UAV distances without collision. This is beneficial for allowing more UAVs to exchange their local states through better channel quality, thereby contributing also to collision avoidance. 
The formation of a flock as mentioned in \cite{b5} is a result of three components: a) separation, i.e. steer to avoid crowding; b) alignment, i.e. steer toward the average heading of neighbors; c) cohesion, i.e. steer toward the average position of neighbors. 
In view of this, we adopt the Cucker-Smale flocking~\cite{b3,b1} that reduces the relative speeds for the UAVs. The relative speed \small$\|v_j(\tau)-v_i(\tau)\|$\normalsize\, and the inter-UAV distance \small$\|r_j(\tau)-r_i(\tau)\|$\normalsize\, are thus incorporated in the numerator and denominator of \small$\phi_G(s_{i}(\tau);s_{-i}(\tau))$\normalsize, respectively.
In addition, inspired by \cite{b13}, the velocity alignment \small$ \phi_\text{A}(t) $\normalsize\, and  number of collision risks \small$ \phi_\text{C}(t)  $\normalsize\, as metrics to compare different algorithms, are defined as 

\vspace{-0pt}
\fontsize{8}{0}\selectfont
\begin{align}
\phi_\text{A}(t) = \frac{1}{t N^2}  \int_{0}^{t} { \sum_{\uav_i\in \mathcal{N}}  \sum_{\uav_j\in \mathcal{N}}   {  \| v_{j}(\tau) - v_i(\tau) \|  }\text{d}\tau  }, \label{Eq:VelocityAlignment}
\end{align}
\normalsize 

\vspace{-10pt}
\fontsize{8}{0}\selectfont
\begin{align}
\phi_\text{C}(t) =  \frac{1}{t N^2} \int_{0}^{t}  { \sum_{\uav_i\in \mathcal{N}}  \sum_{\uav_j\in \mathcal{N}}   {  \mathbbm{1}_{\| r_j(\tau) - r_i(\tau) \|\leq r_\text{C}}  } \text{d}\tau   }, \label{Eq:NoCollision} 
\end{align}
\normalsize 
where the hazard radius \small$ r_\text{C} $\normalsize\, defines a dangerous potential collision zone around the UAV. Lower values of \small$ \phi_\text{A}(t) $\normalsize\, means that the amplitudes of velocity differences between UAVs are small and hence they have made a better flock to travel together. Lower values of \small$ \phi_\text{C}(t) $\normalsize\, mean that the UAVs do not tend to be too close to each other and the risk of them colliding each other is smaller.

Incorporating the cost \eqref{Eq:LRA} under the temporal dynamics \eqref{Eq:StateDy_02}, the control problem of UAV $\uav_i$ at time~\small$t$\normalsize\, is formulated as:

\vspace{-0pt}
\fontsize{8}{0}\selectfont
\begin{align} \label{Eq:psi}
&\psi (s_i,t; s_{-i}) = \min_{a_i}\; \psi^{a_i}(s_i,t;s_{-i} )  \\
&\text{s.t.}\quad \text{d} s_i(t)= \left( f(s_i(t))+ B a_i(t)  \right) \textup{d}t + G \text{d} W_i(t), \label{Eq:StateDyn}
\end{align}\normalsize
The minimum cost \small$\psi (s_i,t; s_{-i})$\normalsize\, is referred to as the \emph{value function} of the optimal control, and should be derived to obtain the optimal action \small$ a_i(t) $\normalsize\, for UAV $\uav_i$. 
The methods to encounter this problem will be introduced in the following sections.

\subsection{UAV to UAV Wireless Channel Model}

In many multi-UAV control problems, communication among the UAVs is a critical condition. However, in the problem of this paper the UAVs will be required to communicate their data with each other while they are moving at a height $ \mathsf{h} $. Following \cite{Goddemeier} for the UAV to UAV communication channels, the Rice model can be used to model both dominant LOS and NLOS paths. The Rice distribution is given by:

\vspace{-0pt}
\fontsize{8}{0}\selectfont
\begin{align}\label{Eq:Rice}
p_{z}(\zeta)=\frac{\zeta}{\chi^{2}}\exp\left(\frac{-\zeta^{2}-\xi^{2}}{2\chi^{2}}\right)I_{0}\left(\frac{z \xi}{\chi^{2}}\right),
\end{align} \normalsize
\noindent where \small$ \zeta\geq 0 $\normalsize, and \small$ \xi $\normalsize\, and \small$ \chi $\normalsize\, are the strength of LOS and NLOS paths respectively.  However, when there is no LOS path between UAVs, this model is reduced to Rayleigh model by setting \small$ \zeta = 0 $\normalsize. Therefore, depending on the scenario, the channel model parameters can be set.
Assuming the frequency division multiple access (FDMA) is used for each UAV to UAV communication,  with the transmission power \small$ P_o $\normalsize, and the distance \small$ r_d $\normalsize\, from a UAV to another UAV, the received signal-to-noise (SNR) at each time is

\vspace{-5pt}
\fontsize{8}{0}\selectfont
\begin{align}
	\text{SNR} = \frac{P_o z r_d^{-\alpha}}{W_o \sigma_n} ,
\end{align} \normalsize
\noindent where \small$ \sigma_n $\normalsize\, is the noise power, \small$ W_o $\normalsize\,  is the bandwidth,  $ \alpha \geq 2 $  is the path loss exponent, and $ z $ is the Rice random variable defined in \eqref{Eq:Rice} and is assumed to be independent and identically distributed (i.i.d) across the different UAVs and times. 

Another parameter which we will use in this paper is the communication latency. A signal is successfully decoded if the SNR at time \small$ t $\normalsize\, is greater than a target SNR \small$ \eta$\normalsize, i.e.,  \small$ \text{SNR}(t) \geq \eta $\normalsize. The number of bits \small$ b_i(D_0) $\normalsize, transmitted  during \small$ D_0 $\normalsize\, time slots, is given as:

\vspace{-10pt}
\fontsize{8}{0}\selectfont
\begin{align}
b_i(D_0) = \theta \sum_{t=1}^{D_0}  \mathbbm{1}_{\text{SNR}(t)  \geq \eta}  W_o \log_2(1+\eta),
\end{align} \normalsize
\noindent where \small$ \theta $\normalsize\, is the channel coherence time. 
The latency of transmitting \small$ b $\normalsize\, bits is the minimum \small$ D_0 $\normalsize, i.e. \small$ D_m $\normalsize\, that satisfies \small$ b_i(D_0) \geq b $\normalsize. A latency outage occurs when \small$  D_m $\normalsize\, is greater than a predefined threshold \small$  D_M $\normalsize, i.e., \small$  D_m >  D_M $\normalsize\, in an algorithm.

\section{HJB and MFG Control} \label{SE:03}
At each time instant, each UAV seeks a solution for the defined objective function. The first intuitive method is to analyze the problem directly as solving an HJB equation and then see if there is a need for other alternatives and also have the basis for the proposed methods. However, in the following, it will be clarified that when the number of UAVs is high, the communications and processing complexity will increase to the extent that the real-time implementation will not be possible. Therefore, an alternative MFG method which can reduce the number of communications significantly will be explained.

\subsection{HJB Control}
In this method, 
we assume that all \small$ N $\normalsize\, UAVs perform their optimal action based on the observed local states of all other UAVs. 
Then according to Bellman's principle for optimal control, for each time \small$ t \leq T_f $\normalsize\, and state \small$ s_i $\normalsize\, of agent \small$ i $\normalsize, we can rewrite \eqref{Eq:psi} as

\vspace{-4pt}\fontsize{8}{0}\selectfont
\begin{align} \label{Eq:psi2}
\hspace{-5pt} \psi (s_i,t; s_{-i}) &\!=\! \min_{a_i}\; \mathbb{E} \[\; \int_{t}^{T_f}  g( a_{i}(\tau), s_{i}(\tau); s_{-i}(\tau)) \textup{d}\tau \] \\
&\!=\! \min_{a_i}\; \{ g(a_{i}(t), s_{i}; s_{-i}) \!+\! { \mathbb{E} \[ \psi (s_i + \textup{d}s_i,t + \textup{d}t; s_{-i}) \] }\} \label{Eq:psi2b}
\end{align}\normalsize
By utilizing \eqref{Eq:StateDyn}, the second order Taylor expansion of the second term in \eqref{Eq:psi2b} is calculated as

\vspace{-0pt}\fontsize{8}{0}\selectfont
\begin{align} \label{Eq:psi3}
\hspace{-5pt}  \mathbb{E} \[ \psi (s_i + \textup{d}s_i,t + \textup{d}t; s_{-i}) \]  =&   \left\{ \right. \![\nabla_{s_{i}} \psi (s_i,t; s_{-i}) ]^\T \left(  f(s_i)+ B a_i(t) \right)   \nonumber \\ 
 \!+\! \frac{1}{2}\textup{tr}(G G^{\T}  [\Delta_{s_{i}}  \psi (s_i,t; s_{-i}) ]  ) &+ \dot{\psi} (s_i,t; s_{-i}) \left. \right\}\textup{d}t + \psi (s_i,t; s_{-i}).
\end{align}\normalsize
Therefore, the HJB equation with this method is obtained as 

\vspace{-0pt}
\fontsize{8}{0}\selectfont 
\begin{align}
\dot{\psi} (s_i,t;&s_{-i})  
\!+\! \min_{a_{i}}  \! \left\{ \right. \! g( a_{i}(t), s_{i}; s_{-i}) 
\!+\!  [\nabla_{s_{i}} \psi (s_i,t; s_{-i}) ]^\T 
\nonumber \\
& 
\! \times \! \left(  f(s_i) \!+\! B a_i(t) \right)   
\!+\! \frac{1}{2}\textup{tr}(G G^{\T}  [\Delta_{s_{i}}  \psi (s_i,t; s_{-i}) ]  ) \!  \left. \right\}  \!=\! 0  \text{,} \label{Eq:HJB00}
\end{align} \normalsize
where  $ \nabla $ and $ \Delta $ denote the gradient and Laplacian operators, respectively. 
The optimal action at UAV $ i $ is obtained as

\vspace{-5pt}
\fontsize{8}{0}\selectfont
\begin{align} 
a_{i}(t) = - \frac{1}{2 c_3} B^{\T} \nabla_{s_i} \psi (s_i,t; s_{-i}) \text{.} \label{Eq:Action01}
\end{align} \normalsize
Therefore, substituting \eqref{Eq:Action01} in \eqref{Eq:HJB00} yields the HJB equation

\vspace{-0pt}
\fontsize{8}{0}\selectfont 
\begin{align} 
\dot{\psi}&(s_i,t; s_{-i})  \!    
+   \left( \! f(s_i)  - \frac{1}{4 c_3} B B^{\T} \nabla_{s_i} \psi (s_i,t; s_{-i}) \right)^\T  [\nabla_{s_{i}} \psi (s_i,t; s_{-i}) ] \!
\nonumber \\
& 
\!+\! \frac{1}{2}\textup{tr}(G G^{\T}  [\Delta_{s_{i}}  \psi (s_i,t; s_{-i}) ]  )  \!+\!  \phi_L(s_i) \! + \! c_2 \phi_G(s_{i};s_{-i})  = 0 \text{.}   \label{Eq:HJB01}
\end{align} \normalsize
Solving this HJB equation requires an enormous exchange of states among the UAVs, which becomes impossible when the number of UAVs, i.e. \small$ N $\normalsize, is high. In order to address this challenge, we leverage the capabilities of the MFG framework explained in the next subsection. Therefore, the UAVs will need to exchange the states only at the beginning of the mission, and after that, they will calculate the optimal actions based on their own state.

\subsection{MFG Control}
Another method to encounter this problem is to use MFG framework when the number of UAVs is very high.
Let \small$ m_{N}(s,t) $\normalsize\, be the empirical state distribution function of the UAVs at time instant \small$ t $\normalsize\, defined as

\vspace{-0pt}
\fontsize{8}{0}\selectfont
\begin{align}
m^{}_{N}(s,t) \delequal \frac{1}{N} \sum_{j=1}^{N} \delta(s - s_{j}(t)) \text{,}
\end{align} \normalsize
where \small$ \delta(\cdot) $\normalsize\, is the Dirac delta function. Then, the interaction term \eqref{Eq:InterAction01} can be rewritten as

\vspace{-0pt}
\fontsize{8}{0}\selectfont\begin{align}
\phi_G(s_{i}(t);s_{-i}(t)) = \int_{s}^{} \! m^{}_{N}(s,t) ~ \frac{  \left\| v_{i}(t) - v \right\|^{2} }{\left( \varepsilon + \left \| r_{i}(t) - r \right \|)^2  \right)^\beta}  \text{d}s \text{.}
\end{align}\normalsize

Since the states \small$ {s}_{i}(t) $\normalsize\, for \small$ i = 1,\ldots,N $\normalsize\, are independent and identically distributed which evolve according to SDE \eqref{Eq:StateDyn}, utilizing the ergodic theory gives

\vspace{-5pt}
\fontsize{8}{0}\selectfont 
\begin{align}
\lim\limits_{N \to \infty } m_{N}(s,t) = m(s,t) \text{,} \label{Eq:DistApp}
\end{align}\normalsize
where \small$ m(s,t) $\normalsize\, is the distribution of generic UAV's state, i.e., \small$ s $\normalsize, corresponding to the SDE \eqref{Eq:StateDyn} with optimal policy \small$ a(t) $\normalsize\, obtained by \eqref{Eq:Action01}. The distribution \small$ m(s,t) $\normalsize, which is called mean field (MF), is the solution of the Fokker-Plank-Kolmogorov (FPK) equation as

\vspace{-5pt}
\fontsize{8}{0}\selectfont 
\begin{align}  \label{Eq:FPK_01}
\hspace{-5pt} \dot{m}(s,t) \!+\! \nabla_{s} \cdot \![(f(s) \!+\! B a(t)  ) m(s,t)] \!-\! \frac{1}{2}\textup{tr}(G G^{\T} \Delta_s  m(s,t)) \!=\! 0 \text{,}
\end{align}\normalsize
where \small$  \nabla_{s} \cdot $\normalsize\, denotes the divergence operator, and the initial distribution of the UAVs is given as \small$ m(s,0) = \frac{1}{N} \sum_{j=1}^{N} \delta(s - s_{j}(0)) $\normalsize. 
For an optimal action rule \small$ a(t) $\normalsize, the FPK equation \eqref{Eq:FPK_01} can give the distribution $ m_{}(s,t) $ of state of a typical agent \small$ s $\normalsize\, at each time \small$ t $\normalsize. In MFG framework, FPK and HJB equations are coupled through the optimal action $ a_{}(t) $ obtained by \eqref{Eq:Action01} and the interaction term \small$ \phi_G(s_{i};s_{-i}) $\normalsize\, which can be approximated by  \small$ \phi_G(s_{i}(t);m(s,t)) $\normalsize\, defined in \eqref{Eq:InterAction03} (since according to \eqref{Eq:DistApp}, for large $ N $, the actual distribution \small$ m_{N}(s,t) $\normalsize\, can be approximated by \small$ m_{}(s,t) $\normalsize\,).

\vspace{-5pt}\fontsize{8}{0}\selectfont 
\begin{align}
\phi_G(s_{i}(t);m(s,t)) \delequal \int_{s} \! m(s,t) ~ \frac{  \left\| v_{i}(t) - v \right\|^{2} }{\left( \varepsilon + \left \| r_{i}(t) - r \right \|)^2  \right)^\beta}  \text{d}s  \label{Eq:InterAction03}
\end{align}\normalsize
By this definition, the HJB equation \eqref{Eq:HJB01} can be rewritten as

\vspace{-5pt}
\fontsize{8}{0}\selectfont \begin{align} 
\dot{\psi}(s_i,t;& m(s,t) ) + \frac{1}{2}\textup{tr}(G G^{\T}  [\Delta_{s_{i}}  \psi (s_i,t; m(s,t) ) ]  )   
\nonumber \\ 
&
\! + \! \left( \!  f(s_i) \! - \! \frac{1}{4 c_3} B B^{\T} \nabla_{s_i} \psi (s_i,t; m(s,t) ) \right)^\T 
\! \nabla_{s_{i}} \psi(s_i,t; m(s,t) ) 
\nonumber \\
& \!+\!  \phi_L(s_i) \! + \! c_2 \phi_G(s_{i}; m(s,t)) \!  = 0 \text{,}  \label{Eq:HJB02}
\end{align} \normalsize
and the corresponding action is

\vspace{-5pt}
\fontsize{8}{0}\selectfont
\begin{align} \label{Eq:Action02}
a_i(t) =  - \frac{1}{2 c_3} B^{\T} \nabla_{s_i} \psi (s_i,t; m(s,t) )  \text{,}
\end{align} \normalsize
Therefore, by substituting \eqref{Eq:Action02} in \eqref{Eq:FPK_01} the FPK equation can also be rewritten in the following form:

\vspace{-5pt}
\fontsize{8}{0}\selectfont
\begin{align}  \label{Eq:FPK_02}
\dot{m}(s,t) & + \nabla_{s} \cdot [(f(s)- \frac{1}{2 c_3} B B^{\T} \nabla_{s} \psi (s,t; m(s,t) )  ) m(s,t)]  
\nonumber \\
&- \frac{1}{2}\textup{tr}(G G^{\T} \Delta_s  m(s,t)) = 0
\end{align}\normalsize

By solving HJB and FPK equation pairs, i.e., \eqref{Eq:HJB02} and \eqref{Eq:FPK_02}, the optimal action for each UAV can be calculated.

\vspace{-10pt}

\section{NN-Based HJB and MFG Learning Control - State Sharing Methods} \label{SE:04}

\vspace{-5pt}
In this section, inspired by \cite{b2}, we discuss NN-based methods to obtain approximate solutions for HJB and FPK equations for the multiple UAV control application. 
However, to have better readability and due to lack of space, we will use the simplifications of Table I unless there is a need for complete forms to avoid confusion.

\vspace{-10pt}
\fontsize{9}{0}\selectfont
\begin{table} 
	\label{TB:SNotations}
	\caption{List of Simplified notations.}
	\centering
	\begin{tabular}{||l l | l l||} 
		\hline
		\multirow{2}{*}{Notation} & Simplified  & \multirow{2}{*}{Notation} & Simplified \\ [0.5ex] 
		&  Definition &   &  Definition  \\ [0.5ex] 
		\hline\hline
		$ \psi (s_i,t; m(s,t) ) $ & \multirow{2}{*}{$ \psi $} & $ m(s,t)  $ & $ m $  \\ \cline{3-4}
		$ \psi (s_i,t; s_{-i} ) $ &   & $ {w}_{\textsf{H}_d}(t)  $  & $ {w}_{\textsf{H}_d}  $ \\ 
		\hline
		$ \phi_L(s_i) $ & $ \phi_L  $  & $ \sigma_\textsf{H}\!\(s_{i}; m(s,t) \) $ & \multirow{2}{*}{$ \sigma_\textsf{H} $}  \\ \cline{1-2}
		$ \phi_G(s_{i}; m(s,t)) $ & \multirow{2}{*}{$ \phi_G $} & $ \sigma_\textsf{H}\!\(s_{i}; s_{-i} \) $ &   \\ \cline{3-4}
		$ \phi_G(s_{i}; s_{-i}) $ &   & $ \sigma_\textsf{F}\!\(s \) $ & {$ \sigma_\textsf{F} $} \\ 
		\hline
		$ f(s_i) $ & $ f $  & $ e_\textsf{H}(s_i,t) $ & $ e_\textsf{H} $  \\ 
		\hline
		$ \nabla_{s_i} $ & \multirow{2}{*}{$ \nabla $} & $ e_\textsf{F}(s,t) $ & $ e_\textsf{F} $  \\ \cline{3-4}
		$ \nabla_{s} $ &   & $ J_\textsf{H}(\hat{w}_{\textsf{H}_0}, \hat{w}_{\textsf{H}_1}) $ & $ J_\textsf{H} $ \\ 
		\hline
		$ \Delta_{s}  $ & \multirow{2}{*}{$ \Delta $} &  $ J_\textsf{F}(\hat{w}_{\textsf{F}_0}, \hat{w}_{\textsf{F}_1}) $ & $ J_\textsf{F} $   \\ \cline{3-4}
		$ \Delta_{s} $ &   & 	$ a_i(t) $ & $ a_i $  \\ 
		\hline
		$ \nabla_{s} \cdot $ & {$ \nabla \cdot $} & $ \varepsilon_{\textsf{H} }(s_i, t) $ & $ \varepsilon_{\textsf{H} } $   \\ 
		\hline
		$ \textsf{H} (s_i,t; m(s,t) ) $ & \multirow{2}{*}{$ \textsf{H} $} & 	$ \varepsilon_{\textsf{F}}(s, t) $ & $ \varepsilon_{\textsf{F}} $  \\ \cline{3-4}
		$ \textsf{H} (s_i,t; s_{-i} ) $ &   & 	$ {L}_s(s_i(t))  $ & $ {L}_s $  \\ 
		\hline
		$ \textsf{F} (s,t; a(t) ) $ & {$ \textsf{F} $} & 	$ R_i(t)  $ & $ R_i $  \\ 
		\hline
	\end{tabular}
\end{table} 
\normalsize

\subsection{HJB Learning Control} 
Here, following our previous work \cite{shiri2019massive}, we find an approximate solution to the HJB equation to obtain the corresponding action. Any approximate solution will result in some error, and the approximated HJB equation may not be exactly equal to zero.
Therefore, first, based on the defined simplifications above, we represent the HJB equation \eqref{Eq:HJB01} and \eqref{Eq:HJB02} by $ \textsf{H} $ as

\vspace{-5pt}
\fontsize{7.5}{0}\selectfont
\small \begin{align} 
\textsf{H} \delequal  \dot{\psi}  \!+\!   \left(  f  \!-\! \frac{B B^{\T} \nabla \psi}{4 c_3}  \right)^\T  \nabla \psi \! +  \phi_L \! + \! c_2 \phi_G \!  + \frac{1}{2}\textup{tr}(G G^{\T}  \Delta  \psi  )  \!=\! 0, \label{Eq:HJB03}
\end{align} \normalsize
where we obtain the $ \phi_G $ empirically by \eqref{Eq:InterAction01} in this subsection.

Similar to \cite{b2}, given the state distribution  of UAVs at each time \small$ t $\normalsize, let the function \small$ \psi (s_i,t; s_{-i} ) $\normalsize\, 
and its derivative correspondingly be approximated by functions as

\vspace{-0pt}
\fontsize{8}{0}\selectfont
\begin{align}  
\hat{\psi} (s_i,t; s_{-i} ) \delequal  \hat{w}_{\textsf{H}_0}\!(t)^\T  \sigma_\textsf{H}\!\(s_{i}; s_{-i} \) \text{,} \label{Eq:HJBmodel_1} \\
\hat{\dot{\psi}} (s_i,t; s_{-i} ) \delequal  \hat{w}_{\textsf{H}_1}\!(t)^\T  \sigma_\textsf{H}\!\(s_{i}; s_{-i} \) \text{,} \label{Eq:HJBmodel_2}
\end{align} \normalsize
where vector functions \small$ \hat{w}_{\textsf{H}_0}\!(t) $\normalsize\, and  \small$ \hat{w}_{\textsf{H}_1}\!(t) $\normalsize\, are  approximations to the optimal vector weight functions \small$ {w}_{\textsf{H}_0}\!(t) $\normalsize\, and \small$ {w}_{\textsf{H}_1}\!(t) $\normalsize, respectively, and the value error of these approximations are

\vspace{-0pt}
\fontsize{8}{0}\selectfont
\begin{align}
	\varepsilon_{\textsf{H}_0}\!(s_i,t) &\delequal {\psi}  (s_i,t; s_{-i} ) - {w}_{\textsf{H}_0}\!(t)^\T\sigma_\textsf{H}\!\(s_{i};s_{-i} \), \\
	\varepsilon_{\textsf{H}_1} (s_i,t) &\delequal \dot{\psi} (s_i,t; s_{-i} ) - {w}_{\textsf{H}_1}\!(t)^\T \sigma_\textsf{H}\!\(s_{i};s_{-i} \).
\end{align} \normalsize
Then, using these definitions, and notation simplifications as in Table I,  the HJB equation \eqref{Eq:HJB03} and its approximatation are written as

\vspace{-5pt}
\fontsize{8}{0}\selectfont 
\begin{align} 
{\textsf{H} }  &=  {w}_{\textsf{H}_1}^\T \sigma_\textsf{H}    \! + \!  \left(  f \!-\! \frac{1}{4 c_3} B B^{\T} ([\nabla \sigma_\textsf{H}]^{\T} {w}_{\textsf{H}_0}  ) \right)^\T  [\nabla \sigma_\textsf{H}]^{\T} {w}_{\textsf{H}_0}  
\nonumber \\
&  \quad 
\!+\!  \frac{1}{2}\! \left[\sum_{k=1}^{N} \textup{tr}(G G^{T}  [\Delta  \sigma_\textsf{H}^{[k]}]  ) {\mathbf{e}}_{k} \right]^{\T} \! {w}_{\textsf{H}_0} \! + \! \phi_L \! + \! c_2 \hat{\phi}_G \!+\! \varepsilon_{\textsf{H} } \!=\! 0 \text{,} \\
\hat{\textsf{H} }  &=  \hat{w}_{\textsf{H}_1}^\T \sigma_\textsf{H}  \! + \!  \left(  f \!-\! \frac{1}{4 c_3} B B^{\T} ([\nabla \sigma_\textsf{H}]^{\T} \hat{w}_{\textsf{H}_0}  ) \right)^\T [\nabla \sigma_\textsf{H}]^{\T} \hat{w}_{\textsf{H}_0}  
\nonumber \\
& \quad 
\!+\!  \frac{1}{2} \left[\sum_{k=1}^{N} \textup{tr}(G G^{T}  [\Delta  \sigma_\textsf{H}^{[k]}]  ) {\mathbf{e}}_{k} \right]^{\T} \! \hat{w}_{\textsf{H}_0} \!+\!  \phi_L \! + \! c_2 \hat{\phi}_G \text{,}
\end{align}\normalsize
where the superscript $ [k] $ shows the $ k $'s element of the corresponding vector, $ {\mathbf{e}}_{k} $ is a vector with $ k $'s element equal to $ 1 $ and other elements equal to zero, and $ \varepsilon_{\textsf{H}} $ is the error of HJB equation with the function approximator defined as

\vspace{-0pt}
\fontsize{8}{0}\selectfont 
\begin{align}
\varepsilon_{\textsf{H}} \delequal
&
 c_2 \varepsilon_{\phi_G} \!+\! \varepsilon_{\textsf{H}_1} 
\!-\! \frac{1}{4 c_3} [ \nabla \varepsilon_{\textsf{H}_0}]^{\T} B B^{\T} [\nabla \sigma_\textsf{H}]^{\T} {w}_{\textsf{H}_0} \!-\! \frac{1}{4 c_3} [ \nabla \varepsilon_{\textsf{H}_0}]^{\T} B B^{\T} \nabla \varepsilon_{\textsf{H}_0} 
 \nonumber \\
& 
\!+\! \left(  f \!-\! \frac{1}{4 c_3} B B^{\T} [\nabla \sigma_\textsf{H}]^{\T} {w}_{\textsf{H}_0}  \right)^\T   \nabla \varepsilon_{\textsf{H}_0}  \!+\! \frac{1}{2} \textup{tr}(G G^{\T} \Delta \varepsilon_{\textsf{H}_0} )
\end{align}\normalsize
where $ \varepsilon_{\phi_G} $ is the uncertainty of the interaction term.
Then, the corresponding approximate action can be obtained by
\vspace{-0pt}
\fontsize{8}{0}\selectfont
\begin{align} \label{Eq:Action03}
{a} &=  - \frac{1}{2 c_3} B^{\T} [\nabla \sigma_\textsf{H}]^{\T} {w}_{\textsf{H}_0} - \frac{1}{2 c_3} B^{\T}  [ \nabla \varepsilon_{\textsf{H}_0}] \text{,} \\
\hat{a} &=  - \frac{1}{2 c_3} B^{\T} [\nabla \sigma_\textsf{H}]^{\T} \hat{w}_{\textsf{H}_0}  \text{.}
\end{align} \normalsize

\noindent Therefore, the error of approximating HJB by NNs is

\vspace{-0pt}\fontsize{8.0}{0}\selectfont  
\begin{align}
e_\textsf{H} \! &\delequal \hat{\textsf{H} }  \!-\! {\textsf{H} }  \nonumber \\
&= 
\frac{1}{2 c_3}  \tilde{w}_{\textsf{H}_0}^{\T}  [\nabla \sigma_\textsf{H}] B B^{\T} [\nabla \sigma_\textsf{H}]^{\T} {w}_{\textsf{H}_0} \!-\! \frac{1}{2} \! \left[\sum_{k=1}^{N} \textup{tr}(G G^{T}  [\Delta  \sigma_\textsf{H}^{[k]}]  ) {\mathbf{e}}_{k} \right]^{\T} \! \tilde{w}_{\textsf{H}_0}
 \nonumber \\ &~ -\! \tilde{w}_{\textsf{H}_1}^\T \sigma_\textsf{H}  \!-\!  \tilde{w}_{\textsf{H}_0}^{\T} [\nabla \sigma_\textsf{H}] f \!-\! \frac{1}{4 c_3}  \tilde{w}_{\textsf{H}_0}^{\T}  [\nabla \sigma_\textsf{H}] B B^{\T} [\nabla \sigma_\textsf{H}]^{\T} \tilde{w}_{\textsf{H}_0}  \!-\! \varepsilon_{\textsf{H} }  \text{,} \label{Eq:HJBerr}
\end{align} \normalsize

\vspace{-5pt}
\noindent where \small$ \tilde{w}_{\textsf{H}_0} \!\delequal\! {w}_{\textsf{H}_0} \!-\! \hat{w}_{\textsf{H}_0} $\normalsize, and \small$ \tilde{w}_{\textsf{H}_1} \!\delequal\! {w}_{\textsf{H}_1} \!-\! \hat{w}_{\textsf{H}_1} $\normalsize.
The optimal weights \small$ {w}_{\textsf{H}_1} $\normalsize\, and \small$ {w}_{\textsf{H}_0} $\normalsize\, should minimize the loss function defined as

\vspace{-0pt}
\fontsize{8}{0}\selectfont
\begin{align}
J_\textsf{H}(\hat{w}_{\textsf{H}_0}, \hat{w}_{\textsf{H}_1}) &\!\delequal\! \frac{1}{2} e_\textsf{H}^{\T} e_\textsf{H} \!+\! c_{\hspace{.5pt}\textsf{H}} \underbrace{\max\l\{ 0 , \dot{L}_s  \r\} \! \mathbbm{1}_{\| s_i(t) \|\geq s_\text{dest}} }_{R_i}  
\end{align} \normalsize

\vspace{-5pt}
\noindent where $c_{\hspace{.5pt}\textsf{H}}$ is a positive constant, $ {L}_s $ as the simplified notation of $ {L}_s(s_i(t)) $ is a Lyapunov candidate function, and $ \dot{L}_s $ is its derivative with respect to time. The regularizer term shown as as $ R_i $ or $ R_i(t) $ is meant to stop the movement when reaching the destination, i.e., \small$\| s_i(t) \| = \| [r_i(t)^\T, v_i(t)^\T]^\T \|  \! \leq \! \| s_{\text{dest}} \| $\normalsize.
Then, by discretizating the  time with small $ \text{d}t $ steps, the gradient descent updates are written as

\vspace{-0pt} \fontsize{8}{0}\selectfont
\begin{align}\label{Eq:HJB_update0}
\hat{w}_{\textsf{H}_0}{\!(n\!+\!1)}\! &=\! \hat{w}_{\textsf{H}_0}{\!(n)} \! -\!  \mu_{\hspace{.5pt}\textsf{H}} ( {  \nabla_{\hat{w}_{\textsf{H}_0}} \!e_\textsf{H}}) e_\textsf{H} \!-\! \mu_{\hspace{.5pt}\textsf{H}} c_{\hspace{.5pt}\textsf{H}} \! \nabla_{\hat{w}_{\textsf{H}_0}} \! R_i ,  \\
\hat{w}_{\textsf{H}_1}{\!(n\!+\!1)}\! &=\! \hat{w}_{\textsf{H}_1}{(n)}  \!-\!  \mu_{\hspace{.5pt}\textsf{H}} ( {  \nabla_{\hat{w}_{\textsf{H}_1}} \! e_\textsf{H}}) e_\textsf{H}, \label{Eq:HJB_update1}
\end{align}\normalsize
where the gradients $ {  \nabla_{\hat{w}_{\textsf{H}_0}} e_\textsf{H}} $ and $ {  \nabla_{\hat{w}_{\textsf{H}_1}} e_\textsf{H}} $ are obtained as

\vspace{-5pt} \fontsize{8}{0}\selectfont
\begin{align}
{  \nabla_{\hat{w}_{\textsf{H}_0}} e_\textsf{H}} \!=
& \!\frac{1}{2 c_3}   [\nabla \sigma_\textsf{H}] B B^{\T} [\nabla \sigma_\textsf{H}]^{\T} \tilde{w}_{\textsf{H}_0}  \!-\! \frac{1}{2 c_3}   [\nabla \sigma_\textsf{H}] B B^{\T} [\nabla \sigma_\textsf{H}]^{\T} {w}_{\textsf{H}_0} 
\nonumber\\ 
& 
+[\nabla \sigma_\textsf{H}] f + \frac{1}{2} \left[\sum_{k=1}^{N} \textup{tr}(G G^{T}  [\Delta  \sigma_\textsf{H}^{[k]}]  ) {\mathbf{e}}_{k} \right] ,  \\
{  \nabla_{\hat{w}_{\textsf{H}_1}} e_\textsf{H}} =& \sigma_\textsf{H} .
\end{align}\normalsize

The corresponding \textsf{Hjb} learning control based on these update equations is described in Algorithm 1. After initialization of the weights, each UAV has to collect instantaneous states of other UAVs and use it to update the equations  \eqref{Eq:HJB_update0} and \eqref{Eq:HJB_update1}. Then, it  uses the updated model to take the proper action. However, the stability of this algorithm is explored by the following Proposition 1.

\begin{algorithm}[t] \label{alg:01}
	\caption{\textsf{Hjb} control}
	\begin{algorithmic}[1]
		\STATE \textbf{Initialization:} \small $\hat{w}_{\textsf{H}_0}{(0)} = 0$ and $ \hat{w}_{\textsf{H}_1}{(0)} =0 $\normalsize.
		
		\FOR {Each UAV $ i = 1, \ldots, N $,  \textbf{in parallel},}
			
		\STATE \textit{Collect} the states $s_{-i}(t)$ from neighboring UAVs.
		
		\STATE \textit{Update} the weights $\hat{w}_{\textsf{H}_0}{(n)}$ and $ \hat{w}_{\textsf{H}_1}{(n)}  $ by \eqref{Eq:HJB_update0} and \eqref{Eq:HJB_update1}.
		
		\STATE \textit{Calculate} the value $\hat{\psi}  =  \hat{w}_{\textsf{H}_0}^\T  \sigma_\textsf{H} $.
		
		\STATE \textit{Take} the optimal action $\hat{a} \!=\!  - \frac{1}{2 c_3} B^{\T} [\nabla \sigma_\textsf{H}]^{\T} \hat{w}_{\textsf{H}_0}  $.
		
		\ENDFOR

	\end{algorithmic} 
\end{algorithm}

\noindent
\textbf{Proposition 1 (HJB Lyapunov stability):}  
For small uncertainty of interaction term, i.e., \small$ \| \varepsilon_{\phi_G} \| \! \ll \! 1 $\normalsize, and a bounded interaction term, i.e., \small$ \| \hat{\phi}_G \| \leq M_1 $\normalsize,
the system state and the model weights of constructed adaptive HJB neural network obtained by Algorithm 1 are uniformly ultimately bounded (UUB), i.e., there exist $ s_\text{dest} $, $ w_0 $, and $ w_1 $ at time $ T $ such that \small$ \| s(t) \| \!\leq\! s_\text{dest}  $\normalsize, \small$ \| {w}_{\textsf{H}_0}\!(t) - \hat{w}_{\textsf{H}_0}\!(t) \| \!\leq\!  w_0  $\normalsize, and \small$ \| {w}_{\textsf{H}_1}\!(t) - \hat{w}_{\textsf{H}_1}\!(t) \| \!\leq  \! w_1 ~ $\normalsize for all \small$ t \geq T + T' $\normalsize. 
 
\begin{proof}
	See Appendix \ref{APP:A}.
\end{proof}

\subsection{ MFG Learning Control} 
Here, we find an approximate solution to the pair of HJB-FPK equations in MFG framework. Regarding the HJB equation, we  follow the method explained in previous subsection and by considering that the interaction term is obtained using \eqref{Eq:InterAction03}. 
Then, we follow the similar approximation procedure to approximate the solution for FPK equation. Let us first rewrite the FPK equation \eqref{Eq:FPK_02} by using the simplified notations in Table I, and define $ \textsf{F} $ as

\vspace{-5pt} \fontsize{8}{0}\selectfont
\begin{align}  \label{Eq:FPK_03}
\textsf{F}  \delequal \dot{m} + \nabla \cdot [(f- \frac{1}{2 c_3} B B^{\T} \nabla \psi  ) m]  - \frac{1}{2}\textup{tr}(G G^{\T} \Delta  m) = 0
\end{align}\normalsize

Using the equality $ \nabla \cdot [a \vec{b}] = a \nabla \cdot \vec{b} + \vec{b}^\T \nabla a $, where $ \vec{b} $ is a vector and $ a $ is scalar,  we rewrite this FPK equation as

\vspace{-5pt} \fontsize{8}{0}\selectfont
\begin{align}  \label{Eq:FPK_04}
\textsf{F} & = \dot{m} +  m \nabla \cdot [(f- \frac{1}{2 c_3} B B^{\T} \nabla \psi  ) ] 
\nonumber \\
& \quad 
+ [(f- \frac{1}{2 c_3} B B^{\T} \nabla \psi  ) ]^{\T} \nabla m  - \frac{1}{2}\textup{tr}(G G^{\T} \Delta  m) = 0
\end{align}\normalsize
Now, we seek to find an approximate solution to the equation \eqref{Eq:FPK_04}. Let us define the linear function approximator \small$ \hat{m}(s,t) $\normalsize, which approximates the density function \small$ m^{*}(s,t) $\normalsize, as 

\vspace{-0pt}
\fontsize{8}{0}\selectfont
\begin{align} 
\hat{m}(s,t) \delequal  \hat{w}_{\textsf{F}_0}(t)^\T \sigma_\textsf{F}\!\(s\) \text{,} \\
\hat{\dot{m}}(s,t) \delequal  \hat{w}_{\textsf{F}_1}(t)^\T \sigma_\textsf{F}\!\(s\) \text{,}
\end{align} \normalsize
where \small$ \sigma_\textsf{F}\!\(s\) $\normalsize\, is a vector of linear or nonlinear functions, and  \small$ \hat{w}_{\textsf{F}_0}\!(t) $\normalsize\, and \small$ \hat{w}_{\textsf{F}_1}\!(t) $\normalsize\, are the approximation to the optimal weight functions  \small$ {w}_{\textsf{F}_0}\!(t) $\normalsize\, and \small$ {w}_{\textsf{F}_1}\!(t) $\normalsize\, respectively. Then, the errors of approximating the distribution function \small$ {m}(s,t) $\normalsize\, and its derivative \small$ {\dot{m}}(s,t) $\normalsize\, are
\small\begin{align}
\varepsilon_{\textsf{F}_0}(s,t) \delequal {m} (s,t) - {w}_{\textsf{F}_0}(t)^\T \sigma_\textsf{F}\!\(s \) \text{,} \label{Eq:FPKmodel_1} \\
\varepsilon_{\textsf{F}_1}(s,t) \delequal \dot{m} (s,t) - {w}_{\textsf{F}_1}(t)^\T \sigma_\textsf{F}\!\(s \)\text{.} \label{Eq:FPKmodel_2}
\end{align} \normalsize
Considering this definition, and notation simplifications of Table I,  the FPK equation \eqref{Eq:FPK_03} and its corresponding approximatation are written as

\vspace{-0pt}
\fontsize{8}{0}\selectfont
\begin{align}  
\textsf{F} &=  {w}_{\textsf{F}_0}^\T \sigma_\textsf{F} \nabla \cdot [(f- \frac{1}{2 c_3} B B^{\T} \nabla \hat{\psi}  )  ] 
+ [(f- \frac{1}{2 c_3} B B^{\T} \nabla \hat{\psi}  ) ]^{\T} [\nabla \sigma_\textsf{F}]^{\T} {w}_{\textsf{F}_0} 
\nonumber \\
& \quad
- \frac{1}{2} \left[\sum_{k=1}^{N} \textup{tr}(G G^{T}  [\Delta  \sigma_\textsf{F}^{[k]}]  ) {\mathbf{e}}_{k} \right]^{\T} {w}_{\textsf{F}_0} + {w}_{\textsf{F}_1}^\T \sigma_\textsf{F} + \varepsilon_{\textsf{F}} = 0 , 
\\ 
\hat{\textsf{F}} &=  \hat{w}_{\textsf{F}_0}^\T \sigma_\textsf{F} \nabla \cdot [(f- \frac{1}{2 c_3} B B^{\T} \nabla \hat{\psi}  )  ] 
+ [(f- \frac{1}{2 c_3} B B^{\T} \nabla \hat{\psi} ) ]^{\T} [\nabla \sigma_\textsf{F}]^{\T} \hat{w}_{\textsf{F}_0} 
\nonumber \\
& \quad
- \frac{1}{2} \left[\sum_{k=1}^{N} \textup{tr}(G G^{T}  [\Delta  \sigma_\textsf{F}^{[k]}]  ) {\mathbf{e}}_{k} \right]^{\T} \hat{w}_{\textsf{F}_0} + \hat{w}_{\textsf{F}_1}^\T \sigma_\textsf{F},
\end{align}\normalsize
where $ \varepsilon_{\textsf{F}} $ denotes the error of FPK equation caused by the NN, and it is defined as

\vspace{-0pt}
\fontsize{8}{0}\selectfont
\begin{align}
	\varepsilon_{\textsf{F}} \! &\delequal \! \nabla \cdot [(f \!-\! \frac{1}{2 c_3} B B^{\T} \nabla \varepsilon_{\psi} ) (\hat{w}_{\textsf{F}_0}^\T \sigma_\textsf{F} \!+\! \varepsilon_{\textsf{F}_0}) ] \!-\! \frac{1}{2}\textup{tr}( G G^{\T} \Delta  \varepsilon_{\textsf{F}_0} ) \!+\! \varepsilon_{\textsf{F}_1} \nonumber \\
	&\quad  \!+\! \varepsilon_{\textsf{F}_0} \nabla \cdot [(f- \frac{1}{2 c_3} B B^{\T} \nabla \hat{\psi}  )  ]  \!+\! [(f \!-\! \frac{1}{2 c_3} B B^{\T} \nabla \hat{\psi}  ) ]^{\T} [\nabla \varepsilon_{\textsf{F}_0}] \text{,}
\end{align} \normalsize
where \small$ \varepsilon_{\psi} $\normalsize\, is the uncertainty in finding \small$ \psi $\normalsize.
Therefore, the error of approximating FPK equation by neural networks is

\vspace{-0pt}
\fontsize{8}{0}\selectfont
\begin{align}
e_\textsf{F}  &\delequal \hat{\textsf{F}}  - {\textsf{F}}  \nonumber \\
&=  - \tilde{w}_{\textsf{F}_0}^\T \sigma_\textsf{F} \nabla \cdot [(f \!-\! \frac{B B^{\T} \nabla \hat{\psi}}{2 c_3}   )  ]   \!-\!   \tilde{w}_{\textsf{F}_0}^{\T} [\nabla \sigma_\textsf{F}] [(f \!-\! \frac{B B^{\T} \nabla \hat{\psi}}{2 c_3}  ) ] 
\nonumber \\
& 
~~~+ \frac{1}{2}  \tilde{w}_{\textsf{F}_0}^{\T}  \left[\sum_{k=1}^{N} \textup{tr}(G G^{T}  [\Delta  \sigma_\textsf{F}^{[k]}]  ) {\mathbf{e}}_{k} \right]  - \tilde{w}_{\textsf{F}_1}^\T \sigma_\textsf{F}  - \varepsilon_{\textsf{F}}, \label{Eq:FPKerr}
\end{align} \normalsize
where \small$ \tilde{w}_{\textsf{F}_0}\! \delequal \! {w}_{\textsf{F}_0} \!-\! \hat{w}_{\textsf{F}_0} $\normalsize, and \small$ \tilde{w}_{\textsf{F}_1} \!\delequal\! {w}_{\textsf{F}_1}\! - \!\hat{w}_{\textsf{F}_1} $\normalsize.
Based on these definitions, the optimal weights \small$ \hat{w}_{\textsf{F}_1} $\normalsize and \small$ \hat{w}_{\textsf{F}_0} $\normalsize\, should minimize the loss function defined as

\vspace{-0pt}
\fontsize{8}{0}\selectfont
\begin{align}
J_\textsf{F}(\hat{w}_{\textsf{F}_0}, \hat{w}_{\textsf{F}_1}) &= \frac{1}{2} e_\textsf{F}^{\T} e_\textsf{F} 
\end{align} \normalsize

Therefore, by discretizating the time with $ \text{d}t $ time steps,  the gradient descent updates for FPK weights are obtained as

\vspace{-0pt}
\fontsize{8}{0}\selectfont
\begin{align}\label{Eq:FPK_update0}
\hat{w}_{\textsf{F}_0}{\!(n\!+\!1)} \!&=\! \hat{w}_{\textsf{F}_0}{\!(n)} \! - \! \mu_{\hspace{.5pt}\textsf{F}} ( {  \nabla_{\hat{w}_{\textsf{F}_0}} \! e_\textsf{F}}) e_\textsf{F}   \\
\hat{w}_{\textsf{F}_1}{\!(n\!+\!1)} \!&=\! \hat{w}_{\textsf{F}_1}{\!(n)}  \!-\!  \mu_{\hspace{.5pt}\textsf{F}} ( {  \nabla_{\hat{w}_{\textsf{F}_1}} \! e_\textsf{H}}) e_\textsf{F} \label{Eq:FPK_update1}
\end{align}\normalsize
where the gradients  $ {  \nabla_{\hat{w}_{\textsf{F}_0}} e_\textsf{F}} $ and $ {  \nabla_{\hat{w}_{\textsf{F}_1}} \! e_\textsf{F}}  $ are calculated as

\vspace{-5pt}
\fontsize{8}{0}\selectfont
\begin{align}
{  \nabla_{\hat{w}_{\textsf{F}_0}} e_\textsf{F}} =& ~\sigma_\textsf{F} \nabla \cdot [(f \!-\! \frac{1}{2 c_3} B B^{\T} \nabla \hat{\psi}  )  ]  + [\nabla \sigma_\textsf{F}] [(f \!-\! \frac{1}{2 c_3} B B^{\T} \nabla \hat{\psi}  ) ] 
\nonumber \\
& 
- \frac{1}{2}  \left[\sum_{k=1}^{N} \textup{tr}(G G^{T}  [\Delta  \sigma_\textsf{F}^{[k]}]  ) {\mathbf{e}}_{k} \right] \\
{  \nabla_{\hat{w}_{\textsf{F}_1}} \! e_\textsf{F}} =& ~\sigma_\textsf{F}.
\end{align}\normalsize

However, based on these update pairs and update pairs for HJB equation, the \textsf{Mfg} learning algorithm is described as in Algorithm 2.
In this algorithm, first, the UAVs share their states $ s_{j}(0) $ at time $ t=0 $ to obtain the distribution of the population and initial samples. Then, the UAVs start collecting samples and and updating theirs weights until they reach the destination at time $ T = T_0 \text{d}t$. 

\begin{algorithm}[t] \label{alg:02}
	\caption{\textsf{Mfg} control}
	\begin{algorithmic}[1]
		\STATE \textbf{Initialization:} \small$\hat{m}(s,0) \!=\! \frac{1}{N} \sum_{j=1}^{N} \delta(s \!-\! s_{j}(0))$, $\hat{w}_{\textsf{H}_0}{(0)} = 0$, $ \hat{w}_{\textsf{H}_1}{(0)} =0 $, $\hat{w}_{\textsf{F}_0}{(0)} = 0$, $ \hat{w}_{\textsf{F}_1}{(0)} =0 $ \normalsize.
		
		\FOR {Each UAV $ i = 1, \ldots, N $,  \textbf{in parallel},}
		
		
			\FOR {$ n = 1,\ldots,T_0 $}
				\STATE \textit{Update}  weights $\hat{w}_{\textsf{H}_0}{(n)}$ and $ \hat{w}_{\textsf{H}_1}{(n)}  $ by \eqref{Eq:HJB_update0} and \eqref{Eq:HJB_update1}.

				\STATE \textit{Calculate}  value {\small$\hat{\psi}  =  \hat{w}_{\textsf{H}_0}^\T  \sigma_\textsf{H} $\normalsize}.
				
				\STATE \textit{Update}  weight $\hat{w}_{\textsf{F}_0}{(n)}$ and $ \hat{w}_{\textsf{F}_1}{(n)}  $ by \eqref{Eq:FPK_update0} and \eqref{Eq:FPK_update1}.
				
				\STATE \textit{Obtain}  MF distribution~{\small$\hat{m} =  \hat{w}_{\textsf{F}_0}^\T \sigma_\textsf{F}$\normalsize}.
				
			\ENDFOR
		
			\STATE \textit{Take} the optimal action \small$\hat{a} \!=\!  - \frac{1}{2 c_3} B^{\T} [\nabla \sigma_\textsf{H}]^{\T} \hat{w}_{\textsf{H}_0}  $\normalsize.

		\ENDFOR
		
	\end{algorithmic} 
\end{algorithm}

We have shown in our previous works \cite{shiri2019massive} that \textsf{Mfg} control algorithm provides better results in terms of energy consumption, communications cost, and flocking of UAVs when sufficient samples are used in model training. However, there are stability concerns  which is analyzed in this section. The MFG learning solution consists of two coupled NNs coined HJB NN and FPK NN, which should be stable to ensure stability of MFG learning. Proving stability of the coupled HJB-FPK equations is a challenging issue. To simplify the stability analysis, we decouple the HJB and FPK equations and separately obtain the stability conditions for each of them. Then  stability can be proved, when the initial conditions of HJB NN and FPK NN  meet the stability conditions required for each of them separately, and if the output of each HJB NN (and FPK NN) falls into the stability space of the other FPK NN (and HJB NN). Starting with the HJB NN, we obtained the stability condition for HJB NN in Proposition 1. In Propositions 2 and 3, we show the stability and convergence conditions required for FPK NN. Then we conclude the stability of MFG NN in Corollary 1.

\noindent
\textbf{Proposition 2 (FPK Lyapunov stability):} 
For almost certain $ \psi $, i.e., \small$ \| \varepsilon_{\psi} \| \! \ll \! 1 $\normalsize, and differentiable and bounded value function $ \psi $, i.e., \small$ \| \psi \| \leq M_2 $\normalsize,
the weights of constructed adaptive FPK neural network obtained in Algorithm 2, which is controlled by its corresponding HJB equation, are UUB,i.e., there exist $ w_2 $, and $ w_3 $ at time $ T $ such that \small$ \| {w}_{\textsf{F}_0}\!(t) - \hat{w}_{\textsf{F}_0}\!(t) \| \!\leq\!  w_2  $\normalsize, and \small$ \| {w}_{\textsf{F}_1}\!(t) - \hat{w}_{\textsf{F}_1}\!(t) \| \!\leq  \! w_3 ~ $\normalsize for all \small$ t \geq T + T' $\normalsize. 
\begin{proof}
	See Appendix \ref{APP:B}.
\end{proof}

\noindent
\textbf{Proposition 3 (FPK Convergence):}
Under the assumptions of Proposition 2 and with small step-sizes $ \mu_{\hspace{.5pt}\textsf{F}}  $, the weights of FPK neural network function approximator converges to its optimal weights in mean with no bias and it is stable in mean square deviation sense. 
\begin{proof}
	See Appendix \ref{APP:C}.
\end{proof}

\textbf{Corollary 1:} 
Considering Propositions 1, 2,  and  3, we can conclude that 
the system state and weights of constructed HJB-FPK neural networks obtained by Algorithm 2 are UUB. 
\begin{proof}
	The Algorithm 2 has two parts as HJB part and FPK parts.
	It is initialized by the states of the UAVs at the source region. At the initial iterations the states are used directly in the algorithm to update the weights of HJB and FPK neural networks, and hence the uncertainty of interaction term is small, i.e., \small$ \| \varepsilon_{\phi_G} \| \! \ll \! 1 $\normalsize. Also, by starting with well trained or zero initialized neural network weights, both the interaction term and value functions are upper-bounded, i.e., \small$ \| \hat{\phi}_G \| \leq M_1 $\normalsize\, and \small$ \| \psi \| \leq M_2 $\normalsize. Hence, there is a design, i.e., a choice of parameters in proofs in Appendices \ref{APP:A}, \ref{APP:B}, and \ref{APP:C}, such that all assumptions necessary for Propositions 1, 2, and 3 hold together and completely
\end{proof}

\section{Federated MFG Learning - Model Sharing Methods} \label{SE:05}
In this section, we propose federated  mean field game learning strategy (\textsf{MfgFL}) and its different implementations, i.e.,  \textsf{MfgFL-H},  \textsf{MfgFL-F}, and \textsf{MfgFL-HF}, to make the UAVs' control models close to each other and to use  sample diversity among the UAVs efficiently. In \textsf{MfgFL-H} the model parameters of HJB neural network are shared with central unit to obtain the global HJB NN model, so the action rules of UAVs are close to each other. In \textsf{MfgFL-F} the FPK NN models are transmitted to the central unit to obtain the global FPK NN model, so the estimation of the population density function at UAVs be more accurate. In \textsf{MfgFL-HF}, both HJB and FPK neural network models are averaged to obtain better global online MFG learning model. In the following, we explain general form of \textsf{MfgFL} strategy, which covers three different implementations. 

Although Algorithm 2 can reduce the communications cost of the control algorithm by leveraging the MFG framework, it still requires big sample sets to train and provide conditions of stability. In other words, there is still a need to share a subset of samples among the UAVs or with a central unit, which requires extra {communication} costs in addition to {privacy} concerns. Therefore, instead of state sharing, we adopt the  federated learning method to address these issues.

In the \textsf{MfgFL} algorithm, one UAV out of all is set to act as a control center, which we call is as \textbf{leader (or header)} UAV. This leader UAV depending on the application may be chosen randomly or considering UAVs power consumption or flight time, which is beyond the scope of this work. Then, we simply set one of the UAVs as the leader, i.e., $\uav_h$, and indicate it by index $ h $ in the algorithm. 

\begin{algorithm}[t] \label{alg:03}
	\caption{\textsf{MfgFL} control}
	\begin{algorithmic}[1]
		\STATE \textbf{Initialization:} \small$\hat{m}(s,0) \!=\! \frac{1}{N} \sum_{j=1}^{N} \delta(s \!-\! s_{j}(0))$, $\hat{w}_{\textsf{H}_0}{(0)} = 0$, $ \hat{w}_{\textsf{H}_1}{(0)} =0 $, $\hat{w}_{\textsf{F}_0}{(0)} = 0$, $ \hat{w}_{\textsf{F}_1}{(0)} =0 $\normalsize.
		
		\FOR {\small$ n = 0, 1, 2,  \ldots , T_0$\normalsize}
		
		\IF {$ n = k n_0  $}
		\STATE $ N_h $ UAVs, in parallel, \textit{send} their model \small$ \hat{w}_{i,d}^{}{(k n_0)} $\normalsize ~ to the leader. 
		
		\STATE Leader \textit{updates} the model parameters \small$\hat{w}_{\text{h},d}^{}{(k)}$\normalsize,  via \eqref{Eq:FL_AVG}.

		\STATE Leader \textit{broadcasts} the model  \small$ \hat{w}_{\text{h},d}^{}{(k)}  $\normalsize.
		\ENDIF

		\FOR {each UAV \small$ i = 1, \ldots, N $\normalsize,  \textbf{in parallel},}
		
		\IF {UAV \small$ i $\normalsize\, receives  \small$ \hat{w}_{\text{h},d}^{}{(k)}  $\normalsize}
		
		\STATE \textit{Update} \small$ \hat{w}_{i,d}^{}{(n)} $\normalsize, as \small$ \hat{w}_{i,d}^{}{(n)} \leftarrow  \hat{w}_{\text{h},d}^{}{(k)}   $\normalsize.

		\ENDIF
		
		\STATE \textit{Update} \small$ \hat{w}_{\textsf{H}_0}{\!(n)} $\normalsize, \small$ \hat{w}_{\textsf{H}_1}{\!(n)} $\normalsize, \small$\hat{w}_{\textsf{F}_0}{(n)}$\normalsize, and \small$ \hat{w}_{\textsf{F}_1}{(n)}  $\normalsize\, by \eqref{Eq:HJB_update0} and \eqref{Eq:HJB_update1}, \eqref{Eq:FPK_update0}, and \eqref{Eq:FPK_update1}.
		
		\textit{Take} the optimal action \small$\hat{a} \!=\!  - \frac{1}{2 c_3} B^{\T} [\nabla \sigma_\textsf{H}]^{\T} \hat{w}_{\textsf{H}_0}  $\normalsize.
		
		\ENDFOR
		
		\ENDFOR
	\end{algorithmic} 
\end{algorithm}

The proposed \textsf{MfgFL} learning control is described in Algorithm 3.
Following the FedAvg algorithm, the leader collects models \small$ \hat{w}_{i,d}^{}{(n)} $\normalsize\, of $ N_h $ UAVs at times $ n \!=\! k n_0 $, where \small$ d \! \in \! \{ H, F, H\!F\} $\normalsize\, which corresponds to three types of implementations as 
\begin{itemize}
	\item In \textsf{MfgFL-H} (\small$ d \!=\! H $\normalsize), the model \small$ \hat{w}_{i,d}^{}{(n)} $\normalsize,  which is shared with the leader, is equal to the set \small$ \{ \hat{w}_{\textsf{H}_0}{\!(n)}, \hat{w}_{\textsf{H}_1}{\!(n)} \}  $\normalsize\, at the  UAV $ i $. However, the FPK models \small$ \hat{w}_{\textsf{F}_0}{\!(n)}$, and  $ \hat{w}_{\textsf{F}_1}{\!(n)}  $\normalsize\, are not shared in this implementation.

	\item In \textsf{MfgFL-F} (\small$ d \!=\! F $\normalsize), the model \small$ \hat{w}_{i,d}^{}{(n)} $\normalsize,  which is shared with the leader, is equal to the set \small$ \{ \hat{w}_{\textsf{F}_0}{\!(n)}, \hat{w}_{\textsf{F}_1}{\!(n)} \}  $\normalsize\, at the  UAV $ i $. However, the HJB models \small$ \hat{w}_{\textsf{H}_0}{\!(n)}$, and  $ \hat{w}_{\textsf{H}_1}{\!(n)}  $\normalsize\, are not shared in this implementation.
	\item In \textsf{MfgFL-HF}, (\small$ d \!=\! H\!F $\normalsize), the model \small$ \hat{w}_{i,d}^{}{(n)} $\normalsize, which is shared with the leader, is equal to the set \small$ \{ \hat{w}_{\textsf{F}_0}{\!(n)}, \hat{w}_{\textsf{F}_1}{\!(n)}, \hat{w}_{\textsf{H}_0}{\!(n)}, \hat{w}_{\textsf{H}_1}{\!(n)} \}  $\normalsize\, at the UAV $ i $. In other words, the complete model of MFG is shared with the leader.
\end{itemize}

It should be noted that the model parameters in different implementations of \textsf{MfgFL} are only shared between the leader and swam of UAVs, and only the models are transmitted rather than raw data samples. This saves significant communication energy as we will see  in Section \ref{SE:06}.
At each iteration \small$n$\normalsize\, in Algorithm 3, the leader obtains the average model $ \hat{w}_{\text{h},d} $ after collecting models \small$ \hat{w}_{i,d}^{}{(k n_0)} $\normalsize\, from \small$ N_h $\normalsize\,  UAVs of the swarm as 

\vspace{-5pt}
\fontsize{8}{0}\selectfont
\begin{align} 
		\hat{w}_{\text{h},d}^{}{(k)} \leftarrow \frac{1}{N_h} \sum_{i \in \mathcal{N}_h}^{} \hat{w}_{i,d}^{}{(k n_0)}, \label{Eq:FL_AVG}
\end{align}  \normalsize
However, after the average model is calculated at the leader and broadcasted to the UAVs, the updates are done locally by the set local samples $ S_i $ at each UAV $ i $, which is the set of local states sampled from the mission starting time to the current time \small$ t=n \text{d}t $\normalsize, i.e., \small$ S_i = \{s_i(j)|j=0,\ldots,n \}$\normalsize.
This procedure is repeated until all the UAVs reach the destination at repetition $ T_0 $. 

One main benefit of this procedure is that the UAVs do not need to exchange their local states to update the HJB-FPK models, and communication cost is reduced compared to \textsf{Mfg} and \textsf{Hjb} control methods. An approximate communications payload up to time \small$ t=n \text{d}t $\normalsize\, for the \textsf{MfgFL} methods is \small$ N \times \frac{n}{n_0} \times L(\hat{w}_{i,d}) \times b $\normalsize, where \small$ L(\hat{w}_{i,d}) $\normalsize\, is the size of \small$ \hat{w}_{i,d} $\normalsize\, and \small$ b $\normalsize\, is the resolution in bits, and for \textsf{Mfg} and \textsf{Hjb} methods, it is \small$  N \times n \times L(s_i) \times b \times N_\text{s} $\normalsize, where \small$ L(s_i) $\normalsize\, is the size of states and \small$  N_\text{s} $\normalsize\, is the number of samples at each time interval \small$  \text{d}t $\normalsize. Then, the \textsf{Mfg} control requires less \small$  N_\text{s} $\normalsize\, than \textsf{Hjb} which corresponds to smaller communications cost of \textsf{Mfg} control, and the \textsf{MfgFL} requires smaller payload than \textsf{Mfg} control because \small$  N_\text{s} \gg 1 $\normalsize\, and \small$ n_0 \gg 1 $\normalsize. This result will be evaluated in Section \ref{SE:06}.
In addition, it is not always safe to share the state information, and privacy is also preserved for the UAVs with \textsf{MfgFL} method.

In addition to reducing communication costs  and increasing the privacy of the UAVs, the \textsf{MfgFL} method can provide other benefits as well such ensuring stability conditions for MFG framework and increasing training speed.
One major condition for the MFG based approach is that the UAVs are indistinguishable. It means that the UAVs should have the same action rule, and hence, it is reasonable that they are trained by big enough samples. Nonetheless, due to energy/bandwidth limitations, it is not possible to provide this huge samples for model training. From this viewpoint, FL-based approaches can increase the model similarity among the UAVs and make them indistinguishable  by efficiently using their samples for training. 

Another benefit of using the \textsf{MfgFL} approach is the increased training speed of the models at UAVs. This is closely related to the communication cost of the algorithm, since utilizing model averaging means that the algorithm benefit from the various sample of UAVs in a shorter time span. Therefore, it is safe to say that it can provide higher model training speed.
However, the performance of \textsf{MfgFL} is explored in next section.

\section{Numerical Results}

 \label{SE:06}
\begin{figure*}[t]
	\centering
	\setlength\abovecaptionskip{-0.0\baselineskip}
	\subfigure
	{\hspace{-13pt}\includegraphics[width = 1.0\textwidth]{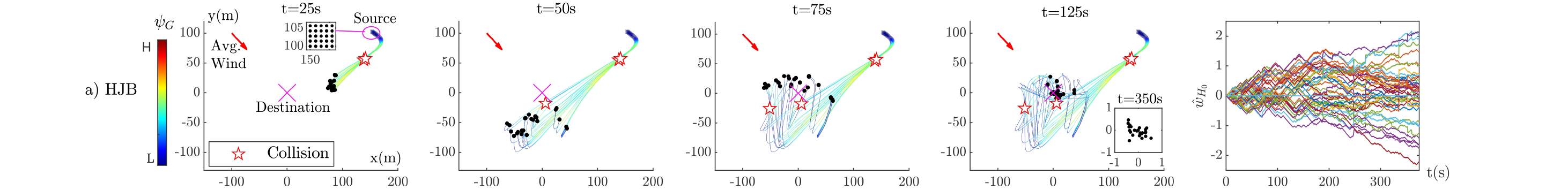}}\vskip -5pt
	\subfigure
	{\hspace{-13pt}\includegraphics[width = 1.0\textwidth]{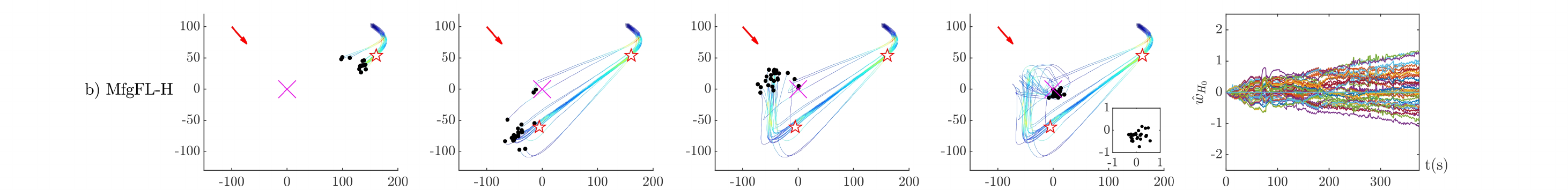}}\vskip -5pt
	\subfigure
	{\hspace{-13pt}\includegraphics[width = 1.0\textwidth]{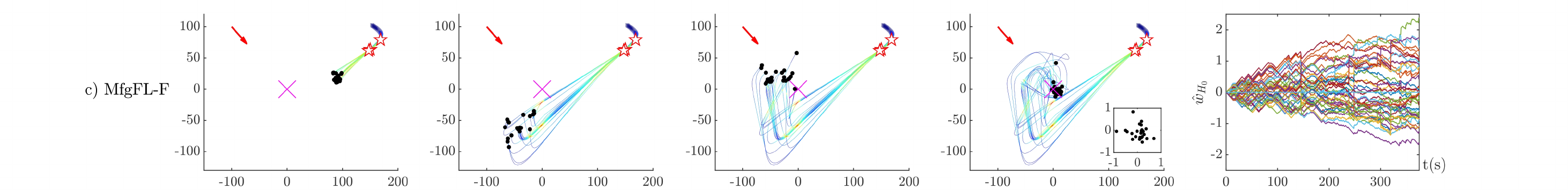}}
	\subfigure
	{\hspace{-13pt}\includegraphics[width = 1.0\textwidth]{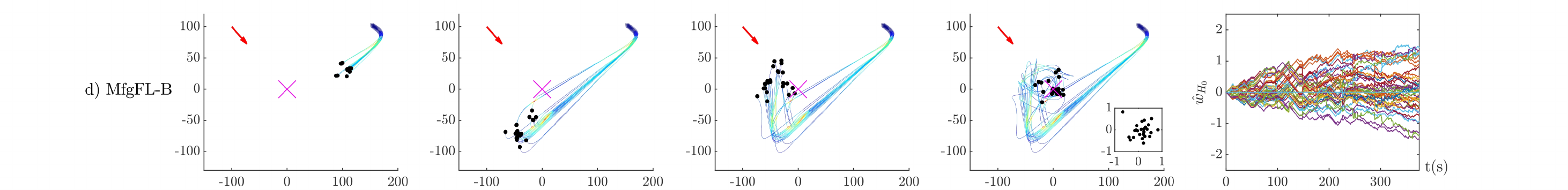}}
	\caption{Trajectory snapshots (left, $4$ subplots for each control method) of 25 UAVs under (a)  \textsf{Hjb}: HJB learning control with the communication range $d=100\text{m}$, (b) $\textsf{MfgFL-H}$: MFG learning control with HJB model averaging, (c) $\textsf{MfgFL-F}$: MFG learning control with FPK model averaging, and (d) $\textsf{MfgFL-HF}$: MFG learning control with both HJB and FPK model averaging. During the travel time $t=0\!\sim\!125$s, $\textsf{MfgFL-HF}$ shows the best flocking behavior and the most stable HJB model parameters $w_{1,\textsf{H}}$ (rightmost subplot for each control method) of a randomly selected reference UAV $\uav_1$. Consequently, $\textsf{MfgFL-HF}$ yields no collision during its entire travel, in sharp contrast to the others.}
	\label{fig:01} \vskip -5pt
\end{figure*}

In this section, we numerically validate the effectiveness of the proposed algorithm \textsf{MfgFL-HF} compared to the baseline methods \textsf{Hjb} control, \textsf{MfgFL-F}, and \textsf{MfgFL-H}, in terms of {travel time, motion energy, collision avoidance, and communications cost}. Throughout the simulations, we consider $N$ UAVs controlled in a two-dimensional plane at the fixed altitude of $ \mathsf{h} = 40\text{m}$. Initially, the UAVs are equally separated with the distance $ \sqrt{2} $m each other, and located at a source, which is a square region centered at (150, 100)m in a 2-dimensional plane (see \figurename{ \ref{fig:01}}-a). Each UAV aims to reach the destination at the origin, under the wind dynamics described by \small$ V_o = \sigma_\text{wind} I $\normalsize\, and \small$ v_o = (1, -1)\text{m/s} $\normalsize\, {(see Sec. \ref{SE:02})}.
 
Following \cite{b14,shiri2019massive}  single hidden layer models \eqref{Eq:HJBmodel_1} and \eqref{Eq:HJBmodel_2} are considered for HJB model, where each hidden node's activation function, i.e., \small$ {\sigma}_{\textsf{H},j}\!\({s_i}(t)\)  $\normalsize\, for \small$ j = 1, \cdots, M_\textsf{H} $\normalsize, corresponds to each scalar term in a polynomial expansion. The polynomial for \small$\sigma_\textsf{H}({s_i}(t))$\normalsize\, is  heuristically chosen as: \small$(1+ x_i(t) + v_{x,i}(t))^6 + (1+ y_i(t) + v_{y,i}(t))^6$\normalsize,  where \small${r}_i(t) = [x_i(t),y_i(t)]^{\T} $\normalsize\, and \small$ {v}_i(t) = [v_{x,i}(t),v_{y,i}(t)]^{\T}$\normalsize, thus the model size for HJB model is \small$ M_\textsf{H}\!=\!  54 $\normalsize.

For the MFG based methods, the same neural network structure described above is considered to approximate  HJB model. In a similar way,  single hidden layer models \eqref{Eq:FPKmodel_1} and \eqref{Eq:FPKmodel_2}  are considered for FPK model, where each hidden node's activation function, i.e., \small$ {\sigma}_{\textsf{F},j}\!\({s_i}(t)\)  $\normalsize\, for \small$ j = 1, \cdots, M_\textsf{F} $\normalsize, corresponds to each scalar term in a polynomial expansion. The polynomial for \small$\sigma_\textsf{F}({s_i}(t))$\normalsize\, is  heuristically chosen as: \small$(1+ x_i(t) + v_{x,i}(t))^6 + (1+ y_i(t) + v_{y,i}(t))^6$\normalsize. Thus the model size for FPK model is \small$ M_\textsf{F}\!=\!  69 $\normalsize.

Unless otherwise stated, the default simulation parameters are: \small{$P_o= 20\text{dBm}$, $W_o=2$MHz, $\sigma_n =-110$ dBm/Hz,  $ \alpha = 0 $, $ \chi = 1.347 $, $ \xi = 6.649 $; $ \sigma_\text{wind}=0.1 $; 
$ N_h = 0.8 N $; $ r_\text{coll} = 0.1 $m, $ r_\text{C} = \sqrt{2}/2 $m; $c_0=0.1$, $c_1 = c_2 = 0.015$, $c_3 = 0.005 $, $\mu_\textsf{H} = \mu_\textsf{F} = 0.01$, $ c_\textsf{H} = 0.5 $, $ n_0  = 100 $ for \textsf{MfgFL-H} and \textsf{MfgFL-F}, $ n_0  = 200 $ for \textsf{MfgFL-HF}, {\normalsize and} \small$ \text{d}t = 0.1\text{s} $}\normalsize~
for the purpose of discretizing time in simulations.
In addition the physical characteristics of UAVs are \small$ \lambda_0 = 0.0049 $, $ \lambda_1 = 0.0887 $ and $ \lambda_2 = 0.0092 $\normalsize, \small$ \omega_\text{tip} = 15$\normalsize m/s, and \small$ \chi_{o} = 1.6120 $\normalsize m/s.

{\figurename{ \ref{fig:01}} shows the trajectories of $ N=25$ UAVs under \textsf{Hjb}, \textsf{MfgFL-H}, \textsf{MfgFL-F}, and \textsf{MfgFL-HF} control methods. With \textsf{Hjb} control, all the UAVs should communicate instantaneous states with each other, and use the received states to update their local HJB model. Therefore, \textsf{Hjb} control is extremely costly to be implement in real-time. However, for comparison purposes, it is assumed that the UAVs communicate their states at each time step to calculate the instantaneous interaction term, but the processing at each UAV is limited to one update of \eqref{Eq:HJB_update0} and \eqref{Eq:HJB_update1} per time step. This results in a fair comparison with FL-based methods, as they are also limited to one update of \eqref{Eq:HJB_update0} and \eqref{Eq:HJB_update1} per time step.

{In all the methods, at first, the untrained UAVs follow the average wind direction while they train the models until the models are trained  to the extend that their output commands  turn the UAVs towards the destination.  Then, the differences among algorithms in terms of collision, model weights, and interaction terms become observable from the trajectory and model weight plots, as explained in the following.}

{Collision occurrences is shown by star marks in the trajectories. 
It can be seen that in the proposed \textsf{MfgFL-HF} method, no collision has happened thanks to more sample utilization for HJB and FPK model training by adopting FL averaging for both models.
Unlike \textsf{MfgFL-HF}, only  one of the HJB or FPK models in \textsf{MfgFL-H} and \textsf{MfgFL-F} methods is trained with enough samples by utilizing FL method.
The less-trained model results in more collisions of \textsf{MfgFL-H} and \textsf{MfgFL-F} methods as seen in the trajectory plots {\figurename{ \ref{fig:01}}-b and {\figurename{ \ref{fig:01}}-c.
In \textsf{Hjb} method, although enough samples are provided, the UAVs can not use them to train the model in real-time due to limited processing power of the UAVs. 
Therefore, the models are not trained with enough samples, and a few collisions occur on the path to the destination as seen in the trajectory plots {\figurename{ \ref{fig:01}}-a. These training behaviors can also  be seen in  HJB model parameters on the most right side of the \figurename{ \ref{fig:01}}, where in comparison to the other control methods,  the model parameters in \textsf{MfgFL-HF} are less divergent after a period of time.

{\figurename{ \ref{fig:01}} shows the interaction term $ \phi_G $ for each UAV using the color map on the trajectories. The bluer trajectories of \textsf{MfgFL-HF} method compared to other methods indicates lower interaction term values and better alignment of UAVs on the path to the destination. The reason is better training of the models in the proposed method as explained above. One main benefit of flocking of the UAVs instead of traveling individually or in different clusters is that it results in better communication channels among the UAV due to shorter distances, which can help in better model training and control. Further features of the proposed  \textsf{MfgFL-HF} method corresponding to \figurename{ \ref{fig:01}} is explained below using \figurename{ \ref{fig:02}} and \figurename{ \ref{fig:05}}. }

\figurename{ \ref{fig:02}} represents the motion energy, communications payload, velocity alignment, number of collision risks, speed, and travel distance of the UAVs corresponding to the scenario and methods in \figurename{ \ref{fig:01}}.
\figurename{ \ref{fig:02}}-a represents the average motion energy and  its variance among the UAVs. The proposed method \textsf{MfgFL-HF} consumes at least $ 16\% $ less energy than the other methods, and requires at least 4 times less communication costs than \textsf{Hjb} method (see \figurename{ \ref{fig:02}}-b) at the cost of  $ 10\% $ and $ 6\% $  more average travel time \small$ T_{avg} $\normalsize\, compared to \textsf{MfgFL-F} and \textsf{MfgFL-FH}, respectively. 
The reason for less energy consumption of \textsf{MfgFL-HF} is that the UAVs can travel in a flock with smaller speed (\figurename{ \ref{fig:02}}-e) and smaller interaction term on the trajectory as we observed in \figurename{ \ref{fig:01}}. 
Furthermore, the reason for less communication costs for \textsf{MfgFL-HF}, \textsf{MfgFL-F}, and \textsf{MfgFL-H} methods is due to adopting the FL method. In FL-based methods, at every $ n_0 = 100 $ time steps, $ 80\% $ of the  UAVs transmit their models to the leader and the leader broadcasts it to all the UAVs. However, in \textsf{Hjb} method, all $ 25 $ UAVs broadcasts their states to all the neighbor UAVs at each time step.

Despite the disadvantage of more travel time for the \textsf{MfgFL-HF} method in this scenario,  it demonstrates better velocity alignment and collision avoidance properties than the other defined FL-based methods as shown in \figurename{ \ref{fig:02}}-c and \figurename{ \ref{fig:02}}-d. 
As explained in definition of metrics $ \phi_\text{A}(t) $  in \eqref{Eq:VelocityAlignment} and $ \phi_\text{C}(t) $ \eqref{Eq:NoCollision}, their smaller values corresponds to better flocking behavior and lower probability of collision occurrence, respectively.
Clearly, the cumulative value of $ \phi_\text{A}(t) $ at time $ T_{avg} = 175 $s for \textsf{MfgFL-HF} method is at least $ 7\% $ less than the other methods, which means better velocity alignment of \textsf{MfgFL-HF}.
Furthermore, the cumulative value of  $ \phi_\text{C}(t) $ at time $ T_{avg} = 175 $s for \textsf{MfgFL-HF} is at least $ 8\% $ less than the other methods, which means lower risk of collision occurrence of the proposed method. This complies with the training discussion above for  \figurename{ \ref{fig:01}}.

Furthermore, the average, maximum and minimum speed of the UAVs are shown in \figurename{ \ref{fig:02}}-e and \figurename{ \ref{fig:02}}-f.
The obtained maximum speed for the mentioned algorithms are: \small$ 12.9 $\normalsize m/s for \textsf{MfgFL-HF}, \small$ 16.0 $\normalsize m/s for \textsf{MfgFL-F},  \small$ 14.3 $\normalsize m/s  for \textsf{MfgFL-H},  \small$ 16.2 $\normalsize m/s  for \textsf{Hjb}. The lower maximum speed of \textsf{MfgFL-HF} is because of less interaction among the UAVs which is a direct result of better flocking. The better flocking is obtained at the cost of increasing the travel distance about $ 10\% $ compared to the \textsf{Hjb} as shown in \figurename{ \ref{fig:02}}-f.
This is also consequently because of better model training of \textsf{MfgFL-HF} by utilizing FL method.

\begin{figure}[t]
	\centering
	\setlength\abovecaptionskip{-0.0\baselineskip}
	\subfigure
	{\hspace{-13pt}\includegraphics[trim=0.0cm 1cm 0cm 1cm, width = 0.2\textwidth]{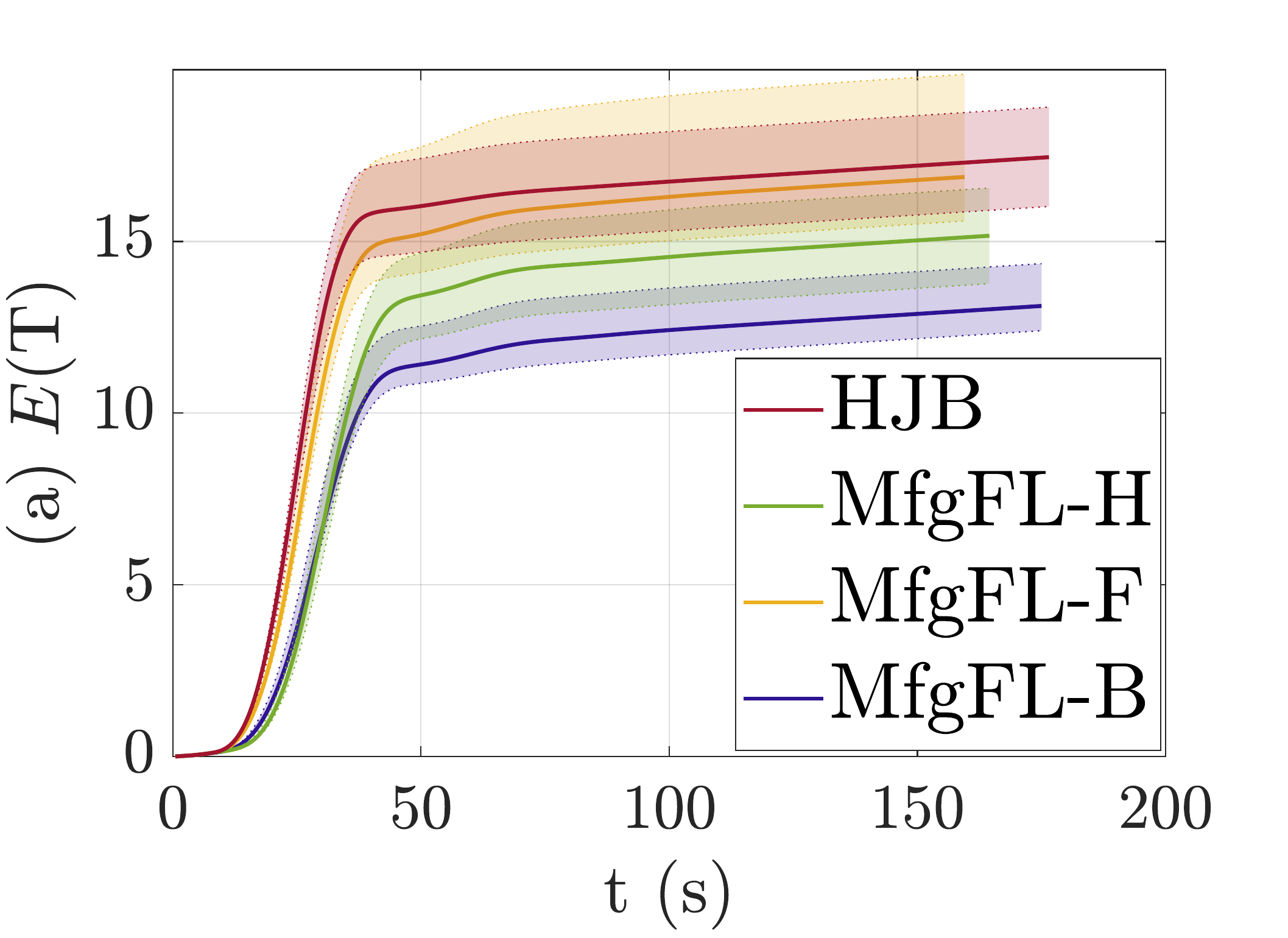}} \quad
	\subfigure
	{\hspace{-13pt}\includegraphics[trim=0.0cm 1cm 0cm 1cm, width = 0.2\textwidth]{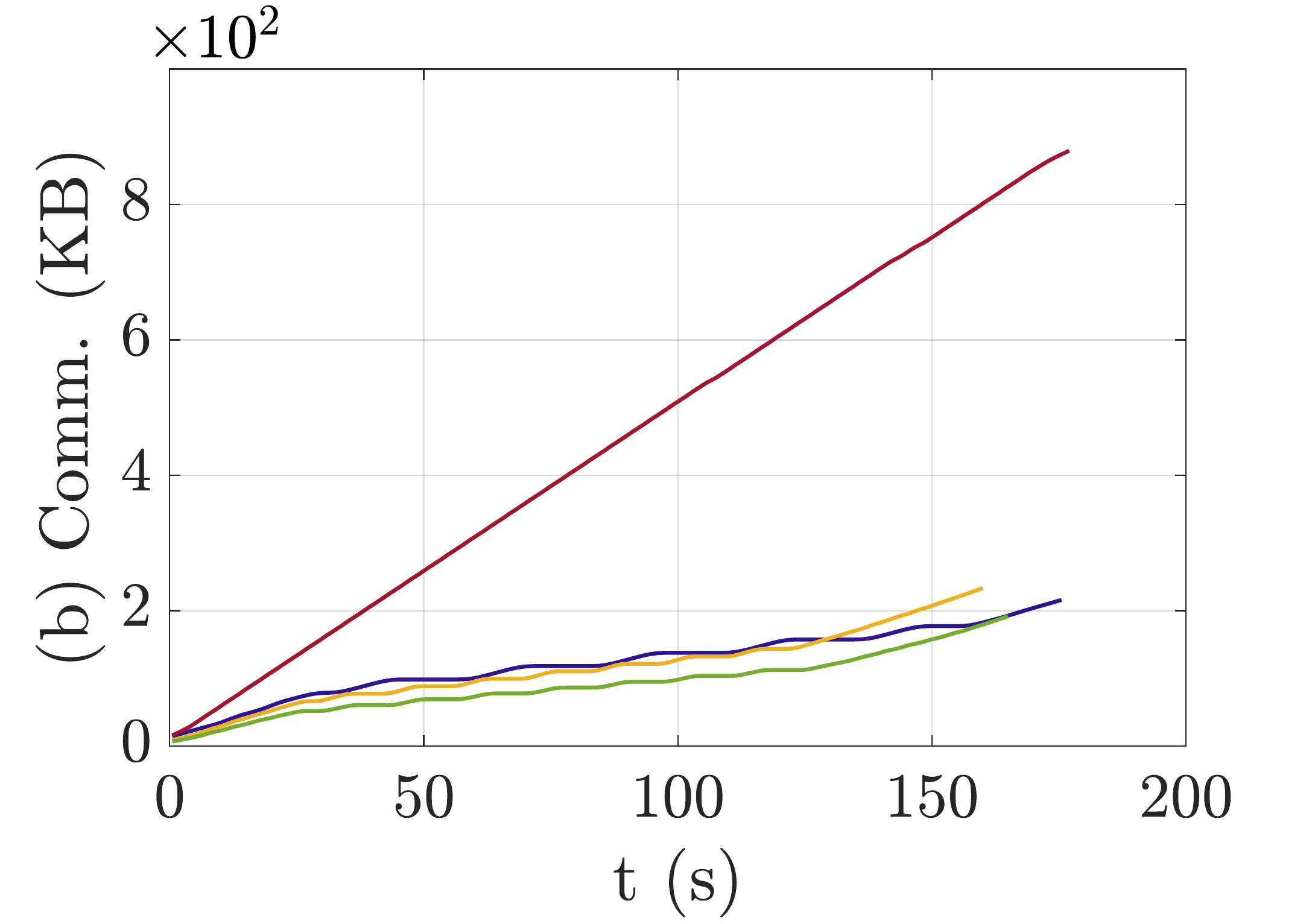}} \quad \\
	\subfigure
	{\hspace{-13pt}\includegraphics[trim=0.0cm 1cm 0cm 1cm, width = 0.2\textwidth]{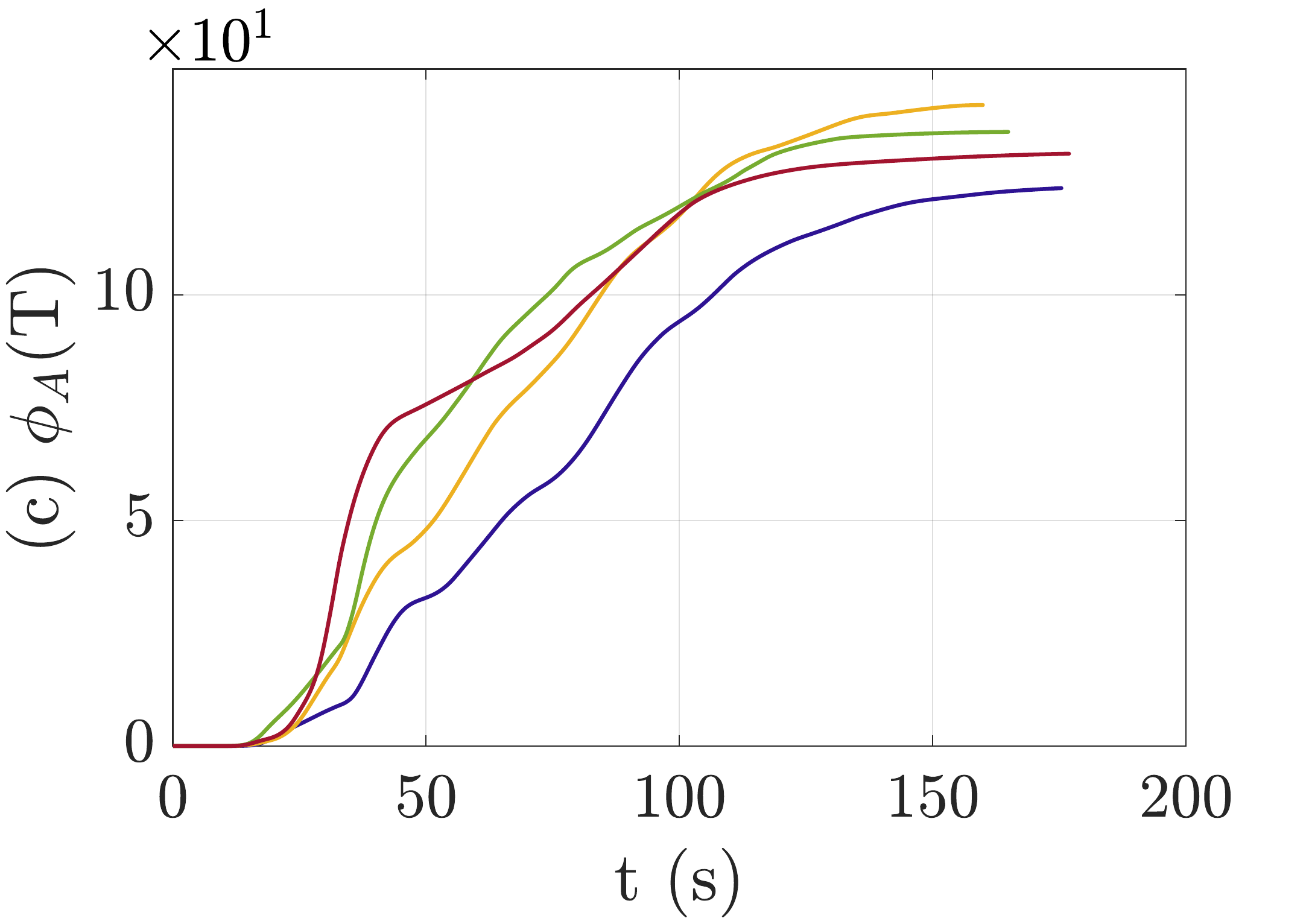}} \quad
	\subfigure
	{\hspace{-13pt}\includegraphics[trim=0.0cm 1cm 0cm 1cm, width = 0.2\textwidth]{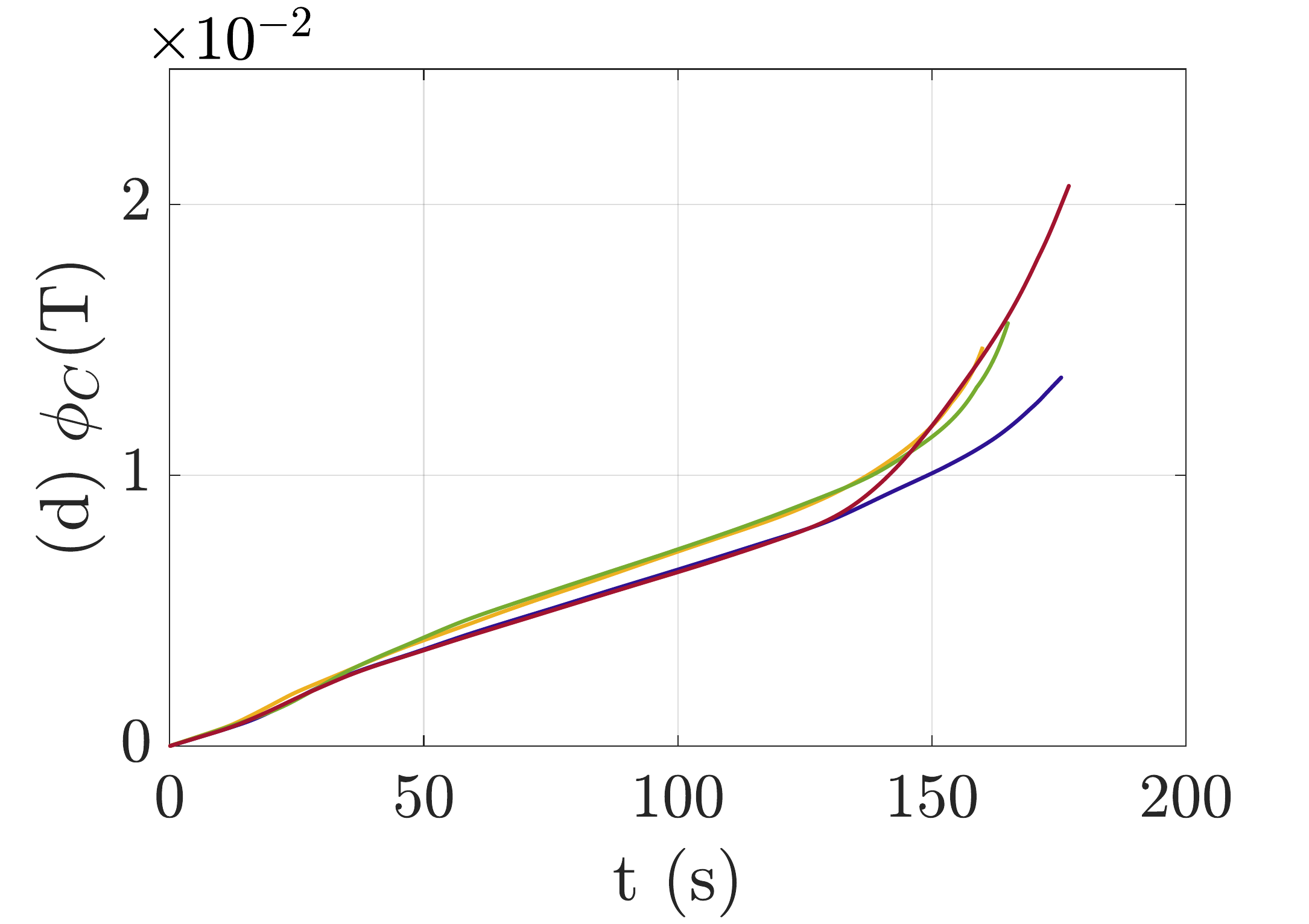}} \quad \\
	\subfigure
	{\hspace{-13pt}\includegraphics[trim=0.0cm 1cm 0cm 1cm, width = 0.2\textwidth]{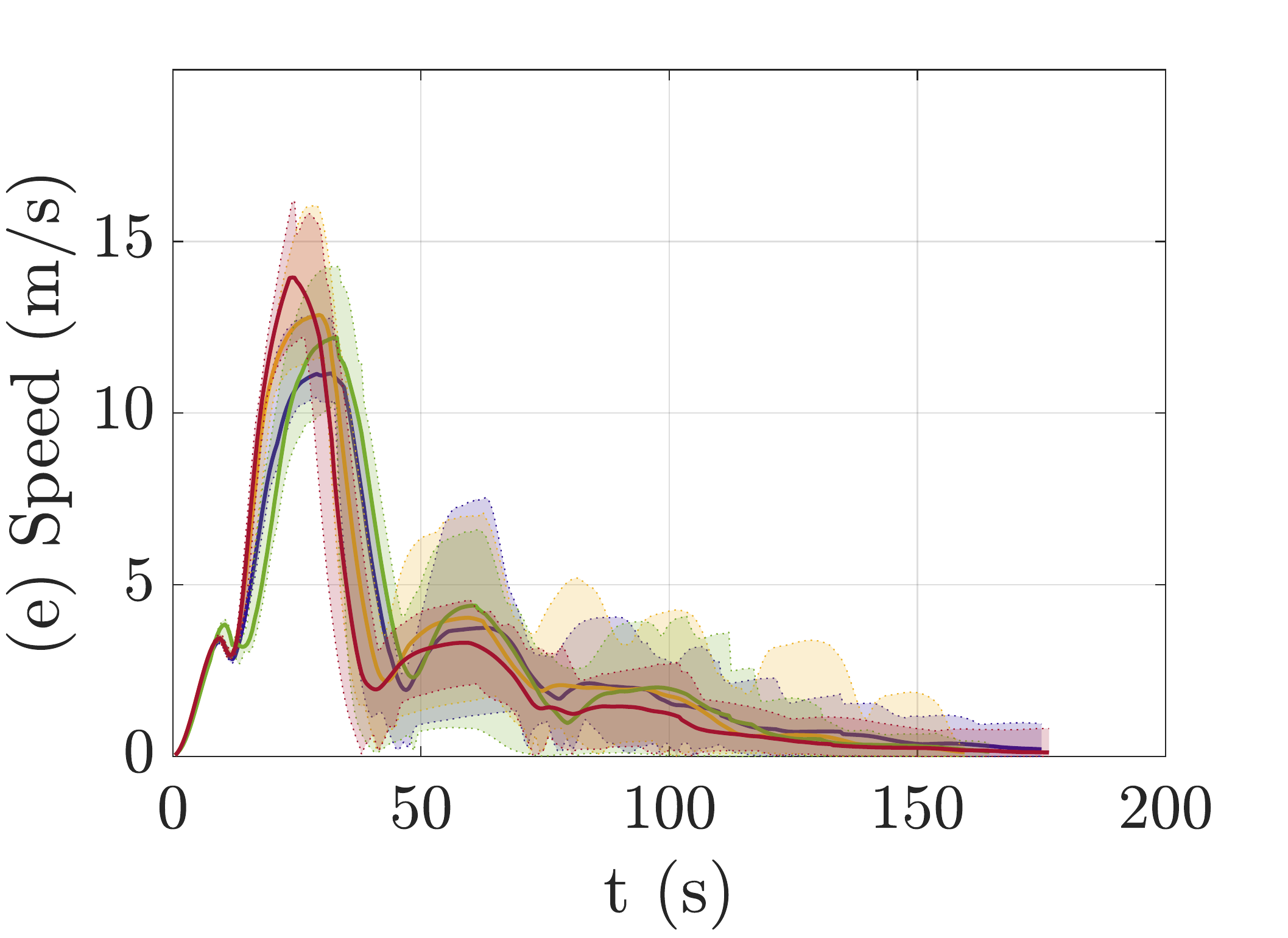}} \quad
	\subfigure
	{\hspace{-13pt}\includegraphics[trim=0.0cm 1cm 0cm 1cm, width = 0.2\textwidth]{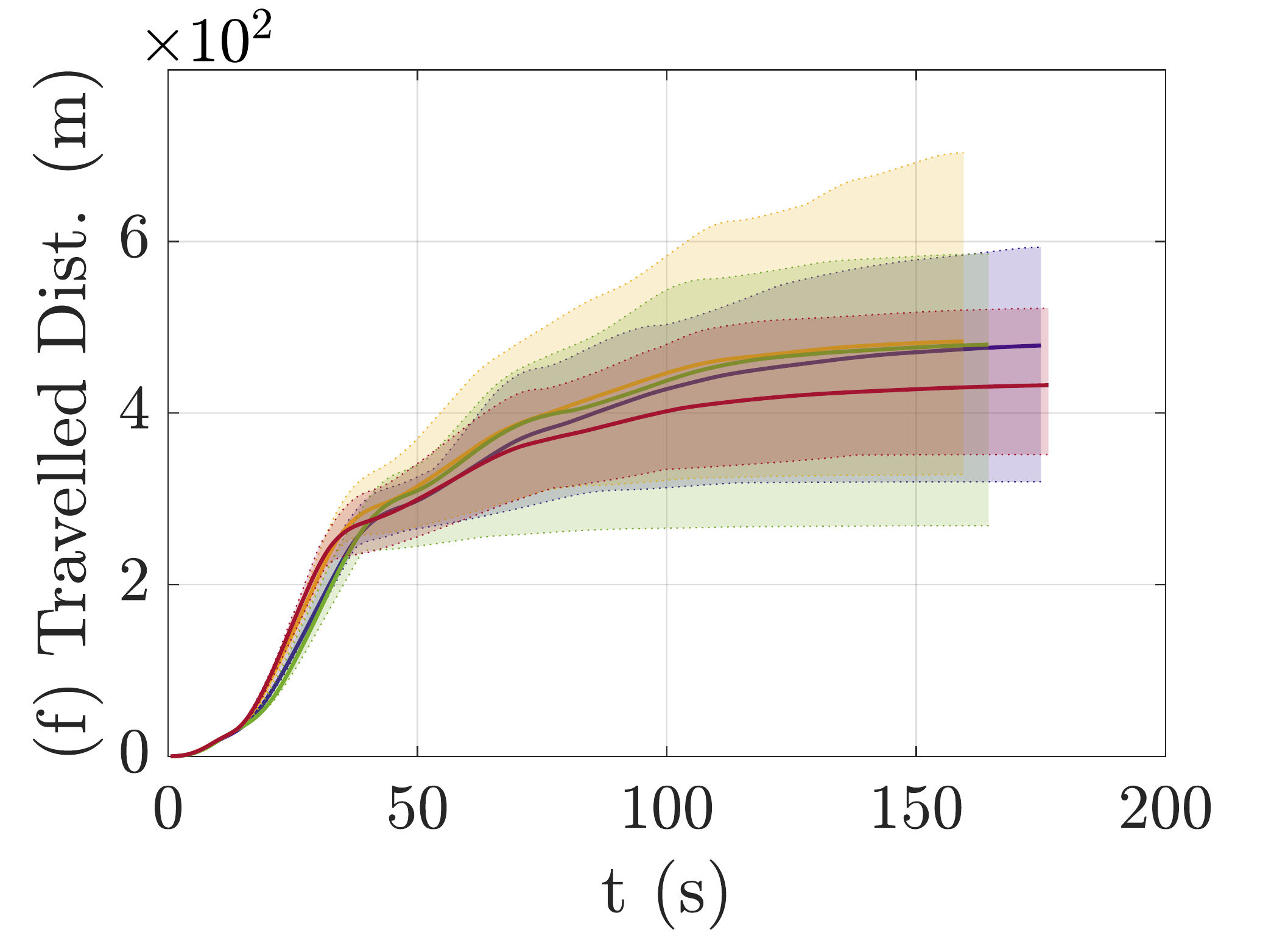}}
	\caption{The comparison of different methods in terms of (a) motion energy, (b) communications payload (c) velocity alignment, (d) number of collision risks, (e) mean/max/min speed, and (f) traveled distance. }
	\label{fig:02} \vskip -5pt
\end{figure}

\figurename{ \ref{fig:05}} represents the absolute values of approximation errors of HJB in \eqref{Eq:HJBerr} and FPK in \eqref{Eq:FPKerr} equations corresponding to the scenario and methods in \figurename{ \ref{fig:01}}. It is noticeable that the values of approximation errors of HJB and FPK, despite not being too small, are acceptably small. 
This is in compliance with the analysis that the model weights are UUB.
The approximation errors of HJB, i.e., $ e_\textsf{H} $ in \eqref{Eq:HJBerr}, depends on the error value of HJB model weights which is proved to be UUB in Proposition 1, and The approximation errors of FPK, i.e., $ e_\textsf{F} $ in \eqref{Eq:FPKerr}, depends on the error value of FPK model weights which is proved to be UUB in Proposition 2. Then, when the error value of model weights is below a threshold, the corresponding absolute values of approximate error of HJB in \eqref{Eq:HJBerr} and/or FPK in \eqref{Eq:FPKerr}  will be  bounded. This can be seen in \figurename{ \ref{fig:05}}-a and \figurename{ \ref{fig:05}}-b that the corresponding absolute error values are below $ 1.5 $ and $ 0.02 $, respectively.

\begin{figure}[t]
	\centering
	\setlength\abovecaptionskip{-0.0\baselineskip}
	\subfigure
	{\hspace{-13pt}\includegraphics[trim=0.0cm 1cm 0cm 1cm, width = 0.2\textwidth]{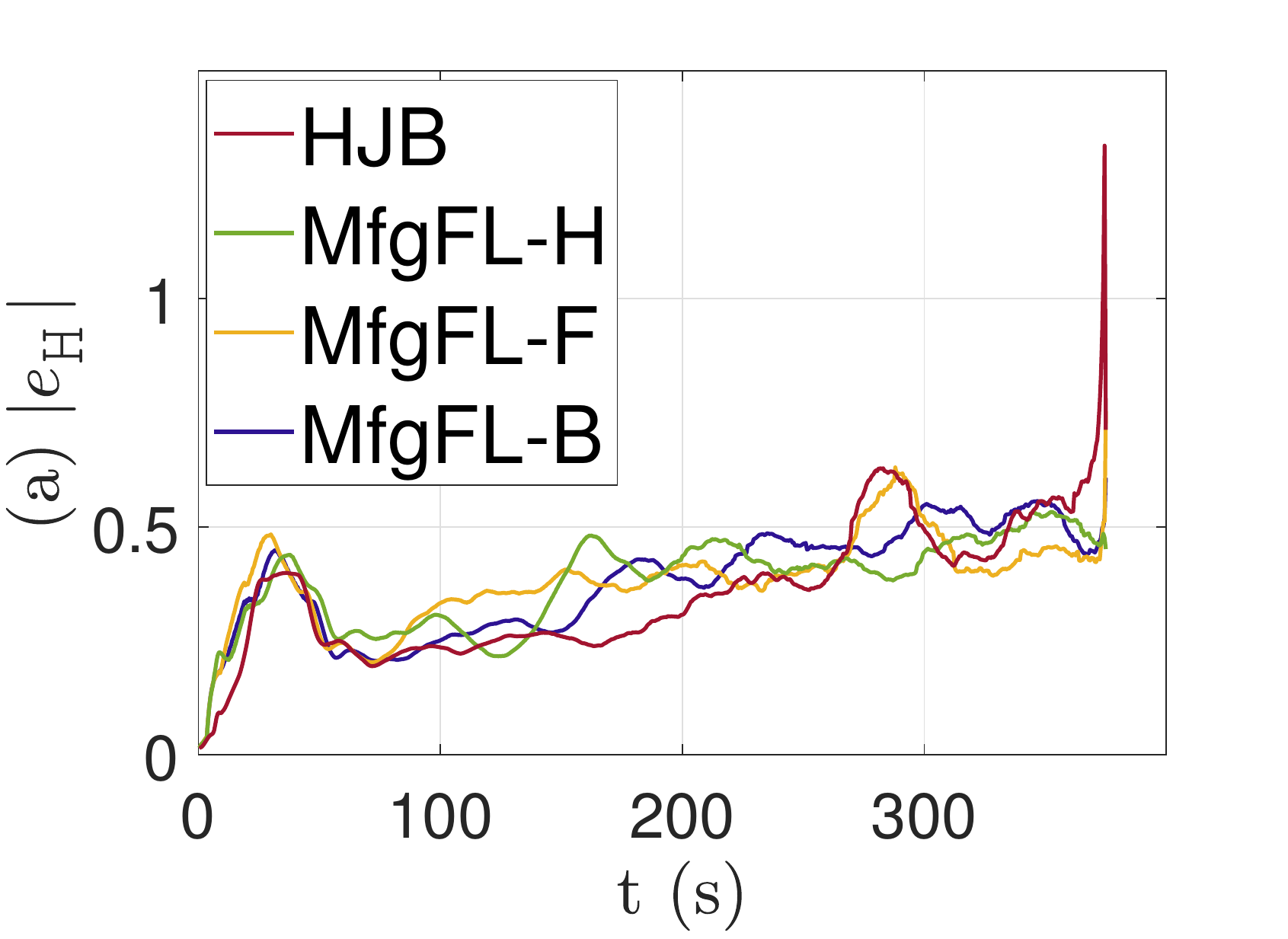}} \quad
	\subfigure
	{\hspace{-13pt}\includegraphics[trim=0.0cm 1cm 0cm 1cm, width = 0.2\textwidth]{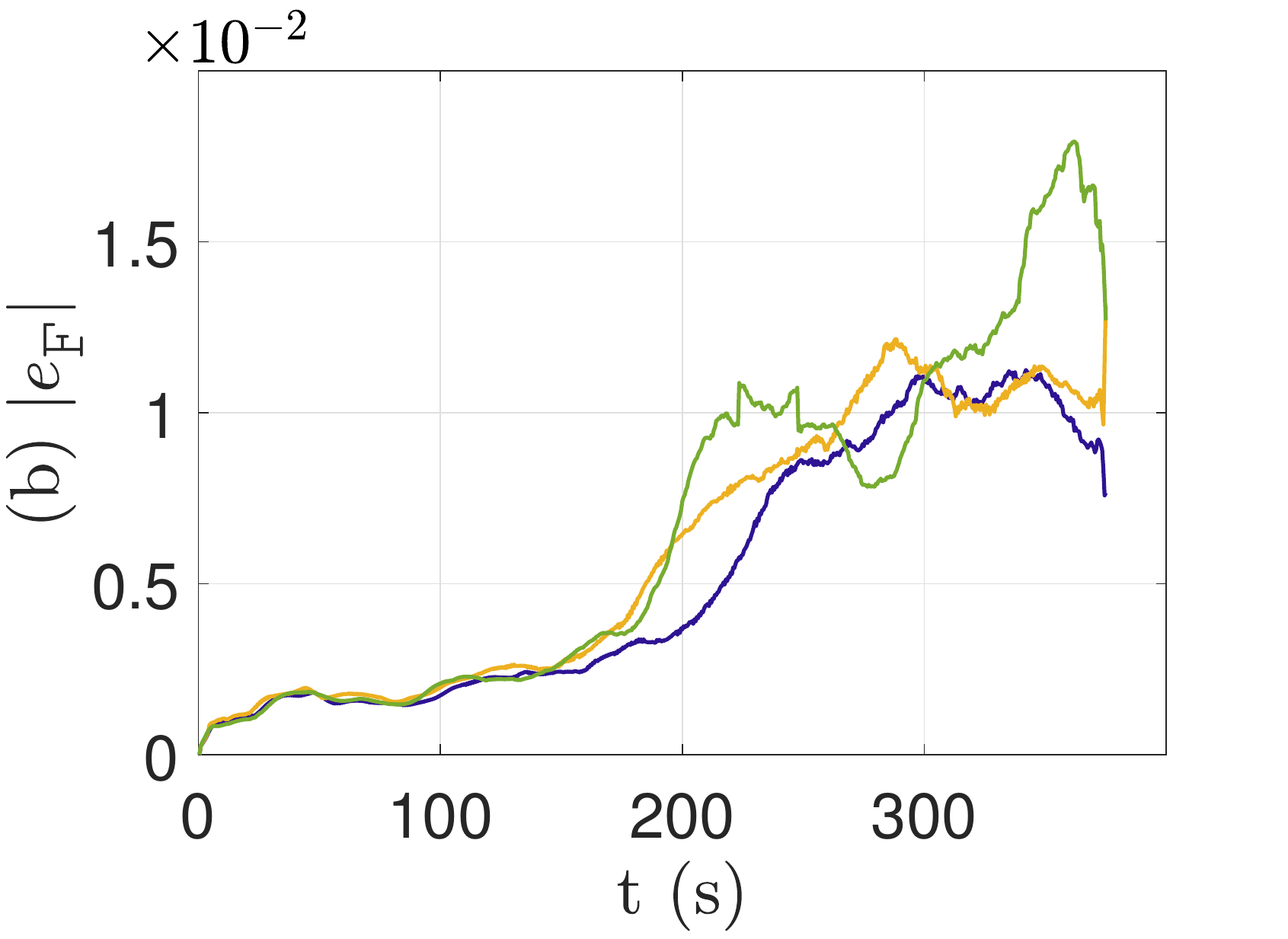}} 
	\caption{Approximation error values of (a) HJB model, (b) FPK model, are  small during the training for all methods.}
	\label{fig:05} \vskip -5pt
\end{figure}

\figurename{ \ref{fig:04}} illustrates the performance of different methods versus $ N  $ number of UAVs. 
Clearly, \textsf{MfgFL-HF} requires less motion energy for $ N = 16,\ldots,64 $, and its performance in terms of travel time \small$ T \leq T_\text{max} $\normalsize, velocity alignment \small$ \phi_\text{A}(T_\text{avg}) $\normalsize, and number of collision risks \small$ \phi_\text{C}(T_\text{avg}) $\normalsize, improves as the number of UAVs increases. This is because, for higher $ N $, more samples can be provided for both HJB and FPK models in \textsf{MfgFL-HF} which results in better training of both models. However, for the other two FL base methods, i.e., \textsf{MfgFL-H} and \textsf{MfgFL-F}, provided samples due to averaging improves only one of the HJB or FPK models, and the other corresponding model still remains less trained. Therefore, the coupled HJB-FPK equation in these two methods still is not well trained and non of \textsf{MfgFL-H} and \textsf{MfgFL-F} can benefit much when number of UAVs increases.
Regarding \textsf{Hjb}, increasing the number of UAVs does not improve the performance  much since it does not utilize the more provided samples for training the model due to the  processing power limitations of the UAVs. 
\begin{figure}[t]
	\centering
	\setlength\abovecaptionskip{-0.0\baselineskip}
	\subfigure
	{\hspace{-13pt}\includegraphics[trim=0.0cm 1cm 0cm 1cm, width = 0.2\textwidth]{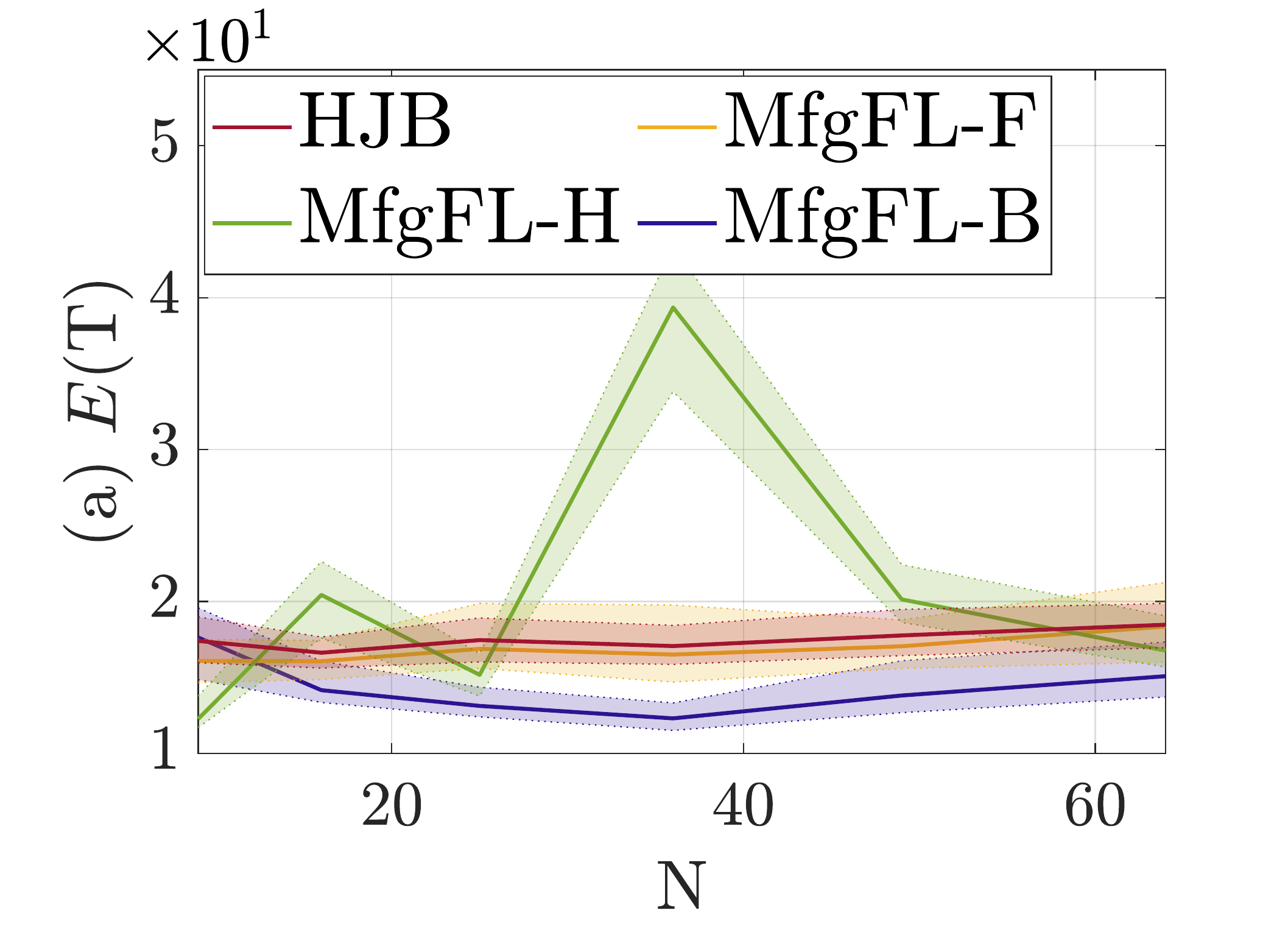}} \quad
	\subfigure
	{\hspace{-13pt}\includegraphics[trim=0.0cm 1cm 0cm 1cm, width = 0.2\textwidth]{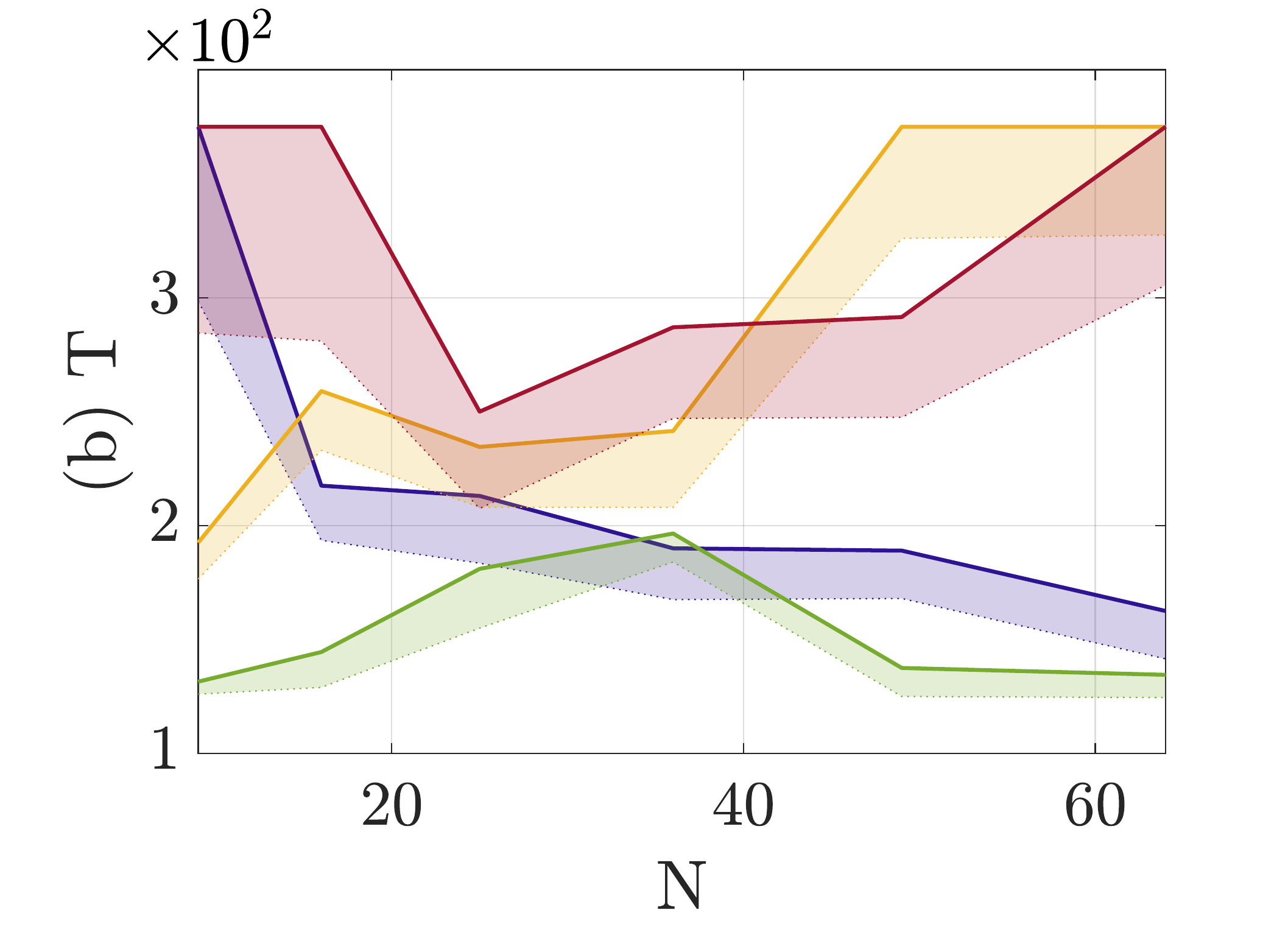}} \quad
	\subfigure
	{\hspace{-13pt}\includegraphics[trim=0.0cm 1cm 0cm 1cm, width = 0.2\textwidth]{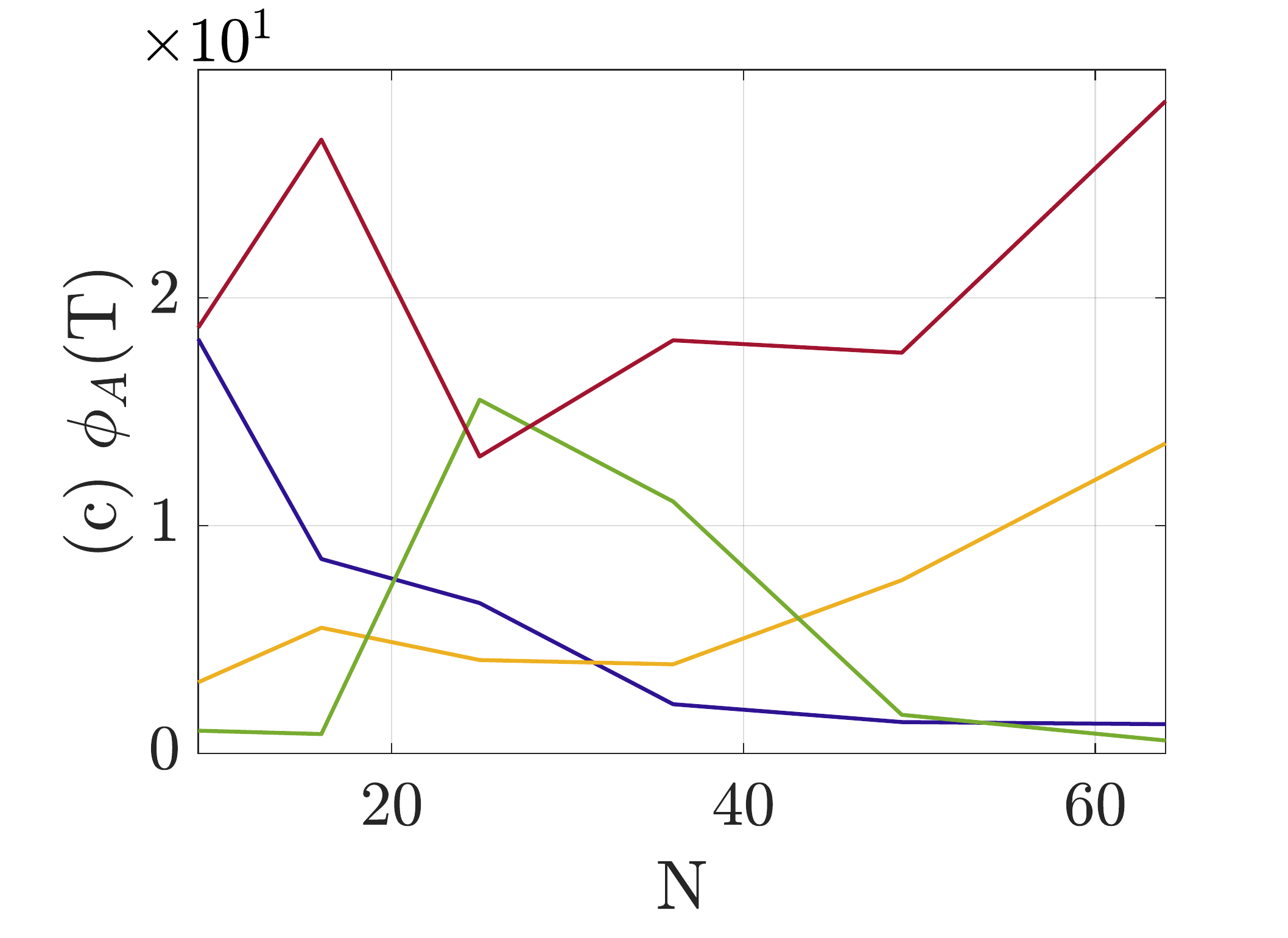}} \quad
	\subfigure
	{\hspace{-13pt}\includegraphics[trim=0.0cm 1cm 0cm 1cm, width = 0.2\textwidth]{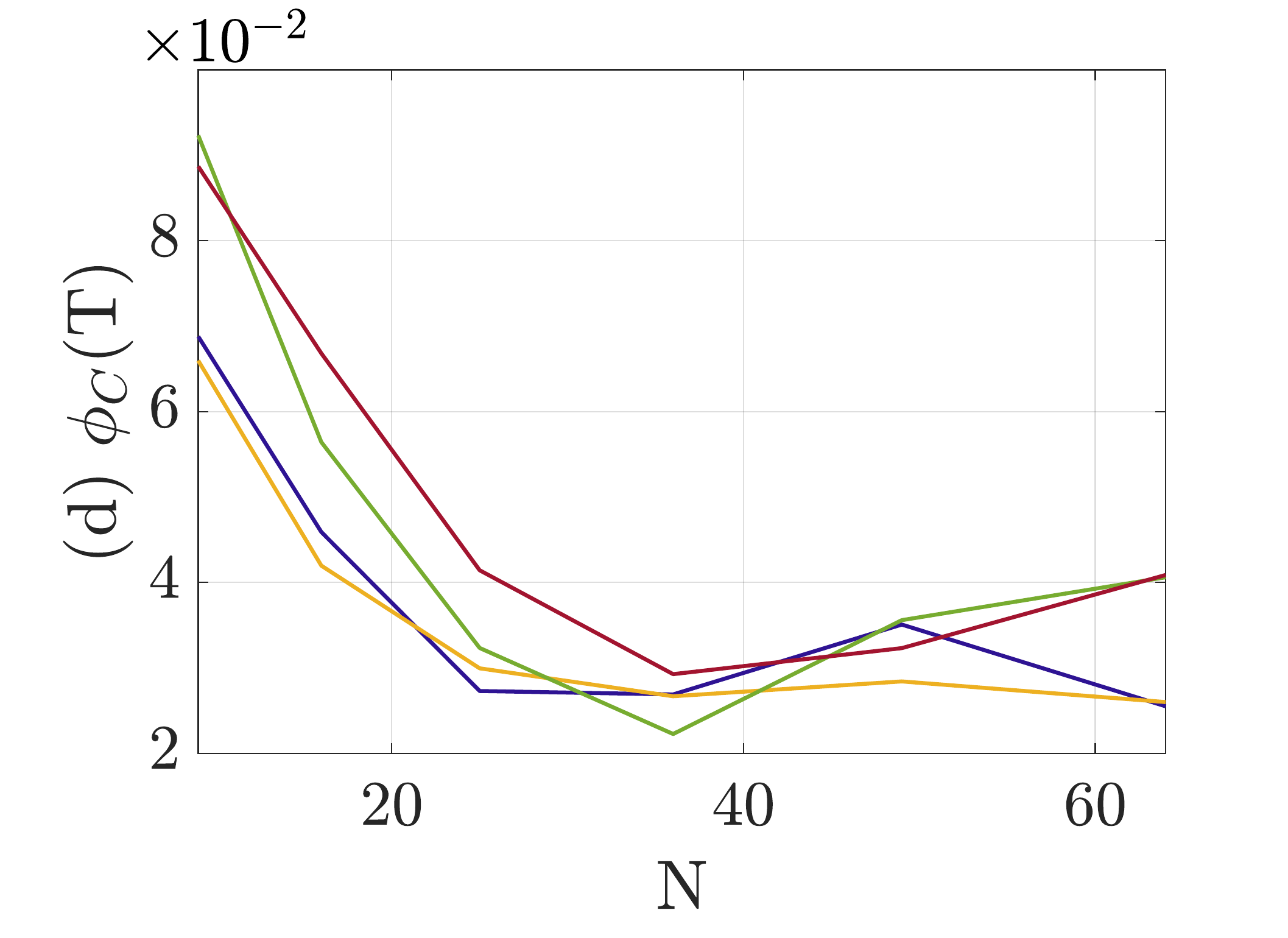}} 
	\caption{Performance of different methods vs number of UAVs in terms of (a) motion energy, (b) travel time (c) velocity alignment, and (d) number of collision risks.}
 	\label{fig:04} \vskip -5pt
\end{figure}

\figurename{ \ref{fig:03}} shows the impact of model update period $ n_0 $ on the performance  criteria
for the FL-based methods.
For $ n_0 $ larger than $ 100 $, the \textsf{MfgFL-HF} method consumes less energy than the other FL-based methods to complete the travel (see \figurename{ \ref{fig:03}}-a), while its travel time is only more than \textsf{MfgFL-H} for most of the choices of $ n_0 $ (see \figurename{ \ref{fig:03}}-b). 
Regarding the velocity alignment \small$ \phi_\text{A}(T_\text{avg}) $\normalsize\,  and number of collision risks \small$ \phi_\text{C}(T_\text{avg}) $\normalsize, for $ n_0 $ in the interval $ n_0 \in \{100,\cdots, 400\} $,  \textsf{MfgFL-HF} method has lower velocity alignment and number of collision risks than other FL-based methods (see \figurename{ \ref{fig:03}}-c and \figurename{ \ref{fig:03}}-d). Additionally, there is an acceptable trade-off among the performance criteria for \textsf{MfgFL-HF} method in this interval. This is due to the fact that, for small values of $ n_0 $, e.g., $ 50 $, fewer UAVs can successfully transmit their data to the leader because of communication costs such as limited transmission power of UAVs. On the other hand, for very high values of $ n_0 $, e.g., $ 500 $, the algorithms cannot benefit  from adopting the FL method in real-time application, because when $ n_0 $ increases the models at UAVs rely mostly on local samples for larger amount of time and become less trained.

\begin{figure}[t]
	\centering
	\setlength\abovecaptionskip{-0.0\baselineskip}
	\subfigure
	{\hspace{-13pt}\includegraphics[trim=0.0cm 1cm 0cm 1cm, width = 0.2\textwidth]{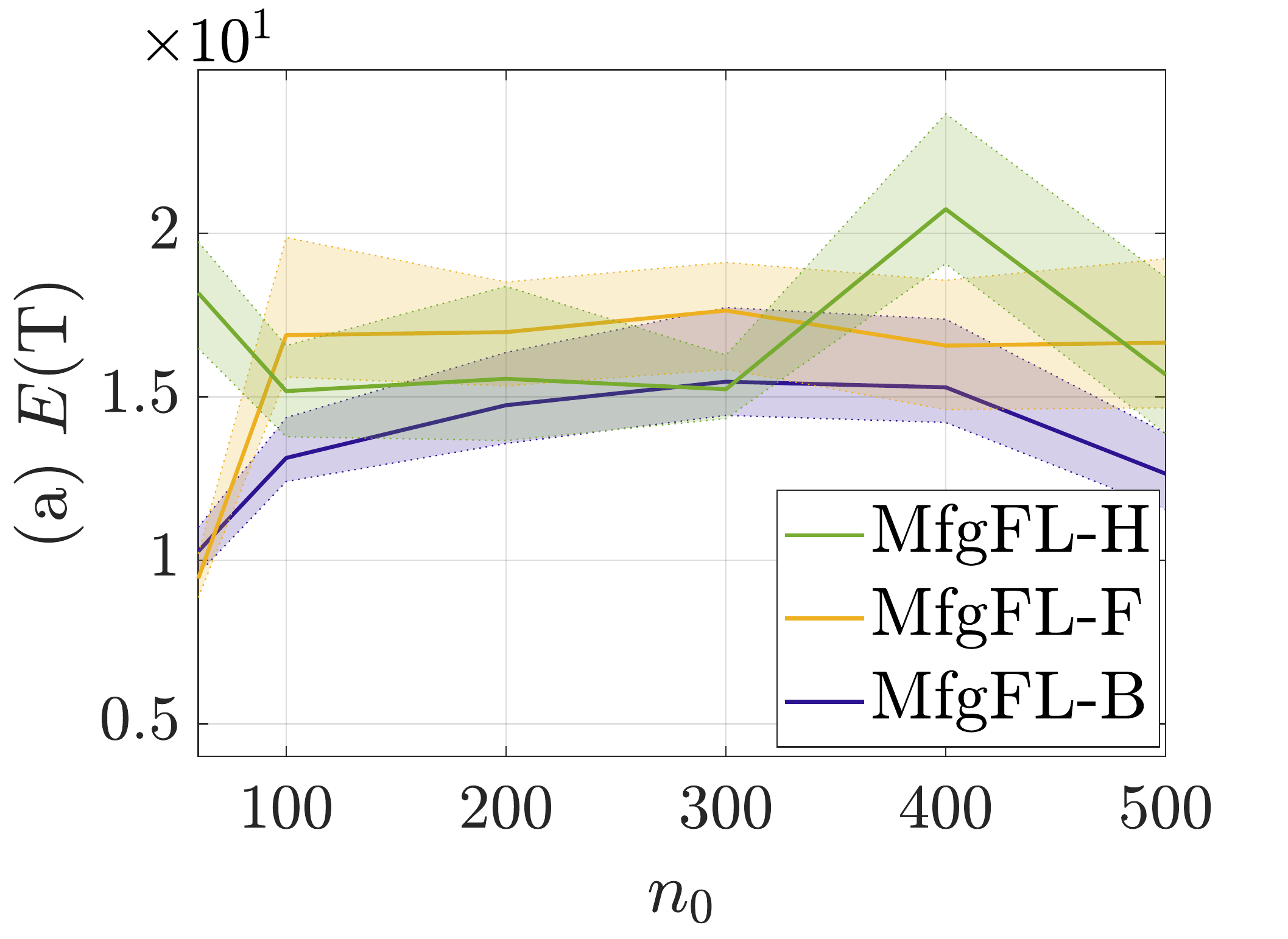}} \quad
	\subfigure
	{\hspace{-13pt}\includegraphics[trim=0.0cm 1cm 0cm 1cm, width = 0.2\textwidth]{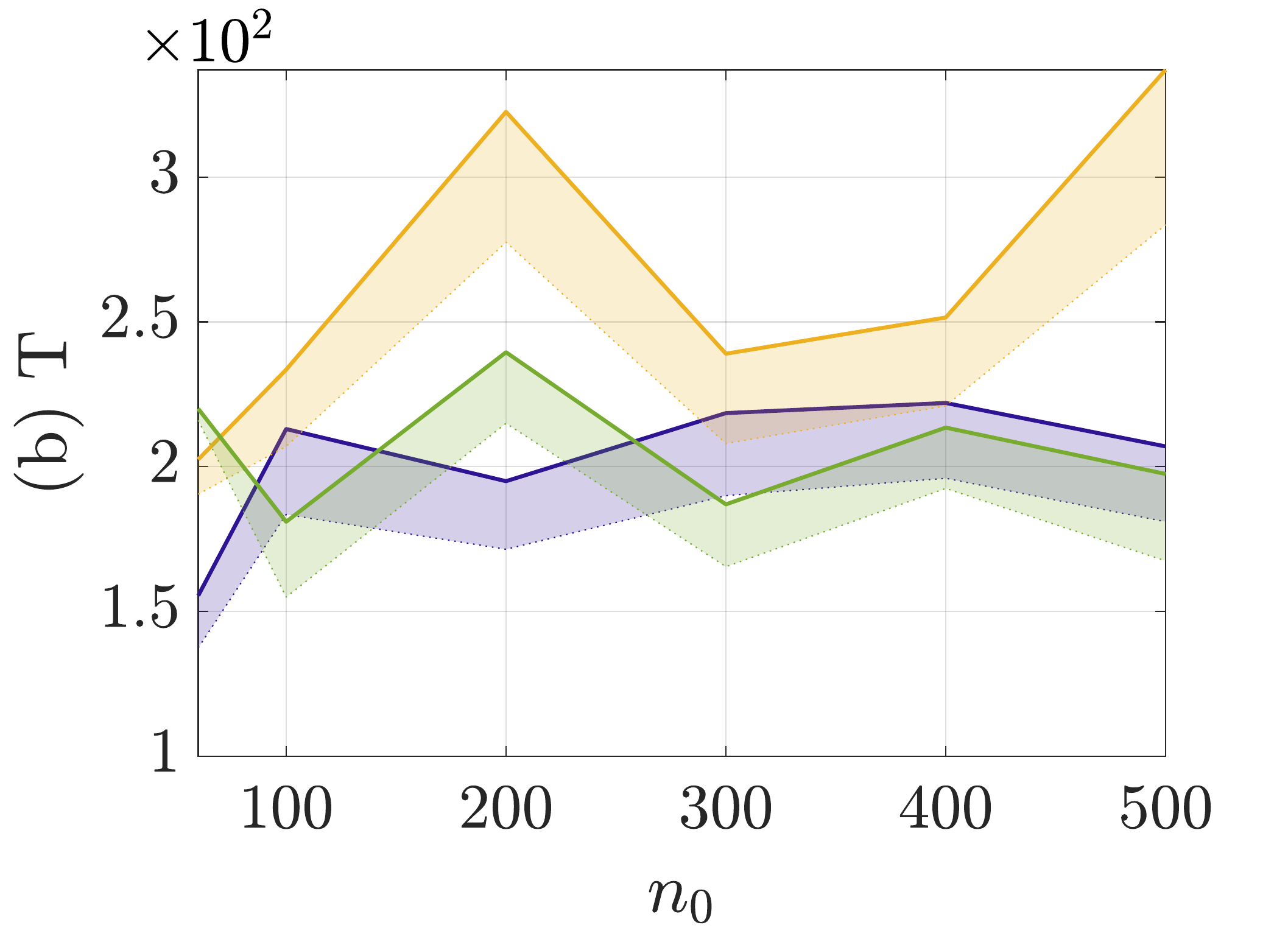}} \quad
	\subfigure
	{\hspace{-13pt}\includegraphics[trim=0.0cm 1cm 0cm 1cm, width = 0.2\textwidth]{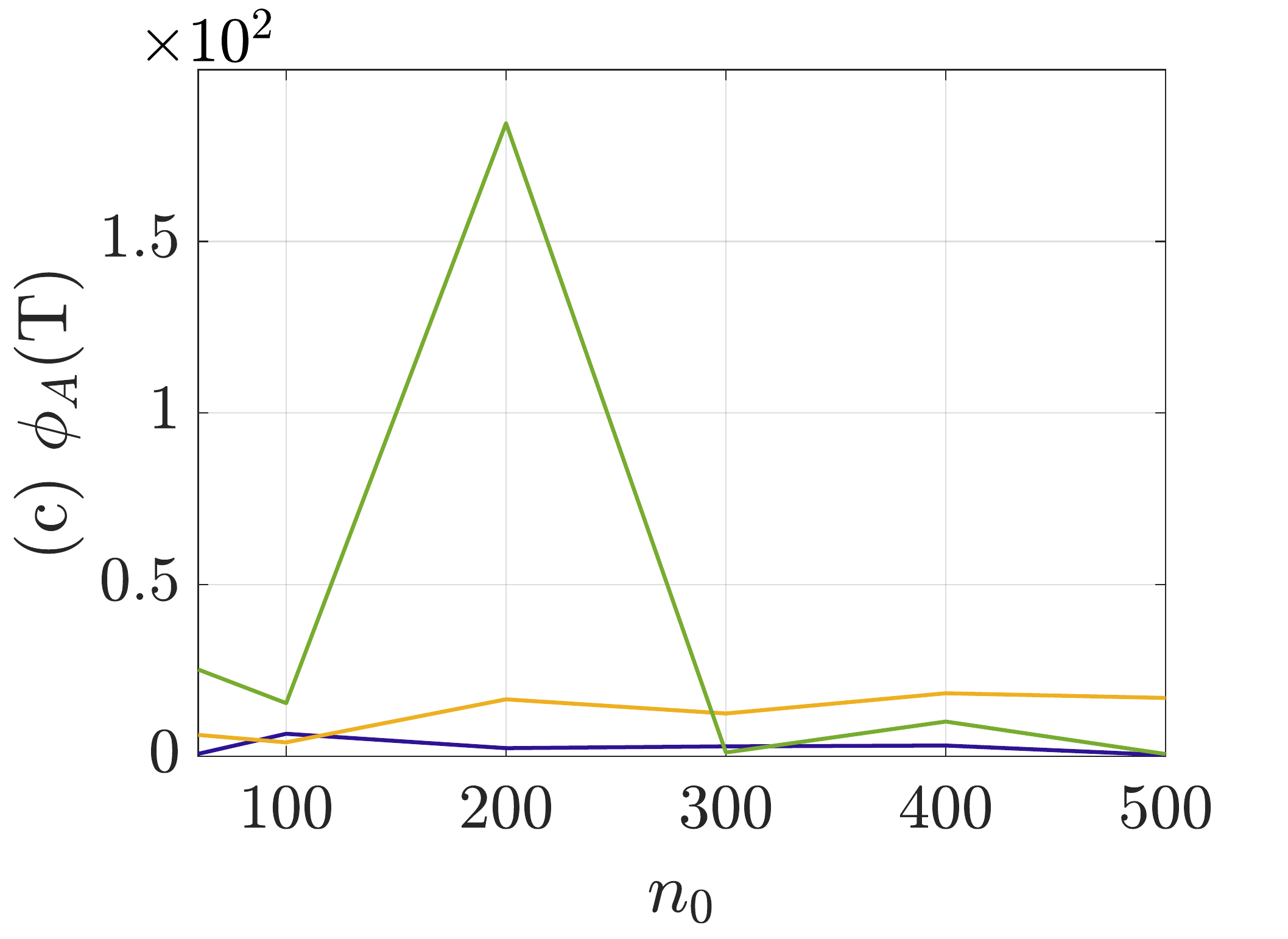}} \quad
	\subfigure
	{\hspace{-13pt}\includegraphics[trim=0.0cm 1cm 0cm 1cm, width = 0.2\textwidth]{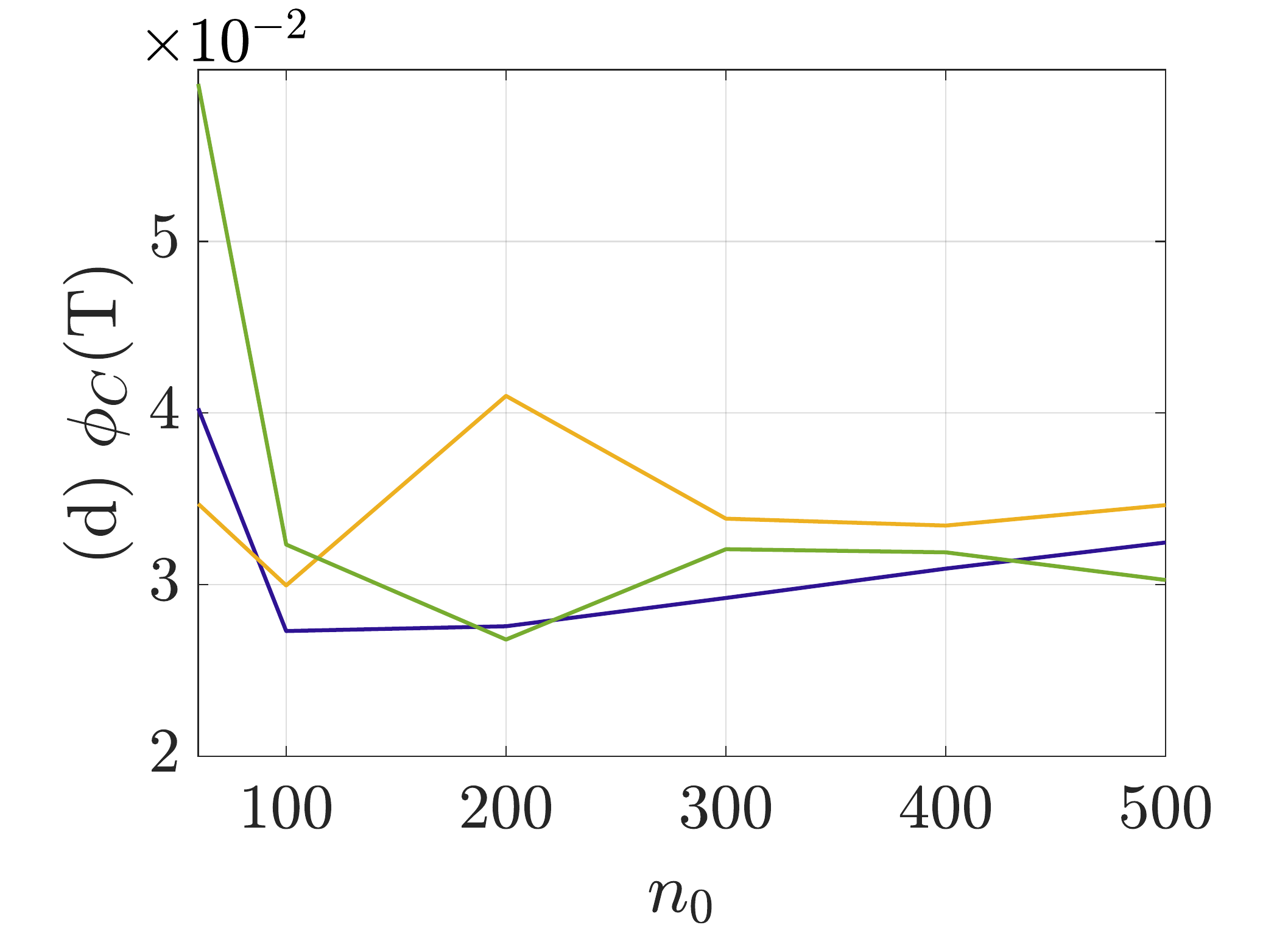}} 
	\caption{Performance of different FL-based methods vs model update period in terms of (a) motion energy, (b) travel time (c) velocity alignment, and (d) number of collision risks.}
	\label{fig:03} \vskip -5pt
\end{figure}

\figurename{ \ref{fig:07}} represents the effect of wind perturbations on the online control of UAVs in the scenario of  \figurename{ \ref{fig:01}}. Here, the UAVs are set to move under different wind perturbation variances \small$ \sigma_\text{wind} = 0.1, \cdots, 0.5 $\normalsize\, and the various comparison criteria are calculated for the algorithms. 
From \figurename{ \ref{fig:07}}-a, the proposed algorithm \textsf{MfgFL-HF} consumes less control energy than the other baselines, and the energy consumption variance is increasing with the wind variance. 
This is because the agents with \textsf{MfgFL-HF} keep a distance away from each other to avoid  collision. Also, \figurename{ \ref{fig:07}}-b to \figurename{ \ref{fig:07}}-d show
smaller travel time of \textsf{MfgFL-HF} than \textsf{MfgFL-F} and \textsf{Hjb}, and better collision avoidance of \textsf{MfgFL-HF} than all other mentioned methods. Overall, these figures emphasize that the algorithm \textsf{MfgFL-HF} is more robust against wind dynamics, thanks to better learning capability of the proposed method.

\begin{figure}[t]
	\centering
	\setlength\abovecaptionskip{-0.0\baselineskip}
	\subfigure
	{\hspace{-13pt}\includegraphics[trim=0.0cm 1cm 0cm 1cm, width = 0.2\textwidth]{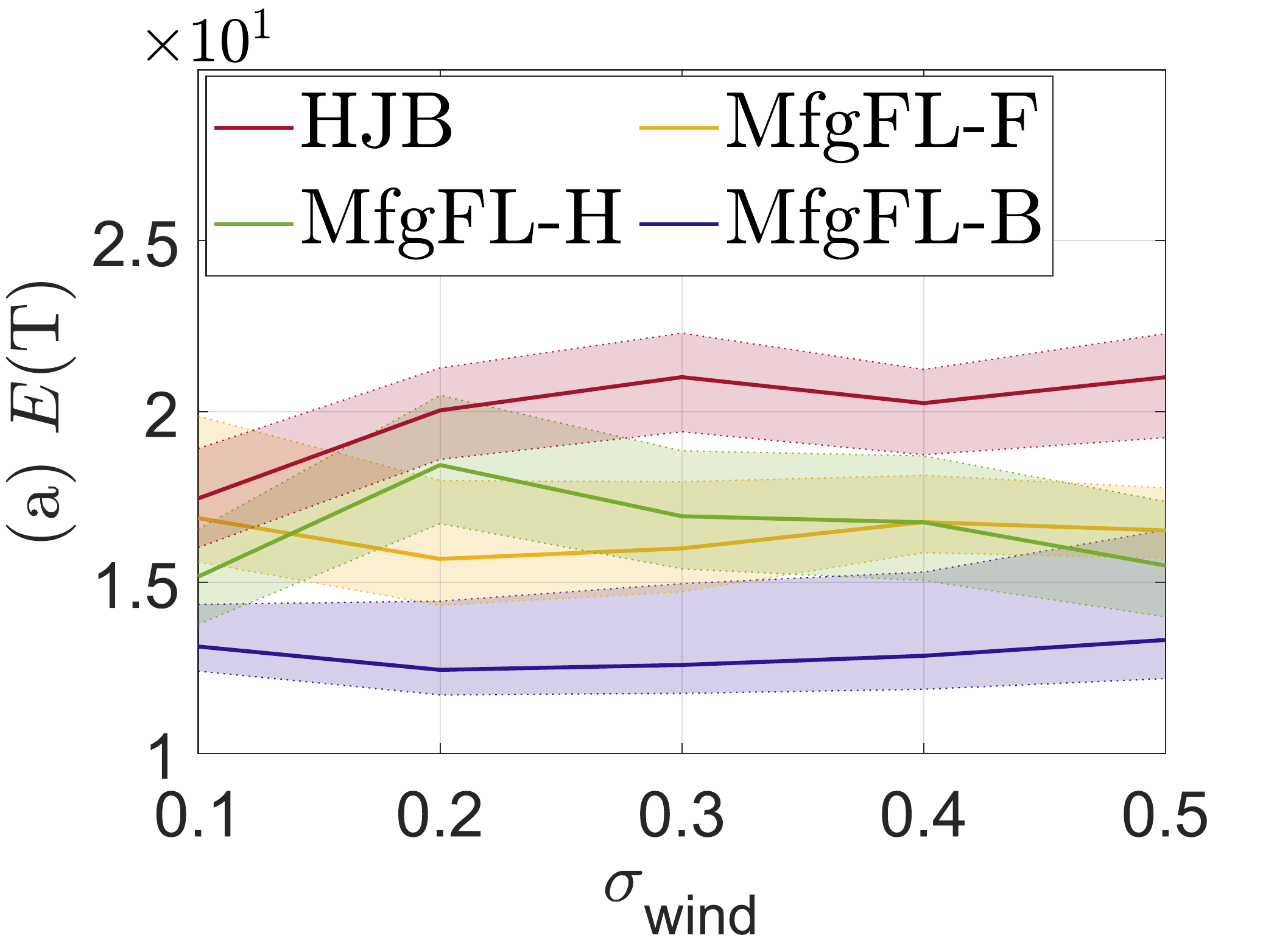}} \quad
	\subfigure
	{\hspace{-13pt}\includegraphics[trim=0.0cm 1cm 0cm 1cm, width = 0.2\textwidth]{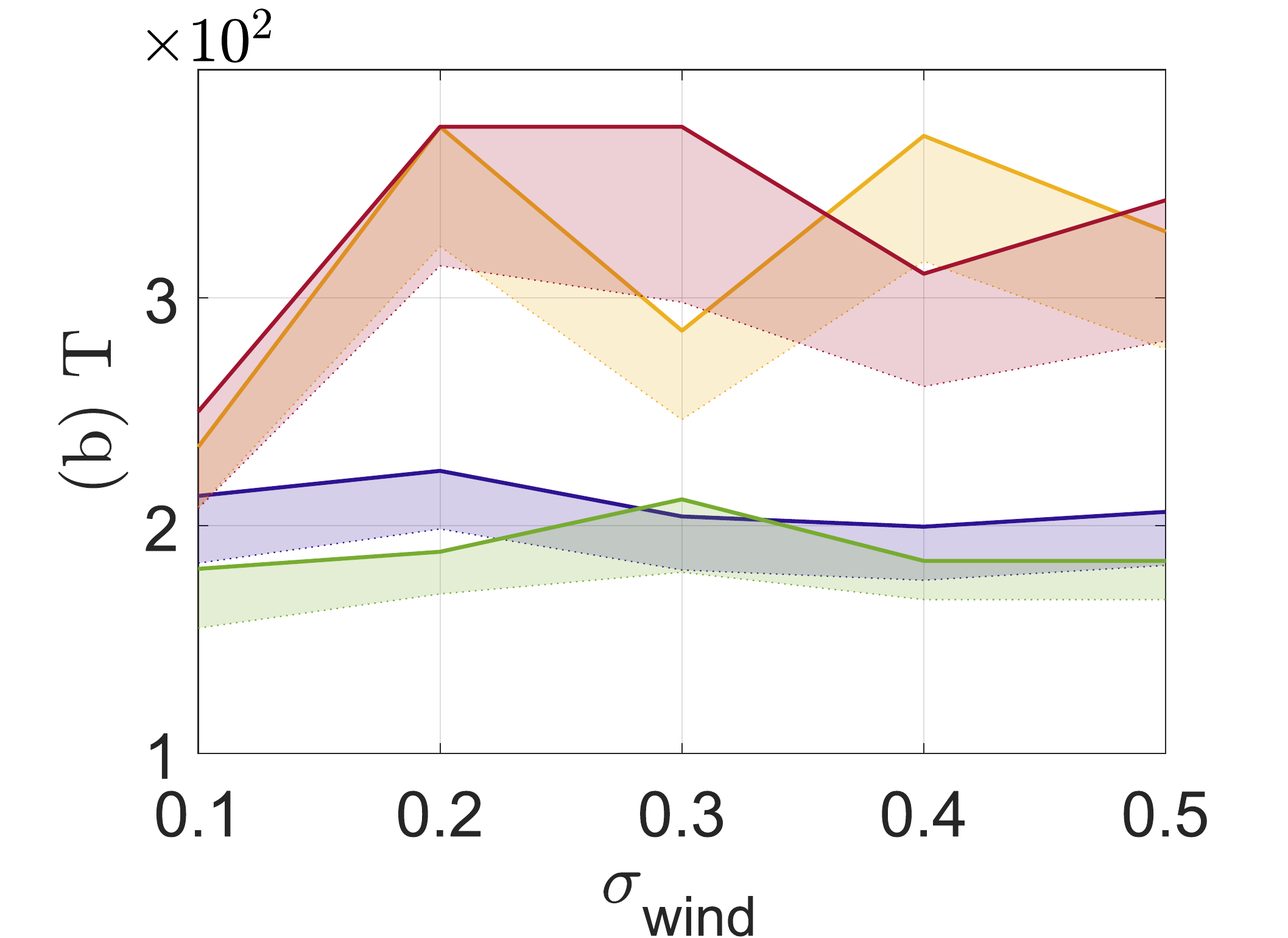}} \quad
	\subfigure
	{\hspace{-13pt}\includegraphics[trim=0.0cm 1cm 0cm 1cm, width = 0.2\textwidth]{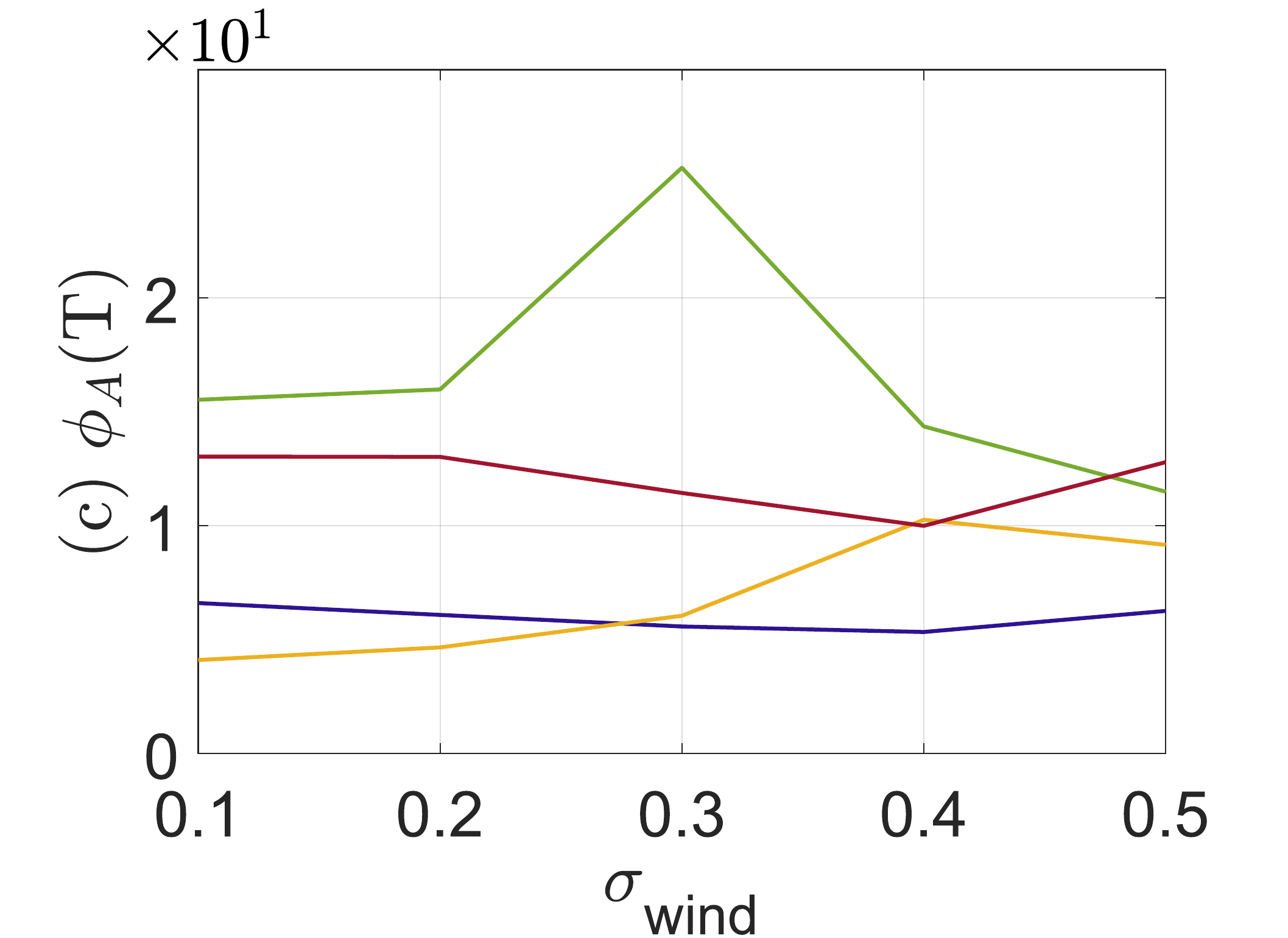}} \quad
	\subfigure
	{\hspace{-13pt}\includegraphics[trim=0.0cm 1cm 0cm 1cm, width = 0.2\textwidth]{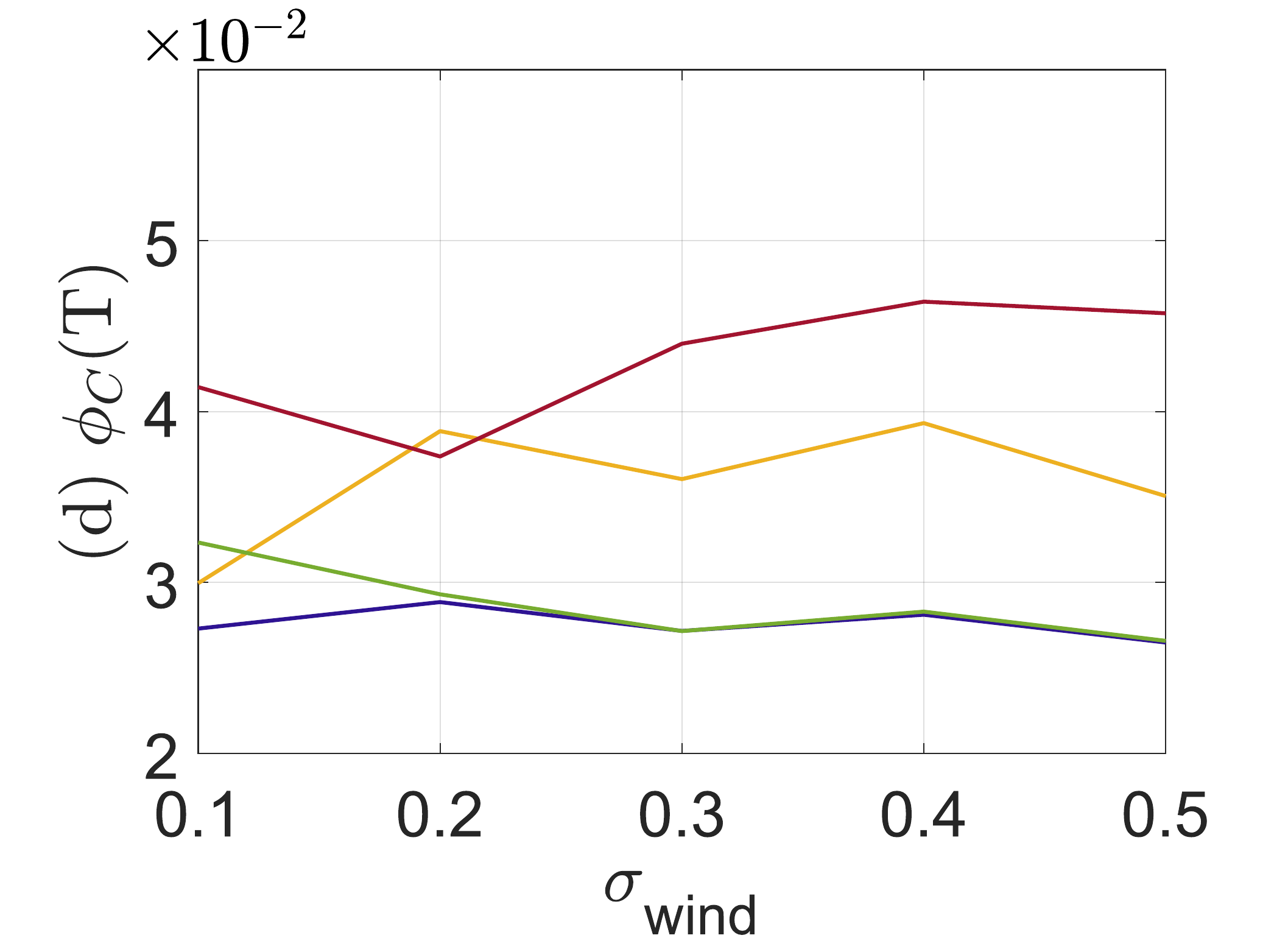}} 
	\caption{Performance of different FL-based methods vs wind variance in terms of (a) motion energy, (b) travel time (c) velocity alignment, and (d) number of collision risks.}
	\label{fig:07}  \vskip -5pt
\end{figure}

All the previous figures show the performance of untrained UAVs in a windy environment. However, there is still an important concern remaining: How the proposed method behaves in comparison to the offline methods in windy environments? This comparison is shown in \figurename{ \ref{fig:08}}.  The term offline method here means that the UAVs are separated enough at the starting point and they are programmed to follow some pre-defined actions to reach the destination when there is no randomness in the environment, i.e., \small$ \sigma_\text{wind} = 0 $\normalsize\, as in \cite{shiller1992computation} and without any collaboration among UAVs. \figurename{ \ref{fig:08a}} shows the optimal shortest path which the UAVs can follow in an imaginary perfect environment without any collision occurrences. Nevertheless, the offline method fails to reach the destination with no collision in the presence of random wind dynamics, i.e., \small$ \sigma_\text{wind} = 0.1 $\normalsize, as shown in \figurename{ \ref{fig:08b}}.  Moreover,  \figurename{ \ref{fig:08d}} shows that the trajectory of the trained UAVs with the proposed method in the windy environment with \small$ \sigma_\text{wind} = 1.5 $\normalsize\,  is  higher than the training environment with \small$ \sigma_\text{wind} = 0.1 $\normalsize. The proposed method \textsf{MfgFL-HF} is much more robust to random wind perturbations and can reach the destination with no collision on the path towards the destination, while there is no such collision avoidance guarantee in the offline method.

 \begin{figure}[t]
 	\centering
 	\setlength\abovecaptionskip{-0.0\baselineskip}
 	\subfigure[Offline, no wind]
 	{\hspace{-10pt}
 	\includegraphics[trim=0.5cm 0.5cm 0.0cm 0.0cm, width = 0.17\textwidth]{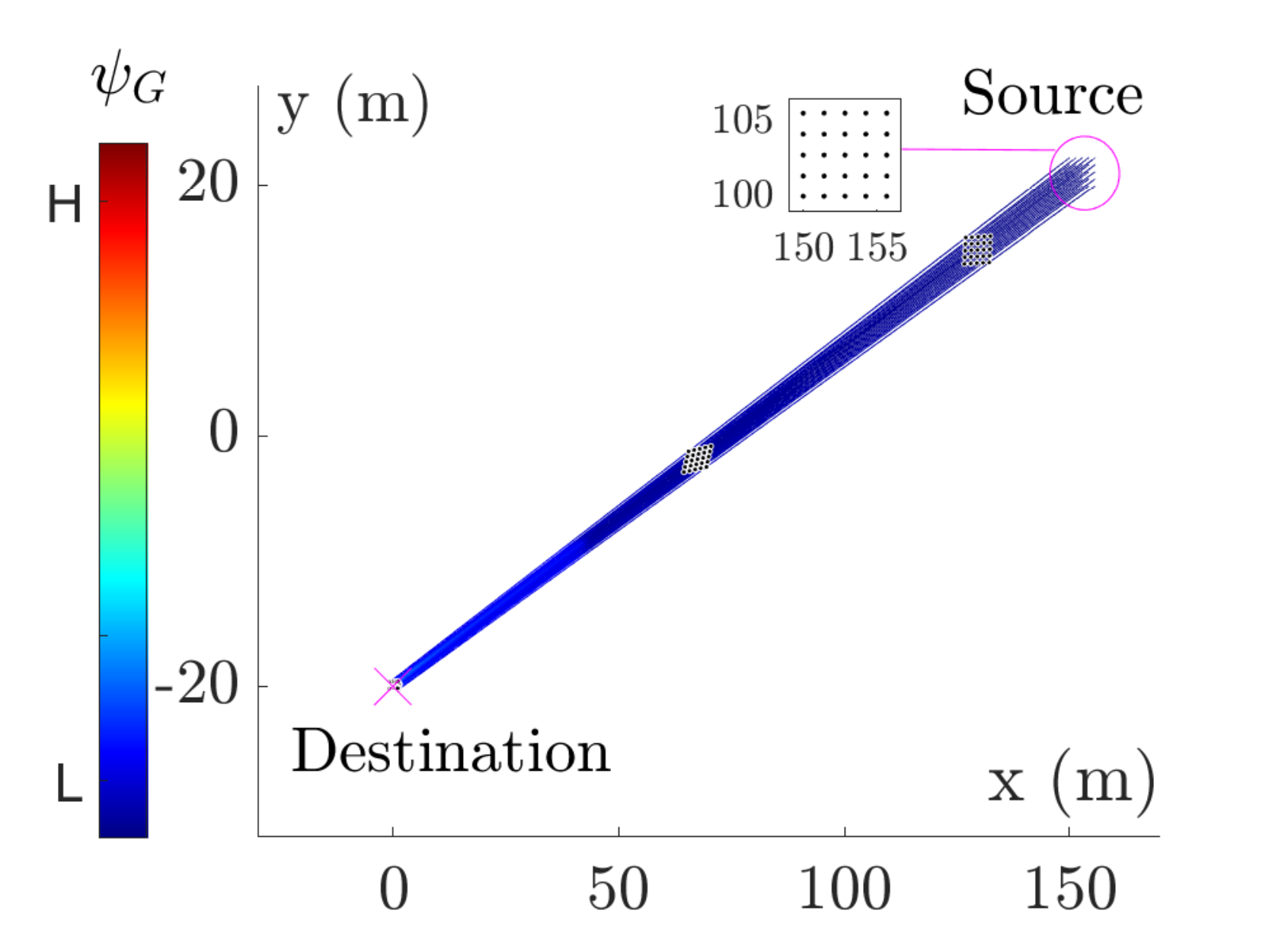}\label{fig:08a}}   
 	\subfigure[Offline, random wind]
 	{\hspace{-8pt}
 	\includegraphics[trim=0.5cm 0.5cm 0.0cm 0.0cm, width = 0.17\textwidth]{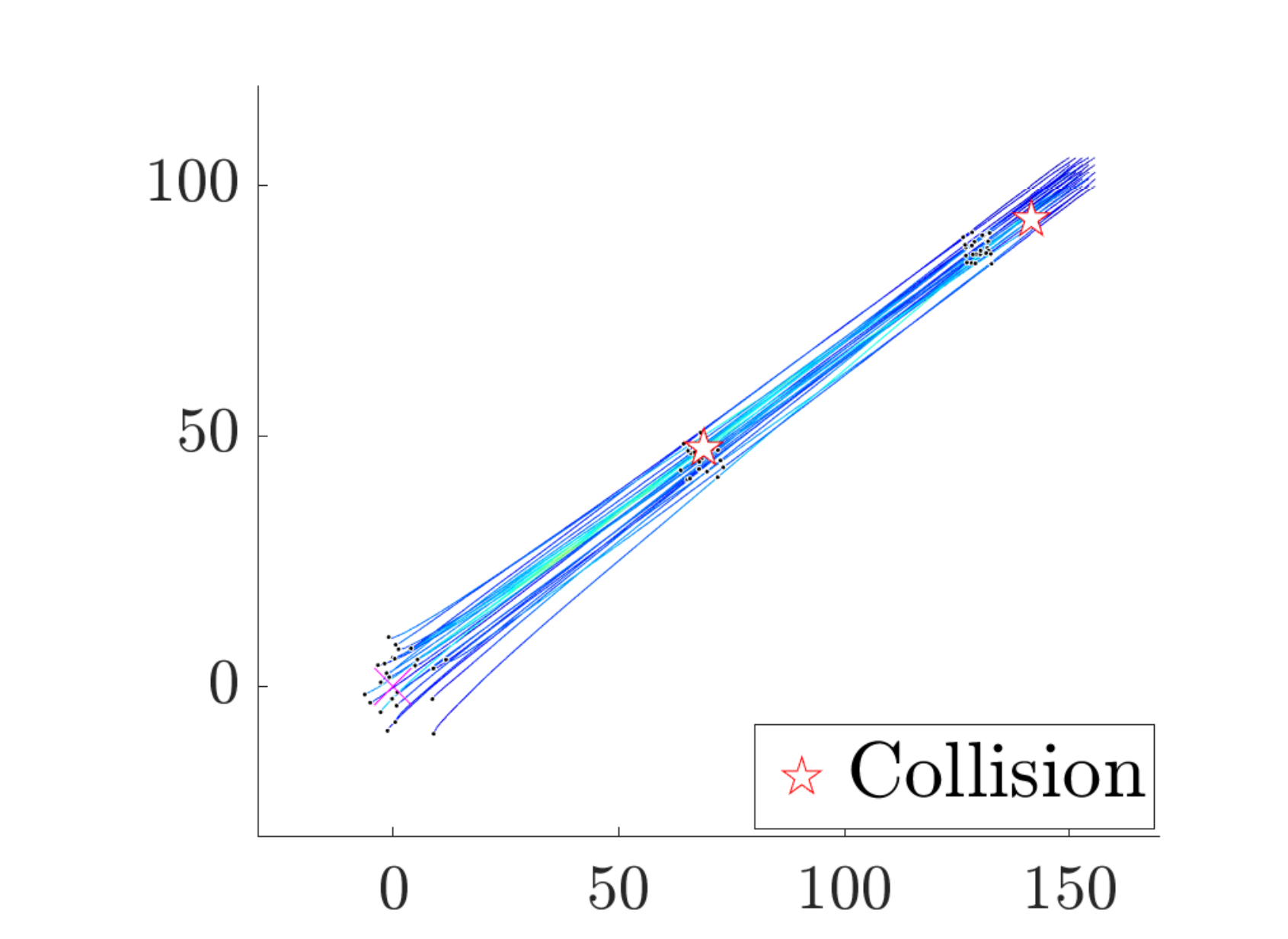}\label{fig:08b}}  
 	\subfigure[Online, random wind]
 	{\hspace{-13pt}
 	\includegraphics[trim=0.5cm 0.5cm 0.5cm 0.cm, width = 0.17\textwidth]{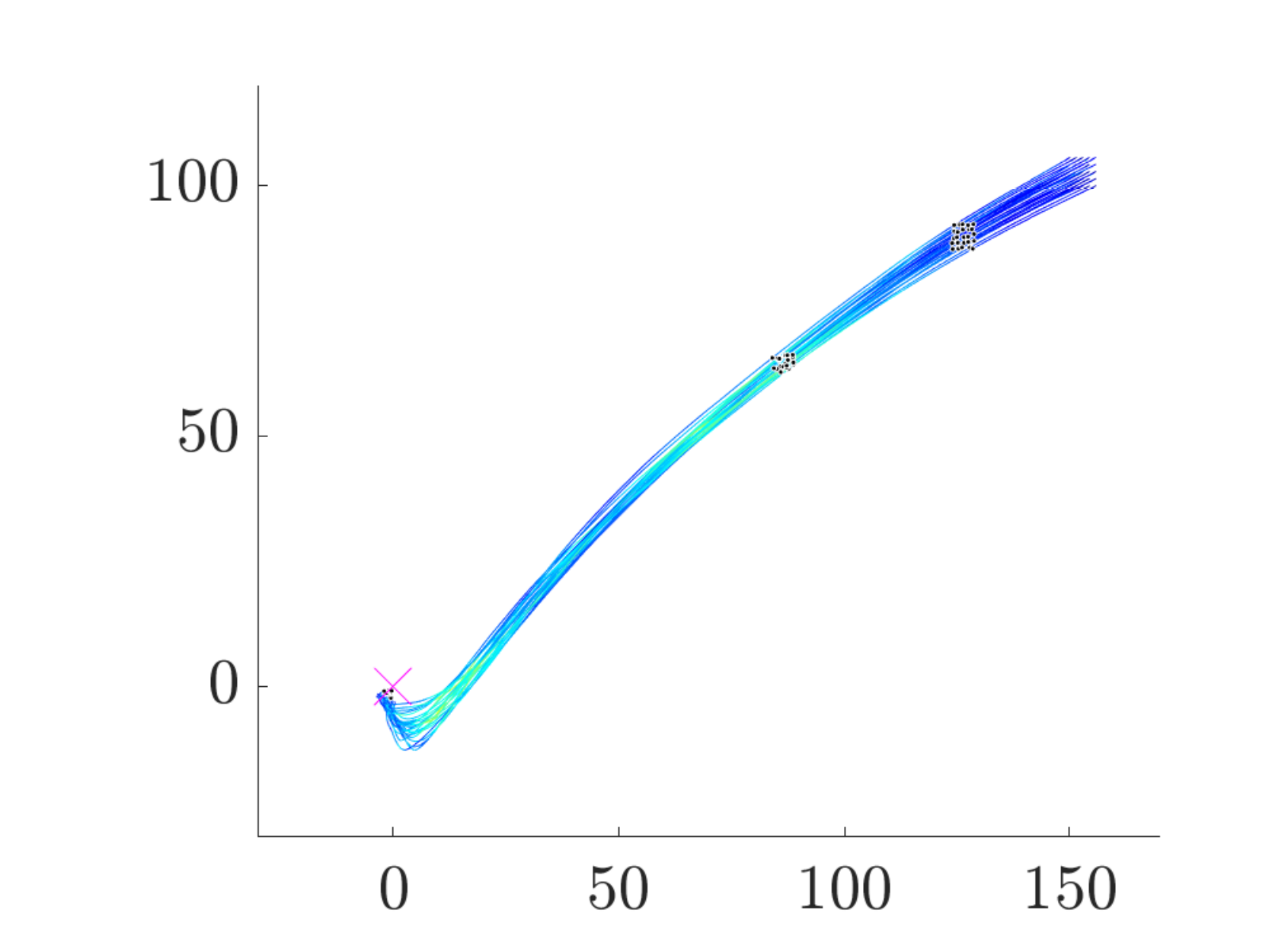}\label{fig:08d}}
 	\caption{The comparison of online training of UAV by \textsf{MfgFL-HF} algorithm with the UAVs pre-programmed for deterministic environment. }
 	\label{fig:08} \vskip -5pt
 \end{figure}

\section{Conclusion}
 \label{SE:07}
In this paper, a novel path planning approach is proposed for a population of UAVs being effected by random wind perturbations in the environment.
To this end, the objective is to minimize the transmission time, motion energy, and the interactions among the UAVs. First, the MFG framework is applied in order to reduce the high amount of communications required to control a massive number of UAVs. Next, a function approximator based on neural networks is proposed to approximate the solution of the HJB and FPK equations. The Lyapunov stability analysis for MFG learning is provided to show that the approximate solution for HJB and FPK equations are bounded. Then, on the bases of these assumptions and analyses, an FL-based MFG learning method named \textsf{MfgFL-HF} is proposed to use the samples of UAVs more efficiently for the purpose of training the model weights of neural networks at UAVs. The numerical results confirm the stability of the proposed method and show that it can be used to control a massive UAV population in a windy environment efficiently.

\appendices  
\numberwithin{equation}{section} 


\section{Proof of Proposition 1.} 
\label{APP:A}
This proof is based on methodology of \cite{b2}, but with a few  differences in system model and the update algorithms.
The candidate Lyapunov function is chosen as

\vspace{-5pt}
\fontsize{8}{0}\selectfont
\begin{align}
L(t) \!=\! \frac{1}{2 \mu_{\hspace{.5pt}\textsf{H}} }  \tilde{w}_{\textsf{H}_0}^{\T} \tilde{w}_{\textsf{H}_0} \!+\! \frac{1}{2 \mu_{\hspace{.5pt}\textsf{H}}  }  \tilde{w}_{\textsf{H}_1}^{\T} \tilde{w}_{\textsf{H}_1} \!+\! c_{\hspace{.5pt}\textsf{H}} {L}_s \mathbbm{1}_{\| s_i \|\geq s_\text{dest}} .
\end{align} \normalsize
Then, the corresponding derivative function is

\vspace{-5pt}
\fontsize{8}{0}\selectfont
\begin{align}
\dot{L}(t) \!=\!  \frac{1}{ \mu_{\hspace{.5pt}\textsf{H}} }  \tilde{w}_{\textsf{H}_0}^{\T} \! \dot{\tilde{w}}_{\textsf{H}_0}   \!+ \! \frac{1}{ \mu_{\hspace{.5pt}\textsf{H}} }  \tilde{w}_{\textsf{H}_1}^{\T}\! \dot{\tilde{w}}_{\textsf{H}_1}  \!+\! c_{\hspace{.5pt}\textsf{H}}  [\nabla {L}_s ]^\T  \dot{s} \mathbbm{1}_{\| s_i \|\geq s_\text{dest}} ,
\end{align} \normalsize
which can be rewritten in the following form

\vspace{-0pt}
\fontsize{8}{0}\selectfont
\begin{align}
\dot{L}(t) &= \tilde{w}_{\textsf{H}_0}^{\T} [ ( {  \nabla_{\hat{w}_{\textsf{H}_0}} \! e_\textsf{H}}) e_\textsf{H} \!+\!  c_{\hspace{.5pt}\textsf{H}} \nabla_{\hat{w}_{\textsf{H}_0}} \! R_i] 
\nonumber \\
&\quad
+  \tilde{w}_{\textsf{H}_1}^{\T} [ ( {  \nabla_{\hat{w}_{\textsf{H}_1}} \! e_\textsf{H}}) e_\textsf{H} ] \!+\! c_{\hspace{.5pt}\textsf{H}}  [\nabla {L}_s ]^\T \dot{s} \mathbbm{1}_{\| s_i \|\geq s_\text{dest}} 
 \nonumber \\
&= [\tilde{w}_{\textsf{H}_0}^{\T}  ( {  \nabla_{\hat{w}_{\textsf{H}_0}} \! e_\textsf{H}})  \!+\!  \tilde{w}_{\textsf{H}_1}^{\T}  ( {  \nabla_{\hat{w}_{\textsf{H}_1}} \! e_\textsf{H}})  ]e_\textsf{H}
 \nonumber \\
&\quad
+  c_{\hspace{.5pt}\textsf{H}} \tilde{w}_{\textsf{H}_0}^{\T} \nabla_{\hat{w}_{\textsf{H}_0}} R_i + c_{\hspace{.5pt}\textsf{H}}  [\nabla  {L}_s ]^\T \dot{s} \mathbbm{1}_{\| s_i \|\geq s_\text{dest}} \nonumber \\
&= -[ \tilde{w}_{\textsf{H}_0}^{\T}  b_\textsf{H} + \tilde{w}_{\textsf{H}_0}^{\T} 	R_\textsf{H} \tilde{w}_{\textsf{H}_0} +  \tilde{w}_{\textsf{H}_1}^{\T}   \sigma_\textsf{H}  ] 
\nonumber \\
&\quad
 \times [ \tilde{w}_{\textsf{H}_0}^{\T}  b_\textsf{H} + \frac{1}{2} \tilde{w}_{\textsf{H}_0}^{\T} 	R_\textsf{H} \tilde{w}_{\textsf{H}_0} +  \tilde{w}_{\textsf{H}_1}^{\T}   \sigma_\textsf{H} + \! \varepsilon_{\textsf{H} }] 
 \nonumber \\
&\quad  
+  c_{\hspace{.5pt}\textsf{H}} \tilde{w}_{\textsf{H}_0}^{\T} \nabla_{\hat{w}_{\textsf{H}_0}} R_i + c_{\hspace{.5pt}\textsf{H}}  [\nabla {L}_s ]^\T \dot{s} \mathbbm{1}_{\| s_i \|\geq s_\text{dest}} \label{Eq:Der_Lyap}
\end{align} \normalsize 
where $ b_\textsf{H} $ and matrix $ R_\textsf{H} $ are defined as

\vspace{-0pt}
\fontsize{8}{0}\selectfont
\begin{align}
\hspace{-5pt}	b_\textsf{H} &= [\nabla \sigma_\textsf{H}] (f  \!-\! \frac{1}{2 c_3} B B^{\T} [\nabla \sigma_\textsf{H}]^{\T} {w}_{\textsf{H}_0}  \!-\! \frac{1}{2 c_3} B B^{\T}  [ \nabla \varepsilon_{\textsf{H}_0}] 
	\nonumber \\
	&\quad
	+ \frac{1}{2 c_3} [\nabla \sigma_\textsf{H}] B B^{\T}  [ \nabla \varepsilon_{\textsf{H}_0}] )   \!+\! \frac{1}{2} \sum_{k=1}^{N} \textup{tr}(G G^{T}  [\Delta  \sigma_\textsf{H}^{[k]}]  ) {\mathbf{e}}_{k}  \nonumber  \\
	& \!=\! [\nabla \sigma_\textsf{H}] \bar{\dot{s}}    \!+\! \frac{1}{2} \sum_{k=1}^{N} \textup{tr}(G G^{T}  [\Delta  \sigma_\textsf{H}^{[k]}]  ) {\mathbf{e}}_{k}  
	\!+\! \frac{1}{2 c_3} [\nabla \sigma_\textsf{H}] B B^{\T}  [ \nabla \varepsilon_{\textsf{H}_0}], \\
	R_\textsf{H} &= \frac{1}{2 c_3}   [\nabla \sigma_\textsf{H}] B B^{\T} [\nabla \sigma_\textsf{H}]^{\T}  ,
\end{align}\normalsize
and $ \bar{\dot{s}} $ is the dynamics of nominal system defined as

\vspace{-5pt}
\fontsize{7.5}{0}\selectfont
\small\begin{align}
	\bar{\dot{s}} &= f  \!-\! \frac{1}{2 c_3} B B^{\T} [\nabla \sigma_\textsf{H}]^{\T} {w}_{\textsf{H}_0}  \!-\! \frac{1}{2 c_3} B B^{\T}  [ \nabla \varepsilon_{\textsf{H}_0}] .
\end{align}\normalsize

\noindent Using the following relation as

\vspace{-0pt}
\fontsize{8}{0}\selectfont
\begin{align}
ab = \frac{1}{2}\left( -(ha-\frac{b}{h})^2 +h^2 a^2 + \frac{b^2}{h^2}  \right),
\end{align} \normalsize
for scalars $ a $ and $ b $, we obtain the upper-bounds for the terms on the right hand side of \eqref{Eq:Der_Lyap} as

\vspace{-0pt}
\fontsize{7.5}{0}\selectfont
\begin{align}
- \frac{3}{2} ( \tilde{w}_{\textsf{H}_0}^{\T}  b_\textsf{H} ) (\tilde{w}_{\textsf{H}_0}^{\T} 	R_\textsf{H} \tilde{w}_{\textsf{H}_0} )  
&\leq \frac{3}{4} h_1^2 \lambda_{1M}^2 \| \tilde{w}_{\textsf{H}_0} \|^2 \!+\! \frac{3}{4  h_1^2}  \lambda_{2M}^2 \| \tilde{w}_{\textsf{H}_0}  \|^4 ,
\\
-  (\tilde{w}_{\textsf{H}_0}^{\T}  b_\textsf{H} ) ( \varepsilon_{\textsf{H} } )  &\leq \frac{1}{2} h_2^2 \lambda_{1M}^2 \| \tilde{w}_{\textsf{H}_0}  \|^2 + \frac{1}{2 h_2^2}  \lambda_{3M}^2 ,
\\
- (\tilde{w}_{\textsf{H}_0}^{\T} 	R_\textsf{H} \tilde{w}_{\textsf{H}_0}) ( \varepsilon_{\textsf{H} } )  &\leq  \frac{1}{2} h_3^2 \lambda_{2M}^2 \| \tilde{w}_{\textsf{H}_0} \|^4 + \frac{1}{2 h_3^2}  \lambda_{3M}^2 ,
\\
-\frac{3}{2} ( \tilde{w}_{\textsf{H}_1}^{\T}   \sigma_\textsf{H} )(\tilde{w}_{\textsf{H}_0}^{\T} 	R_\textsf{H} \tilde{w}_{\textsf{H}_0} ) &\leq \frac{3}{4} h_4^2 \lambda_{4M}^2 \| \tilde{w}_{\textsf{H}_1} \|^2 + \frac{3}{ 4 h_4^2}  \lambda_{2M}^2 \| \tilde{w}_{\textsf{H}_0} \|^4 ,
\\
-(\tilde{w}_{\textsf{H}_1}^{\T}   \sigma_\textsf{H}) (\varepsilon_{\textsf{H} })  &\leq \frac{1}{2} h_5^2 \lambda_{4M}^2 \| \tilde{w}_{\textsf{H}_1} \|^2  + \frac{1}{2  h_5^2}  \lambda_{3M}^2 ,
\\
-  2 ( \tilde{w}_{\textsf{H}_0}^{\T}  b_\textsf{H}  ) (\tilde{w}_{\textsf{H}_1}^{\T}   \sigma_\textsf{H})  &\leq  h_6^2 \lambda_{1M}^2 \| \tilde{w}_{\textsf{H}_0} \|^2 + \frac{1}{  h_6^2}  \lambda_{4M}^2 \| \tilde{w}_{\textsf{H}_1} \|^2 ,
\\
- ( \tilde{w}_{\textsf{H}_0}^{\T}  b_\textsf{H} )^2 & \leq - \lambda_{1m}^2 \| \tilde{w}_{\textsf{H}_0} \|^2   ,
\\
- \frac{1}{2} (\tilde{w}_{\textsf{H}_0}^{\T} 	R_\textsf{H} \tilde{w}_{\textsf{H}_0} )^2 & \leq - \frac{1}{2} \lambda_{2m}^2 \| \tilde{w}_{\textsf{H}_0} \|^4  ,
\\
-(\tilde{w}_{\textsf{H}_1}^{\T}   \sigma_\textsf{H})^2 & \leq  - \lambda_{4m}^2 \| \tilde{w}_{\textsf{H}_1} \|^2   \text{,}
\end{align} \normalsize
where we assumed that

\vspace{-0pt}
\fontsize{8}{0}\selectfont
\begin{align}
	\lambda_{1m}  \leq  \|  b_\textsf{H} \| &\leq  \lambda_{1M}, \label{Assump:HJB1}  \\
	\lambda_{2m}  \leq  \|  R_\textsf{H} \| &\leq  \lambda_{2M} , \label{Assump:HJB2}  \\
	\| \varepsilon_{\textsf{H} } \| &\leq  \lambda_{3M}, \label{Assump:HJB3}  \\
	\lambda_{4m}  \leq  \| \sigma_\textsf{H} \| &\leq  \lambda_{4M} . \label{Assump:HJB4}
\end{align} \normalsize

\noindent
Therefore, the derivative of Lyapunov function is upper-bounded as

\vspace{-0pt}
\fontsize{8}{0}\selectfont
\begin{align}
\dot{L}(t) & \leq - \lambda_0 \| \tilde{w}_{\textsf{H}_0} \|^4  + \lambda_1 \| \tilde{w}_{\textsf{H}_0} \|^2 + \lambda_2^2 - \lambda_3  \| \tilde{w}_{\textsf{H}_1} \|^2 
\nonumber \\
&\quad
+  c_{\hspace{.5pt}\textsf{H}} \tilde{w}_{\textsf{H}_0}^{\T} \nabla_{\hat{w}_{\textsf{H}_0}} \! R_i + c_{\hspace{.5pt}\textsf{H}}  [\nabla {L}_s ]^\T \dot{s}  \mathbbm{1}_{\| s_i \|\geq s_\text{dest}} , \label{Eq:Lyap_HJB}
\end{align} \normalsize
where \small$ \lambda_0 $, $ \lambda_1 $, $ \lambda_2 $, and $ \lambda_3 $\normalsize\, are defined as

\vspace{-0pt}
\fontsize{8}{0}\selectfont
\begin{align}
\lambda_0 &= -\frac{3}{4  h_1^2}  \lambda_{2M}^2 - \frac{1}{2} h_3^2 \lambda_{2M}^2 - \frac{3}{ 4 h_4^2}  \lambda_{2M}^2  + \frac{1}{2} \lambda_{2m}^2 , \nonumber \\
\lambda_1 &= \frac{3}{4 } h_1^2 \lambda_{1M}^2  + \frac{1}{2} h_2^2 \lambda_{1M}^2 +   h_6^2 \lambda_{1M}^2 - \lambda_{1m}^2 , \nonumber \\
\lambda_3 &= - \frac{3}{4} h_4^2 \lambda_{4M}^2 - \frac{1}{2} h_5^2 \lambda_{4M}^2 - \frac{1}{  h_6^2}  \lambda_{4M}^2 + \lambda_{4m}^2 , \nonumber \\
\lambda_2^2 &= \frac{1}{ 2  h_2^2}  \lambda_{3M}^2 + \frac{1}{ 2 h_3^2}  \lambda_{3M}^2  + \frac{1}{ 2 h_5^2}  \lambda_{3M}^2  .
\end{align} \normalsize

Depending on the state of the UAV, three cases can occur in \eqref{Eq:Lyap_HJB} as

\textbf{Case 1}: \small $ \mathbbm{1}_{\| s_i \|\geq s_\text{dest}}  = 0  $\normalsize.
With this condition, we can conclude that the UAVs are in destination, and we focus only on the weights of the models

\vspace{-0pt}
\fontsize{8}{0}\selectfont
\begin{align}
\dot{L}(t)
& \leq - \lambda_0 \| \tilde{w}_{\textsf{H}_0} \|^4  + \lambda_1 \| \tilde{w}_{\textsf{H}_0} \|^2 + \lambda_2^2 - \lambda_3  \| \tilde{w}_{\textsf{H}_1} \|^2  .
\end{align} \normalsize
Then, when the following conditions hold, i.e., 

\vspace{-0pt}
\fontsize{8}{0}\selectfont
\begin{align}
\| \tilde{w}_{\textsf{H}_0} \| &\geq  \sqrt{\frac{\lambda_1 + \sqrt{\lambda_1^2 + 4 \lambda_0 \lambda_2^2 }}{2 \lambda_0 } } \delequal \omega_{0,1},  \\
\| \tilde{w}_{\textsf{H}_1} \| &\geq \sqrt{ \frac{4  \lambda_2^2 \lambda_0 + \lambda_1^2}{4 \lambda_3 \lambda_0} } \delequal \omega_{1,1} ,
\end{align} \normalsize
the stability condition $ \dot{L}(t) < 0 $ is satisfied.

\textbf{case 2}: \small$ \mathbbm{1}_{\| s_i \|\geq s_\text{dest}}  = 1 $ and $  \dot{L}_s  \leq  0 $\normalsize.
In this case, the regulizer term  is inactive, and the upper-bound for derivative of Lyapunov is reduced to

\vspace{-0pt}
\fontsize{7.5}{0}\selectfont
\begin{align}
\dot{L}(t) & \leq \!-\! \lambda_0 \| \tilde{w}_{\textsf{H}_0} \|^4  \!+\! \lambda_1 \| \tilde{w}_{\textsf{H}_0} \|^2 \!+\! \lambda_2^2 \!-\! \lambda_3  \| \tilde{w}_{\textsf{H}_1} \|^2 
+ c_{\hspace{.5pt}\textsf{H}}  [\nabla {L}_s ]^\T \dot{s} \mathbbm{1}_{(\| s_i \|\geq s_\text{dest})}  \nonumber \\
& \leq - \lambda_0 \| \tilde{w}_{\textsf{H}_0} \|^4  + \lambda_1 \| \tilde{w}_{\textsf{H}_0} \|^2 + \lambda_2^2 - \lambda_3  \| \tilde{w}_{\textsf{H}_1} \|^2 
+ c_{\hspace{.5pt}\textsf{H}} \lambda_{4} \| \nabla {L}_s \|
\end{align} \normalsize
where $ \lambda_{4} $ is a number such that \small$ 0 < \lambda_{4} \| \nabla {L}_s \| \leq  \!-\! [\nabla {L}_s ]^\T \dot{s}  $ \normalsize.
Therefore, when the following inequalities as

\vspace{-0pt}
\fontsize{8}{0}\selectfont
\begin{align}
\| \tilde{w}_{\textsf{H}_0} \| &\geq  \sqrt{\frac{\lambda_1 + \sqrt{\lambda_1^2 + 4 \lambda_0 \lambda_2^2 }}{2 \lambda_0 } } \delequal \omega_{0,2} \\
\| \tilde{w}_{\textsf{H}_1} \| &\geq \sqrt{ \frac{4  \lambda_2^2 \lambda_0 + \lambda_1^2}{4 \lambda_3 \lambda_0} } \delequal \omega_{1,2}  \\
\| \nabla {L}_s(s_i(t)) \| &\geq { \frac{4  \lambda_2^2 \lambda_0 + \lambda_1^2}{4 \lambda_{4} \lambda_0} } \delequal \gamma_{2}
\end{align} \normalsize
occur, the stability condition $ \dot{L}(t) < 0 $ holds.

\textbf{case 3}: \small$ \mathbbm{1}_{\| s_i \|\geq s_\text{dest}}  = 1 $ and $ \dot{L}_s  \geq  0 $\normalsize. In this case, we find the upper-bound for the $ \dot{L}(t) $ as

\vspace{-0pt}
\fontsize{8}{0}\selectfont
\begin{align}
\dot{L}(t) & \leq - \lambda_0 \| \tilde{w}_{\textsf{H}_0} \|^4  + \lambda_1 \| \tilde{w}_{\textsf{H}_0} \|^2 + \lambda_2^2 - \lambda_3  \| \tilde{w}_{\textsf{H}_1} \|^2 
\nonumber \\
&\qquad
+  c_{\hspace{.5pt}\textsf{H}} \tilde{w}_{\textsf{H}_0}^{\T} \nabla_{\hat{w}_{\textsf{H}_0}} \! R_i + c_{\hspace{.5pt}\textsf{H}}  [\nabla  {L}_s ]^\T \dot{s} \mathbbm{1}_{\| s_i \|\geq s_\text{dest}}   \nonumber \\
& = - \lambda_0 \| \tilde{w}_{\textsf{H}_0} \|^4  + \lambda_1 \| \tilde{w}_{\textsf{H}_0} \|^2 + \lambda_2^2 - \lambda_3  \| \tilde{w}_{\textsf{H}_1} \|^2 
\nonumber \\
&\qquad
+  c_{\hspace{.5pt}\textsf{H}} \tilde{w}_{\textsf{H}_0}^{\T} \nabla_{\hat{w}_{\textsf{H}_0}} [[\nabla  {L}_s ]^\T \dot{s}] + c_{\hspace{.5pt}\textsf{H}}  [\nabla  {L}_s ]^\T \dot{s}   \nonumber \\
& \stackrel{(1)}{=} - \lambda_0 \| \tilde{w}_{\textsf{H}_0} \|^4  + \lambda_1 \| \tilde{w}_{\textsf{H}_0} \|^2 + \lambda_2^2 - \lambda_3  \| \tilde{w}_{\textsf{H}_1} \|^2 
\nonumber \\
&\qquad
+ c_{\hspace{.5pt}\textsf{H}}  [\nabla  {L}_s ]^\T \bar{\dot{s}} + \frac{c_{\hspace{.5pt}\textsf{H}}}{2 c_3}   [\nabla  {L}_s ]^\T  B B^{\T}  [ \nabla \varepsilon_{\textsf{H}_0}]   \nonumber \\
& \stackrel{(2)}{\leq}  - \lambda_0 \| \tilde{w}_{\textsf{H}_0} \|^4  + \lambda_1 \| \tilde{w}_{\textsf{H}_0} \|^2 + \lambda_2^2 - \lambda_3  \| \tilde{w}_{\textsf{H}_1} \|^2 
\nonumber \\
&\qquad
- c_{\hspace{.5pt}\textsf{H}} \lambda_{5m} \| \nabla  {L}_s \|^2 + \frac{c_{\hspace{.5pt}\textsf{H}}}{2 c_3} \lambda_{6M} \| \nabla  {L}_s \|  ,
\end{align} \normalsize
where equality (1) is obtained by the calculations as

\vspace{-0pt}
\fontsize{8}{0}\selectfont
 \begin{align}
	 &c_{\hspace{.5pt}\textsf{H}} \tilde{w}_{\textsf{H}_0}^{\T} \nabla_{\hat{w}_{\textsf{H}_0}} [[\nabla  {L}_s ]^\T \dot{s}] = \frac{1}{2 c_3} c_{\hspace{.5pt}\textsf{H}} [\tilde{w}_{\textsf{H}_0}]^{\T} [\nabla \sigma_\textsf{H}] B B^{\T}  [\nabla  {L}_s ] ,
\\
&c_{\hspace{.5pt}\textsf{H}} [\nabla  {L}_s ]^\T \dot{s} =  c_{\hspace{.5pt}\textsf{H}} [\nabla  {L}_s ]^\T (f - \frac{1}{2 c_3} B B^{\T} [\nabla \sigma_\textsf{H}]^{\T} \hat{w}_{\textsf{H}_0} ) ,
\end{align} \normalsize
and inequality (2) is based on the following assumptions,

\vspace{-0pt}
\fontsize{8}{0}\selectfont
\begin{align}
	[\nabla  {L}_s ]^\T \bar{\dot{s}}  &= - [\nabla  {L}_s ]^\T A [\nabla {L}_s ] , \label{Assump:HJB5}  \\
	& \leq - \lambda_{5m} \| \nabla {L}_s \|^2 , \label{Assump:HJB6}
\end{align} \normalsize

\vspace{-10pt}\small \begin{align}
	B B^{\T}  [ \nabla \varepsilon_{\textsf{H}_0}]  &\leq  \lambda_{6M}. \label{Assump:HJB7}
\end{align} \normalsize
where $ \lambda_{5m} $ is the minimum eigenvalue of matrix $ A $.

Therefore, when the following conditions occur, i.e., 

\vspace{-0pt}
\fontsize{8}{0}\selectfont
\begin{align}
\| \tilde{w}_{\textsf{H}_0} \| &\geq  \sqrt{\frac{\lambda_1 + \sqrt{\lambda_1^2 + 4 \lambda_0 (\lambda_2^2 + \frac{c_{\hspace{.5pt}\textsf{H}} \lambda_{6M}^2}{16 c_3^2 \lambda_{5m} }) }}{2 \lambda_0 } } \delequal \omega_{0,3}  \\
\| \tilde{w}_{\textsf{H}_1} \| &\geq \sqrt{ \frac{4  (\lambda_2^2 + \frac{c_{\hspace{.5pt}\textsf{H}} \lambda_{6M}^2}{16 c_3^2 \lambda_{5m} }) \lambda_0 + \lambda_1^2}{4 \lambda_3 \lambda_0} } \delequal \omega_{1,3}  \\
\| \nabla {L}_s(s_i(t)) \| &\geq \frac{\frac{c_{\hspace{.5pt}\textsf{H}}}{2 c_3} \lambda_{6M} \!+\! \sqrt{(\frac{c_{\hspace{.5pt}\textsf{H}}}{2 c_3} \lambda_{6M} )^2 \!+\! 4 c_{\hspace{.5pt}\textsf{H}} \lambda_{5m} (\lambda_2^2 \!+\! \frac{\lambda_1^2}{4 \lambda_0})  } }{c_{\hspace{.5pt}\textsf{H}} \lambda_{5m}} \delequal \gamma_{3} 
\end{align} \normalsize
the Lyapunov stability condition holds, i.e., $ \dot{L}(t) < 0 $.

In summary, when $ \| \tilde{w}_{\textsf{H}_0} \| \geq \omega_{0} = \max \{\omega_{0,1}, \omega_{0,2}, \omega_{0,3}  \} $, or $ \| \tilde{w}_{\textsf{H}_1} \| \geq \omega_{1} = \max \{\omega_{1,1}, \omega_{1,2}, \omega_{1,3}  \} $, or $  \| \nabla {L}_s(s_i(t)) \| \geq  \max \{ \gamma_{2} , \gamma_{3}  \} $ occurs, then the Lyapunov stability condition holds, i.e., $ \dot{L}(t) \! < \! 0 $.
Considering all the cases 1-3, we can conclude that there exist $ s_\text{dest} $, $ w_0 $, and $ w_1 $ at time $ T $ such that \small$ \| s(t) \| \!\leq\! s_\text{dest}  $\normalsize, \small$ \| {w}_{\textsf{H}_0}\!(t) - \hat{w}_{\textsf{H}_0}\!(t) \| \!\leq\!  w_0  $\normalsize, and \small$ \| {w}_{\textsf{H}_1}\!(t) - \hat{w}_{\textsf{H}_1}\!(t) \| \!\leq  \! w_1 ~ $\normalsize for all \small$ t \geq T + T' $\normalsize.

\section{Proof of Proposition 2.}
\label{APP:B}
The candidate Lyapunov function is chosen as

\vspace{-5pt}
\fontsize{8}{0}\selectfont
\begin{align}
L(t) = \frac{1}{2 \mu_{\hspace{.5pt}\textsf{F}}  }  \tilde{w}_{\textsf{F}_0}^{\T} \tilde{w}_{\textsf{F}_0} + \frac{1}{2 \mu_{\hspace{.5pt}\textsf{F}}  }  \tilde{w}_{\textsf{F}_1}^{\T} \tilde{w}_{\textsf{F}_1} .
\end{align} \normalsize
Then, the derivative of Lyapunov function is obtained as

\vspace{-0pt}
\fontsize{8}{0}\selectfont
\begin{align}
\dot{L}(t) &=  \tilde{w}_{\textsf{F}_0}^{\T} ( {  \nabla_{\hat{w}_{\textsf{F}_0}} e_\textsf{F}}) e_\textsf{F}  + \tilde{w}_{\textsf{F}_1}^{\T} ( {  \nabla_{\hat{w}_{\textsf{F}_1}} e_\textsf{F}}) e_\textsf{F} , \nonumber \\
& = - [\tilde{w}_{\textsf{F}_0}^\T [{  \nabla_{\hat{w}_{\textsf{F}_0}} e_\textsf{F}} ]   +   \tilde{w}_{\textsf{F}_1}^\T [{  \nabla_{\hat{w}_{\textsf{F}_1}} e_\textsf{F}} ]  ] 
, \nonumber \\
&\qquad 
\times [\tilde{w}_{\textsf{F}_0}^\T [{  \nabla_{\hat{w}_{\textsf{F}_0}} e_\textsf{F}} ]   +   \tilde{w}_{\textsf{F}_1}^\T [{  \nabla_{\hat{w}_{\textsf{F}_1}} e_\textsf{F}} ] + \varepsilon_{\textsf{F}}  ] , \nonumber \\
& = - [\tilde{w}_{\textsf{F}}^\T [{  \nabla_{\hat{w}_{\textsf{F}}} e_\textsf{F}} ]   ]  [\tilde{w}_{\textsf{F}}^\T [{  \nabla_{\hat{w}_{\textsf{F}}} e_\textsf{F}} ] + \varepsilon_{\textsf{F}}  ]  .
\end{align} \normalsize

\noindent Each term of the derivative of Lyapunov function has an upper-bound obtained as

\vspace{-0pt}
\fontsize{8}{0}\selectfont
\begin{align}
-(\tilde{w}_{\textsf{F}}^\T {  \nabla_{\hat{w}_{\textsf{F}}} \! e_\textsf{F}} )^2 \leq - \lambda_{7m}^2 \| \tilde{w}_{\textsf{F}} \|^2 ,
\end{align} \normalsize

\vspace{-10pt}
\fontsize{8}{0}\selectfont
\begin{align}
-(\tilde{w}_{\textsf{F}}^\T [{  \nabla_{\hat{w}_{\textsf{F}}} e_\textsf{F}} ]) (\varepsilon_{\textsf{F}} )  &\leq  \lambda_{8M} \lambda_{7M} \| \tilde{w}_{\textsf{F}}^\T  \| ,
\end{align} \normalsize
where it is assumed that

\vspace{-0pt}
\fontsize{8}{0}\selectfont
\begin{align}
\lambda_{7m} \leq  \| {  \nabla_{\hat{w}_{\textsf{F}}}   e_\textsf{F}} \| \leq \lambda_{7M} , \label{Assump:FPK1} \\
\| { \varepsilon_{\textsf{F}} } \| \leq \lambda_{8M} . \label{Assump:FPK2}
\end{align}\normalsize

\noindent
Then, the derivative of Lyapunov is upper-bounded as

\vspace{-0pt}
\fontsize{8}{0}\selectfont
\begin{align}
\dot{L}(t) &\leq - \lambda_{7m}^2 \| \tilde{w}_{\textsf{F}} \|^2 + \lambda_{8M} \lambda_{7M} \| \tilde{w}_{\textsf{F}}^\T  \| .
\end{align} \normalsize

\noindent
Therefore, when the following condition occurs, i.e.,

\vspace{-0pt}
\fontsize{8}{0}\selectfont
\begin{align}
\| \tilde{w}_{\textsf{F}}^\T  \| \geq  \frac{\lambda_{8M} \lambda_{7M}}{ \lambda_{7m}^2 } \text{,} \label{Eq:FPK_stab_cond}
\end{align} \normalsize
the stability condition holds, i.e.,  $ \dot{L}(t) \leq 0 $. 
However, \eqref{Eq:FPK_stab_cond} means that the model which makes the term $  \| { \varepsilon_{\textsf{F}} } \| $  small, can increase the stability of FPK learning algorithm.

\section{Proof of Proposition 3.}
\label{APP:C}
Here we aim to show the convergence and bias of the FPK learning updates following the proof method of \cite{shiri2018distributed}.
Let us first define extended vectors as

\vspace{-0pt}
\fontsize{8}{0}\selectfont
\begin{align}
\hat{w}_{\textsf{F}}{(n)} &= [ [\hat{w}_{\textsf{F}_0}{(n)}]^\T \qquad [\hat{w}_{\textsf{F}_1}{(n)}]^\T ]^\T , \\
\tilde{w}_{\textsf{F}}{(n)} &= [ [\tilde{w}_{\textsf{F}_0}{(n)}]^\T \qquad [\tilde{w}_{\textsf{F}_1}{(n)}]^\T  ]^\T , \\
{\nabla_{\hat{w}_{\textsf{F}}} e_\textsf{F}} &= [ [{\nabla_{\hat{w}_{\textsf{F}_0}} e_\textsf{F}}]^\T \qquad  [{\nabla_{\hat{w}_{\textsf{F}_1}} e_\textsf{F}}]^\T ]^\T .
\end{align} \normalsize
Then, the FPK error vector update is obtained as

\vspace{-0pt}
\fontsize{8}{0}\selectfont
\begin{align}
\tilde{w}_{\textsf{F}}{(n\!+\!1)} \!&=\! [I \!-\! \mu_{\hspace{.5pt}\textsf{F}} [{\nabla_{\hat{w}_{\textsf{F}}} e_\textsf{F}} ]  [{\nabla_{\hat{w}_{\textsf{F}}} e_\textsf{F}} ]^\T ] \tilde{w}_{\textsf{F}}{(n)}   - \mu_{\hspace{.5pt}\textsf{F}} [{\nabla_{\hat{w}_{\textsf{F}}} e_\textsf{F}} ] \varepsilon_{\textsf{F}} . \label{Eq:Weight_err}
\end{align}\normalsize

\subsection{FPK Mean Stability}

Taking the expectation of equation \eqref{Eq:Weight_err} yields

\vspace{-0pt}
\fontsize{8}{0}\selectfont
\begin{align}
\mathbb{E} [\tilde{w}_{\textsf{F}}{(n\!+\!1)}] &= [I \!-\! \mu_{\hspace{.5pt}\textsf{F}} \mathbb{E}[[{\nabla_{\hat{w}_{\textsf{F}}} e_\textsf{F}} ]  [{\nabla_{\hat{w}_{\textsf{F}}} e_\textsf{F}} ]^\T] ] \mathbb{E}[\tilde{w}_{\textsf{F}}{(n)}]  
\nonumber \\ 
&\quad 
- \mu_{\hspace{.5pt}\textsf{F}} \mathbb{E}[[{\nabla_{\hat{w}_{\textsf{F}}} e_\textsf{F}} ] \varepsilon_{\textsf{F}}] , \label{Eq:Weight_err_Exp}
\end{align}\normalsize
We can assume that the vector $ [{\nabla_{\hat{w}_{\textsf{F}}} e_\textsf{F}} ] $ which depends on the inputs, and $ \varepsilon_{\textsf{F}} $ which depends on the neural network design, are independent. Then we can write

\vspace{-0pt}
\fontsize{8}{0}\selectfont
\begin{align}
\mathbb{E}[[{\nabla_{\hat{w}_{\textsf{F}}} e_\textsf{F}} ] \varepsilon_{\textsf{F}}] = 0. \label{Eq:covar_e_eps}
\end{align}\normalsize

\noindent
Let us define the matrix $ R $ as

\vspace{-0pt}
\fontsize{8}{0}\selectfont
\begin{align}
R \delequal \mathbb{E}[[{\nabla_{\hat{w}_{\textsf{F}}} e_\textsf{F}} ]  [{\nabla_{\hat{w}_{\textsf{F}}} e_\textsf{F}} ]^\T] . \label{Eq:R} 
\end{align} \normalsize
By substituting \eqref{Eq:covar_e_eps} and \eqref{Eq:R} in equation \eqref{Eq:Weight_err_Exp}, it can be rewritten in the form

\vspace{-0pt}
\fontsize{8}{0}\selectfont
\begin{align}
\mathbb{E} [\tilde{w}_{\textsf{F}}{(n\!+\!1)}] &= [I - \mu_{\hspace{.5pt}\textsf{F}} R ] \mathbb{E}[\tilde{w}_{\textsf{F}}{(n)}] .
\end{align}\normalsize
Then, the necessary condition for the convergence of this equation is

\vspace{-10pt}
\fontsize{8}{0}\selectfont
\begin{align}
0 < \mu_{\hspace{.5pt}\textsf{F}} < \frac{2}{\lambda_{max}} \text{,} \label{Eq:step_cond}
\end{align} \normalsize
where $ {\lambda_{max}} $ is the largest eigenvalue of the matrix $ R $.

\subsection{Biasness}

Assuming small step-sizes and also the condition \eqref{Eq:step_cond}, the bias of estimation is calculated as

\vspace{-0pt}
\fontsize{8}{0}\selectfont
\begin{align}
\text{bias} &= \lim_{n \to \infty} - \mathbb{E} [\tilde{w}_{\textsf{F}}{(n)}] = 0,
\end{align}\normalsize
which means that if the step size is small enough such that the convergence condition holds, the parameters of the FPK equation tend to its optimal values.

\subsection{Mean Square Convergence Analysis}

The mean square deviation (MSD) of the estimation algorithm is defined as

\vspace{-0pt}
\fontsize{8}{0}\selectfont
\begin{align}
\text{MSD}_{\textsf{F}} &= \lim_{n \to \infty} \mathbb{E}[\| \tilde{w}_{\textsf{F}}{(n)} \|_{}^2] .
\end{align} \normalsize

\noindent
In order to find the MSD, let us first define the weighted MSD of the algorithm as $ \mathbb{E}[\| \tilde{w}_{\textsf{F}}{(n)} \|_{\Sigma}^2] $, which can be obtained by the recursive equation

\vspace{-0pt}
\fontsize{8}{0}\selectfont
\begin{align}
\mathbb{E}[\| \tilde{w}_{\textsf{F}}{(n+1)} \|_{\Sigma}^2] &=  \mathbb{E}[\| \tilde{w}_{\textsf{F}}{(n)} \|_{\Sigma^{'}}^2] +  \mu_{\hspace{.5pt}\textsf{F}}^2 \text{tr}(R_\Sigma) \| \varepsilon_{\textsf{F}} \|^{2} ,         
\end{align}\normalsize
where $ \Sigma $ is a positive definite matrix, and

\vspace{-0pt}
\fontsize{8}{0}\selectfont
\begin{align}
\Sigma{'} &= (I - \mu_{\hspace{.5pt}\textsf{F}} R)^{\T} \Sigma (I - \mu_{\hspace{.5pt}\textsf{F}} R) \text{,} \label{Eq:SigmaP} \\
R_\Sigma &= \mathbb{E}[[{\nabla_{\hat{w}_{\textsf{F}}} e_\textsf{F}} ]^{\T} \Sigma [{\nabla_{\hat{w}_{\textsf{F}}} e_\textsf{F}} ]  ] \text{.}  \label{Eq:R_sigma}
\end{align} \normalsize
We know that $ \text{tr}(\Sigma X) = [\text{vec}(X)]^{\T} \sigma $, and $ \text{vec}(U\Sigma V) = (V^\T \otimes U) \sigma   $, where $ \text{vec}(\cdot) $ is a vectorazation operator, i.e., $ \text{vec}(\Sigma) = \sigma $. Using these equalities, we can obtain

\vspace{-0pt}
\fontsize{8}{0}\selectfont
\begin{align}
\text{tr}(R_\Sigma) &=  [\text{vec}(R)]^{\T} \sigma, \\
\sigma{'} &= \mathcal{F} \sigma, \\
\mathcal{F} &= (I - \mu_{\hspace{.5pt}\textsf{F}} R)^{\T}  \otimes (I - \mu_{\hspace{.5pt}\textsf{F}} R)^{\T} .
\end{align} \normalsize

\noindent
At the convergence stage, the MSD is written as

\vspace{-0pt}
\fontsize{8}{0}\selectfont
\begin{align}
\lim_{n \to \infty} \mathbb{E}[\| \tilde{w}_{\textsf{F}}{(n)} \|_{\Omega}^2] &=  \mu_{\hspace{.5pt}\textsf{F}}^2  \| \varepsilon_{\textsf{F}} \|^{2} [\text{vec}(R)]^{\T} (I -\mathcal{F} )^{-1}  \text{vec}(\Omega)  ,
\end{align}\normalsize
where \small$ \text{vec}(\Omega) = (I -\mathcal{F}) \sigma $ \normalsize
Therefore the steady state MSD is obtained as

\vspace{-0pt}
\fontsize{8}{0}\selectfont
\begin{align}
\text{MSD}_{\textsf{F}}  &=  \mu_{\hspace{.5pt}\textsf{F}}^2  \| \varepsilon_{\textsf{F}} \|^{2} [\text{vec}(R)]^{\T} (I -\mathcal{F} )^{-1}  \text{vec}(I) ,       
\end{align}\normalsize

The value of MSD can be very small by choosing a small value for step sizes, i.e.,  $ \mu_{\hspace{.5pt}\textsf{F}} $, and choosing a model which makes $ \| \varepsilon_{\textsf{F}} \| $ small.

%

\begin{thebibliography}{10}
	\providecommand{\url}[1]{#1}
	\csname url@samestyle\endcsname
	\providecommand{\newblock}{\relax}
	\providecommand{\bibinfo}[2]{#2}
	\providecommand{\BIBentrySTDinterwordspacing}{\spaceskip=0pt\relax}
	\providecommand{\BIBentryALTinterwordstretchfactor}{4}
	\providecommand{\BIBentryALTinterwordspacing}{\spaceskip=\fontdimen2\font plus
		\BIBentryALTinterwordstretchfactor\fontdimen3\font minus
		\fontdimen4\font\relax}
	\providecommand{\BIBforeignlanguage}[2]{{%
			\expandafter\ifx\csname l@#1\endcsname\relax
			\typeout{** WARNING: IEEEtran.bst: No hyphenation pattern has been}%
			\typeout{** loaded for the language `#1'. Using the pattern for}%
			\typeout{** the default language instead.}%
			\else
			\language=\csname l@#1\endcsname
			\fi
			#2}}
	\providecommand{\BIBdecl}{\relax}
	\BIBdecl
	
	\bibitem{b3}
	H.~Kim, J.~Park, M.~Bennis, and S.-L. Kim, ``Massive {UAV}-to-ground
	communication and its stable movement control: A mean-field approach,'' in
	\emph{Proc. IEEE SPAWC, Kalamata, Greece}, Jun. 2018.
	
	\bibitem{Ackerman:18}
	E.~{Ackerman} and E.~{Strickland}, ``Medical delivery drones take flight in
	east africa,'' \emph{IEEE Spectrum}, vol.~55, no.~1, pp. 34--35, Jan. 2018.
	
	\bibitem{b45}
	J.~{Tisdale}, Z.~{Kim}, and J.~K. {Hedrick}, ``Autonomous {UAV} path planning
	and estimation,'' \emph{IEEE Robot. Autom. Mag.}, vol.~16, no.~2, pp. 35--42,
	Jun. 2009.
	
	\bibitem{shiri2019massive}
	H.~Shiri, J.~Park, and M.~Bennis, ``Massive autonomous {UAV} path planning: A
	neural network based mean-field game theoretic approach,'' \emph{arXiv
		preprint arXiv:1905.04152}, 2019.
	
	\bibitem{b10}
	M.~Huang, R.~P. Malham{\'e}, P.~E. Caines \emph{et~al.}, ``Large population
	stochastic dynamic games: closed-loop {M}c{K}ean-{V}lasov systems and the
	{N}ash certainty equivalence principle,'' \emph{Commun. Inf. Syst.}, vol.~6,
	no.~3, pp. 221--252, 2006.
	
	\bibitem{b11}
	M.~Huang, P.~E. Caines, and R.~P. Malham{\'e}, ``Large-population cost-coupled
	{LQG} problems with nonuniform agents: individual-mass behavior and
	decentralized $\varepsilon $-{N}ash equilibria,'' \emph{IEEE Trans. Autom.
		Control}, vol.~52, no.~9, pp. 1560--1571, 2007.
	
	\bibitem{b12}
	J.-M. Lasry and P.-L. Lions, ``Mean field games,'' \emph{Japanese J. math.
		(JJM)}, vol.~2, no.~1, pp. 229--260, 2007.
	
	\bibitem{mcmahan2016communication}
	H.~B. McMahan, E.~Moore, D.~Ramage, S.~Hampson \emph{et~al.},
	``Communication-efficient learning of deep networks from decentralized
	data,'' \emph{arXiv preprint arXiv:1602.05629}, 2016.
	
	\bibitem{shokri2015privacy}
	R.~Shokri and V.~Shmatikov, ``Privacy-preserving deep learning,'' in
	\emph{Proc. 22nd ACM SIGSAC conf. Comput. comm. Secur.}, 2015, pp.
	1310--1321.
	
	\bibitem{b14}
	D.~{Liu}, D.~{Wang}, F.~{Wang}, H.~{Li}, and X.~{Yang}, ``Neural-network-based
	online {HJB} solution for optimal robust guaranteed cost control of
	continuous-time uncertain nonlinear systems,'' \emph{IEEE Trans. Cybern.},
	vol.~44, no.~12, pp. 2834--2847, Dec. 2014.
	
	\bibitem{shiri2019remote}
	H.~{Shiri}, J.~{Park}, and M.~{Bennis}, ``Remote {UAV} online path planning via
	neural network based opportunistic control,'' \emph{IEEE Wireless Commun.
		Lett.}, pp. 1--1, Feb. 2020.
	
	\bibitem{valavanis2015handbook}
	K.~P. Valavanis and G.~J. Vachtsevanos, \emph{Handbook of unmanned aerial
		vehicles}.\hskip 1em plus 0.5em minus 0.4em\relax Springer, 2015, vol.~1.
	
	\bibitem{8329013}
	J.~{Lyu}, Y.~{Zeng}, and R.~{Zhang}, ``{UAV}-aided offloading for cellular
	hotspot,'' \emph{IEEE Trans. Wireless Commun.}, vol.~17, no.~6, pp.
	3988--4001, 2018.
	
	\bibitem{Feng19Coverage}
	A.~A. {Khuwaja}, G.~{Zheng}, Y.~{Chen}, and W.~{Feng}, ``Optimum deployment of
	multiple {UAV}s for coverage area maximization in the presence of co-channel
	interference,'' \emph{IEEE Access}, vol.~7, pp. 85\,203--85\,212, 2019.
	
	\bibitem{Lim10Surveillance}
	C.~W. {Lim}, C.~K. {Ryoo}, K.~{Choi}, and J.~{Cho}, ``Path generation algorithm
	for intelligence, surveillance and reconnaissance of an {UAV},'' in
	\emph{Proc. SICE Annu. Conf.}, 2010, pp. 1274--1277.
	
	\bibitem{lee2020integrating}
	J.-H. Lee, J.~Park, M.~Bennis, and Y.-C. Ko, ``Integrating {LEO} satellite and
	{UAV} relaying via reinforcement learning for non-terrestrial networks,''
	2020.
	
	\bibitem{7888557}
	Y.~{Zeng} and R.~{Zhang}, ``Energy-efficient {UAV} communication with
	trajectory optimization,'' \emph{IEEE Trans. Wireless Commun.}, vol.~16,
	no.~6, pp. 3747--3760, 2017.
	
	\bibitem{CellUAV2019Zhang}
	S.~{Zhang}, Y.~{Zeng}, and R.~{Zhang}, ``Cellular-enabled {UAV} communication:
	A connectivity-constrained trajectory optimization perspective,'' \emph{IEEE
		Trans. Commun.}, vol.~67, no.~3, pp. 2580--2604, 2019.
	
	\bibitem{Walid2019InterferenceUAVrl}
	U.~{Challita}, W.~{Saad}, and C.~{Bettstetter}, ``Interference management for
	cellular-connected {UAV}s: A deep reinforcement learning approach,''
	\emph{IEEE Trans. Wireless Commun.}, vol.~18, no.~4, pp. 2125--2140, 2019.
	
	\bibitem{Yang18}
	Q.~{Yang}, J.~{Zhang}, and G.~{Shi}, ``Path planning for unmanned aerial
	vehicle passive detection under the framework of partially observable markov
	decision process,'' in \emph{Chinese Control Decision Conf. (CCDC)}, Jun.
	2018, pp. 3896--3903.
	
	\bibitem{oatao19443}
	\BIBentryALTinterwordspacing
	J.-A. Delamer, Y.~Watanabe, and C.~P.~C. Chanel, ``Towards a {MOMDP} model for
	{UAV} safe path planning in urban environment,'' \emph{9th Int. Micro Air
		Veh. Conf. Competition (IMAV)}, pp. 1--8, Sep. 2017. [Online]. Available:
	\url{https://oatao.univ-toulouse.fr/19443/}
	\BIBentrySTDinterwordspacing
	
	\bibitem{Comm19UAVmardani}
	A.~{Mardani}, M.~{Chiaberge}, and P.~{Giaccone}, ``Communication-aware {UAV}
	path planning,'' \emph{IEEE Access}, vol.~7, pp. 52\,609--52\,621, Apr. 2019.
	
	\bibitem{zeng2019accessing}
	Y.~Zeng, Q.~Wu, and R.~Zhang, ``Accessing from the sky: A tutorial on {UAV}
	communications for 5{G} and beyond,'' \emph{arXiv preprint arXiv:1903.05289},
	2019.
	
	\bibitem{austin2011unmanned}
	R.~Austin, \emph{Unmanned aircraft systems: {UAVS} design, development and
		deployment}.\hskip 1em plus 0.5em minus 0.4em\relax John Wiley \& Sons, 2011,
	vol.~54.
	
	\bibitem{3gpp15}
	\BIBentryALTinterwordspacing
	``Enhanced {LTE} support for aerial vehicles.'' [Online]. Available:
	\url{Available: ftp://www.3gpp.org/specs/archive/36_series/36.777}
	\BIBentrySTDinterwordspacing
	
	\bibitem{azari2017coexistence}
	M.~M. Azari, F.~Rosas, A.~Chiumento, and S.~Pollin, ``Coexistence of
	terrestrial and aerial users in cellular networks,'' in \emph{IEEE {GLOBECOM}
		Workshops (GC Wkshps)}.\hskip 1em plus 0.5em minus 0.4em\relax IEEE, 2017,
	pp. 1--6.
	
	\bibitem{liu2018multibeam}
	\BIBentryALTinterwordspacing
	L.~Liu, S.~Zhang, and R.~Zhang, ``Multi-beam {UAV} communication in cellular
	uplink: Cooperative interference cancellation and sum-rate maximization,''
	\emph{Trans. Wireless. Comm.}, vol.~18, no.~10, p. 4679â€“4691, Oct. 2019.
	[Online]. Available: \url{https://doi.org/10.1109/TWC.2019.2926981}
	\BIBentrySTDinterwordspacing
	
	\bibitem{Taleb2018GC}
	H.~{Hellaoui}, A.~{Chelli}, M.~{Bagaa}, and T.~{Taleb}, ``Towards mitigating
	the impact of {UAV}s on cellular communications,'' in \emph{IEEE Global
		Commun. Conf. ({GLOBECOM})}, Dec. 2018.
	
	\bibitem{Taleb2019WCNC}
	H.~{Hellaoui}, A.~{Chelli}, M.~{Bagaa}, T.~{Taleb}, and M.~{PÃ¤tzold},
	``Towards efficient control of mobile network-enabled {UAV}s,'' in \emph{IEEE
		Wireless Commun. Netw. Conf. (WCNC)}, Apr. 2019.
	
	\bibitem{RZHANG19}
	S.~{Zhang} and R.~{Zhang}, ``Radio map based path planning for
	cellular-connected {UAV},'' in \emph{IEEE Global Commun. Conf. (GLOBECOM)},
	Dec. 2019, pp. 1--6.
	
	\bibitem{Wang19A2Achannel}
	C.~{Yan}, L.~{Fu}, J.~{Zhang}, and J.~{Wang}, ``A comprehensive survey on {UAV}
	communication channel modeling,'' \emph{IEEE Access}, vol.~7, pp.
	107\,769--107\,792, 2019.
	
	\bibitem{8579209}
	B.~{Li}, Z.~{Fei}, and Y.~{Zhang}, ``Uav communications for 5{G} and beyond:
	Recent advances and future trends,'' \emph{IEEE Internet of Things J.},
	vol.~6, no.~2, pp. 2241--2263, 2019.
	
	\bibitem{8624565}
	S.~{Zhang}, H.~{Zhang}, B.~{Di}, and L.~{Song}, ``Cellular {UAV}-to-{X}
	communications: Design and optimization for multi-{UAV} networks,''
	\emph{IEEE Trans. Wireless Commun.}, vol.~18, no.~2, pp. 1346--1359, 2019.
	
	\bibitem{b20}
	H.~M. La, ``Multi-robot swarm for cooperative scalar field mapping,'' in
	\emph{Handbook of Research on Design, Control, and Modeling of Swarm
		Robotics}.\hskip 1em plus 0.5em minus 0.4em\relax IGI Global, 2016, pp.
	383--395.
	
	\bibitem{b21}
	H.~M. La, W.~Sheng, and J.~Chen, ``Cooperative and active sensing in mobile
	sensor networks for scalar field mapping,'' \emph{IEEE Trans. Syst., Man,
		Cybern. Syst.}, vol.~45, no.~1, pp. 1--12, 2015.
	
	\bibitem{b26}
	P.~Sujit, S.~Saripalli, and J.~B. Sousa, ``Unmanned aerial vehicle path
	following: A survey and analysis of algorithms for fixed-wing unmanned aerial
	vehicless,'' \emph{IEEE Contr. Syst. Mag.}, vol.~34, no.~1, pp. 42--59, 2014.
	
	\bibitem{cano2013quadrotor}
	J.~M. Cano \emph{et~al.}, ``Quadrotor {UAV} for wind profile
	characterization,'' \emph{Proyectos Fin de Carrera, Universidad Carlos III de
		Madrid}, 2013.
	
	\bibitem{abichandani2020wind}
	P.~Abichandani, D.~Lobo, G.~Ford, D.~Bucci, and M.~Kam, ``Wind measurement and
	simulation techniques in multi-rotor small unmanned aerial vehicles,''
	\emph{IEEE Access}, vol.~8, pp. 54\,910--54\,927, Mar. 2020.
	
	\bibitem{Agate13Env}
	H.~{Chen}, K.~{Chang}, and C.~S. {Agate}, ``{UAV} path planning with
	tangent-plus-{L}yapunov vector field guidance and obstacle avoidance,''
	\emph{IEEE Trans. Aerosp. Electron. Syst.}, vol.~49, no.~2, pp. 840--856,
	2013.
	
	\bibitem{cui2018uav}
	J.-h. Cui, R.-x. Wei, Z.-c. Liu, and K.~Zhou, ``{UAV} motion strategies in
	uncertain dynamic environments: A path planning method based on {Q}-learning
	strategy,'' \emph{Appl. Sci.}, vol.~8, no.~11, p. 2169, 2018.
	
	\bibitem{wang20Prediction}
	\BIBentryALTinterwordspacing
	W.~Li, M.~Tan, L.~Wang, and Q.~Wang, ``A cubic spline method combing improved
	particle swarm optimization for robot path planning in dynamic uncertain
	environment,'' \emph{Int. J. Adv. Robot. Syst.}, vol.~17, no.~1, p.
	1729881419891661, 2020. [Online]. Available:
	\url{https://doi.org/10.1177/1729881419891661}
	\BIBentrySTDinterwordspacing
	
	\bibitem{b22}
	H.~X. Pham, H.~M. La, D.~Feil-Seifer, and L.~V. Nguyen, ``Autonomous {UAV}
	navigation using reinforcement learning,'' \emph{arXiv preprint
		arXiv:1801.05086}, 2018.
	
	\bibitem{b23}
	\BIBentryALTinterwordspacing
	W.~Koch, R.~Mancuso, R.~West, and A.~Bestavros, ``Reinforcement learning for
	{UAV} attitude control,'' \emph{ACM Trans. Cyber-Phys. Syst.}, vol.~3, no.~2,
	Feb. 2019. [Online]. Available: \url{https://doi.org/10.1145/3301273}
	\BIBentrySTDinterwordspacing
	
	\bibitem{b27}
	C.~Richter, W.~Vega-Brown, and N.~Roy, ``Bayesian learning for safe high-speed
	navigation in unknown environments,'' in \emph{Robotics Research}.\hskip 1em
	plus 0.5em minus 0.4em\relax Springer, 2018, pp. 325--341.
	
	\bibitem{b6}
	M.~Nourian, P.~E. Caines, and R.~P. Malham{\'e}, ``Mean field analysis of
	controlled {C}ucker-{S}male type flocking: Linear analysis and perturbation
	equations,'' \emph{18th IFAC World Congress, Milan}, vol.~44, no.~1, pp.
	4471--4476, Aug. 2011.
	
	\bibitem{b15}
	C.~{Beck}, S.~{Becker}, P.~{Grohs}, N.~{Jaafari}, and A.~{Jentzen}, ``{Solving
		stochastic differential equations and Kolmogorov equations by means of deep
		learning},'' \emph{arXiv e-prints}, p. arXiv:1806.00421, Jun. 2018.
	
	\bibitem{XU2011279}
	\BIBentryALTinterwordspacing
	H.~Xu, Z.~Sun, and S.~Xie, ``An iterative algorithm for solving a kind of
	discrete {HJB} equation with {M}-functions,'' \emph{Appl. Math. Lett.},
	vol.~24, no.~3, pp. 279 -- 282, 2011. [Online]. Available:
	\url{http://www.sciencedirect.com/science/article/pii/S0893965910003514}
	\BIBentrySTDinterwordspacing
	
	\bibitem{kharroubi2014numerical}
	I.~Kharroubi, N.~Langren{\'e}, and H.~Pham, ``A numerical algorithm for fully
	nonlinear {HJB} equations: an approach by control randomization,''
	\emph{Monte Carlo Meth. Appl.}, vol.~20, no.~2, pp. 145--165, 2014.
	
	\bibitem{b19}
	K.~W. Morton and D.~F. Mayers, \emph{Numerical solution of partial differential
		equations: an introduction}.\hskip 1em plus 0.5em minus 0.4em\relax Cambridge
	university press, 2005.
	
	\bibitem{b8}
	S.~Khardi, ``Aircraft flight path optimization. the
	{H}amilton-{J}acobi-{B}ellman considerations,'' \emph{Appl. Math. Sci.},
	vol.~6, no.~25, pp. pp--1221, 2012.
	
	\bibitem{b9}
	C.~Greif, ``Numerical methods for hamilton-jacobi-bellman equations,''
	\emph{University of Wisconsin-Milwaukee, U.S.A}, 2017.
	
	\bibitem{b25}
	Y.~Tassa and T.~Erez, ``Least squares solutions of the {HJB} equation with
	neural network value-function approximators,'' \emph{IEEE Trans. Neural
		Netw.}, vol.~18, no.~4, pp. 1031--1041, 2007.
	
	\bibitem{b16}
	B.~Luo, H.-N. Wu, T.~Huang, and D.~Liu, ``Reinforcement learning solution for
	{HJB} equation arising in constrained optimal control problem,'' \emph{Neural
		Netw.}, vol.~71, pp. 150--158, 2015.
	
	\bibitem{b17}
	A.~Faust, P.~Ruymgaart, M.~Salman, R.~Fierro, and L.~Tapia, ``Continuous action
	reinforcement learning for control-affine systems with unknown dynamics,''
	\emph{IEEE/CAA J. Automatica Sinica}, vol.~1, no.~3, pp. 323--336, 2014.
	
	\bibitem{b18}
	J.~W. Kim, B.~J. Park, H.~Yoo, J.~H. Lee, and J.~M. Lee, ``Deep reinforcement
	learning based finite-horizon optimal tracking control for nonlinear
	system,'' \emph{IFAC-PapersOnLine}, vol.~51, no.~25, pp. 257--262, 2018.
	
	\bibitem{samarakoon2018federated}
	S.~Samarakoon, M.~Bennis, W.~Saad, and M.~Debbah, ``Federated learning for
	ultra-reliable low-latency {V2V} communications,'' in \emph{IEEE Global
		Commun. Conf. ({GLOBECOM})}, Dec. 2018, pp. 1--7.
	
	\bibitem{zeng2019joint}
	T.~Zeng, O.~Semiari, W.~Saad, and M.~Bennis, ``Joint communication and control
	for wireless autonomous vehicular platoon systems,'' \emph{IEEE Trans.
		Comm.}, vol.~67, no.~11, pp. 7907--7922, 2019.
	
	\bibitem{8593118}
	Y.~Y. {Nazaruddin}, A.~{Widyotriatmo}, T.~A. {Tamba}, M.~S. {Arifin}, and R.~A.
	{Santosa}, ``Communication-efficient optimal-based control of a quadrotor
	{UAV} by event-triggered mechanism,'' in \emph{2018 5th Asian Conf. Def.
		Tech. (ACDT)}, 2018, pp. 96--101.
	
	\bibitem{ShiriSparseIET2018}
	H.~{Shiri}, M.~A. {Tinati}, M.~{Codreanu}, and G.~{Azarnia}, ``Distributed
	sparse diffusion estimation with reduced communication cost,'' \emph{IET
		Signal Process.}, vol.~12, no.~8, pp. 1043--1052, 2018.
	
	\bibitem{Goddemeier}
	N.~{Goddemeier} and C.~{Wietfeld}, ``Investigation of air-to-air channel
	characteristics and a {UAV} specific extension to the {R}ice model,'' in
	\emph{IEEE {GLOBECOM} Workshops (GC Wkshps), San Diego, CA, USA}, Dec. 2015,
	pp. 1--5.
	
	\bibitem{b49}
	R.~Z{\'a}rate-Minano, F.~M. Mele, and F.~Milano, ``{SDE}-based wind speed
	models with {W}eibull distribution and exponential autocorrelation,'' in
	\emph{Proc. IEEE PESGM}, Boston, MA, USA, 2016.
	
	\bibitem{b35}
	H.~Inaltekin, M.~Gorlatova, and M.~Chiang, ``Virtualized control over fog:
	interplay between reliability and latency,'' \emph{IEEE Internet Things J.},
	vol.~5, no.~6, pp. 5030--5045, 2018.
	
	\bibitem{ZengeRotary}
	Y.~{Zeng}, J.~{Xu}, and R.~{Zhang}, ``Energy minimization for wireless
	communication with rotary-wing {UAV},'' \emph{IEEE Trans. Wireless Commun.},
	vol.~18, no.~4, pp. 2329--2345, Apr. 2019.
	
	\bibitem{b1}
	F.~Cucker and J.-G. Dong, ``Avoiding collisions in flocks,'' \emph{IEEE Trans.
		Autom. Control}, vol.~55, no.~5, pp. 1238--1243, 2010.
	
	\bibitem{b5}
	Z.~Lin, M.~Broucke, and B.~Francis, ``Local control strategies for groups of
	mobile autonomous agents,'' \emph{IEEE Trans. Autom. Control}, vol.~49,
	no.~4, pp. 622--629, 2004.
	
	\bibitem{b13}
	\BIBentryALTinterwordspacing
	G.~V{\'a}s{\'a}rhelyi, C.~Vir{\'a}gh, G.~Somorjai, T.~Nepusz, A.~E. Eiben, and
	T.~Vicsek, ``Optimized flocking of autonomous drones in confined
	environments,'' \emph{Sci. Robot.}, vol.~3, no.~20, 2018. [Online].
	Available: \url{http://robotics.sciencemag.org/content/3/20/eaat3536}
	\BIBentrySTDinterwordspacing
	
	\bibitem{b2}
	D.~Liu, D.~Wang, F.-Y. Wang, H.~Li, and X.~Yang, ``Neural-network-based online
	{HJB} solution for optimal robust guaranteed cost control of continuous-time
	uncertain nonlinear systems,'' \emph{IEEE Trans. Cybern.}, vol.~44, no.~12,
	pp. 2834--2847, Dec. 2014.
	
	\bibitem{shiller1992computation}
	Z.~Shiller and H.-H. Lu, ``Computation of path constrained time optimal motions
	with dynamic singularities,'' 1992.
	
	\bibitem{shiri2018distributed}
	H.~Shiri, M.~A. Tinati, M.~Codreanu, and S.~Daneshvar, ``Distributed sparse
	diffusion estimation based on set membership and affine projection
	algorithm,'' \emph{Digital Signal Processing}, vol.~73, pp. 47--61, 2018.
	
\end{thebibliography}

\begin{IEEEbiography}[{\includegraphics[width=1in,height=1.25in,clip,keepaspectratio]{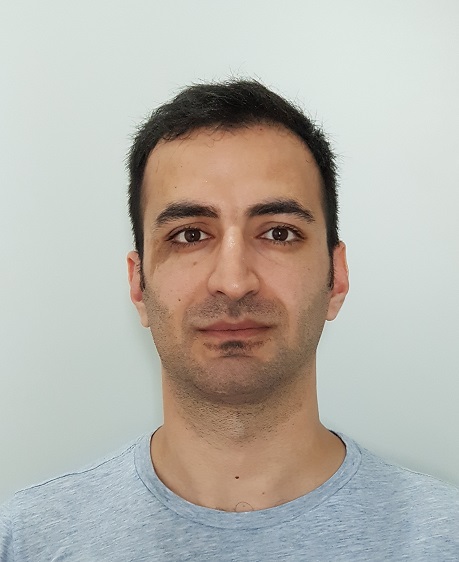}}]{Hamid Shiri} (S’19)
	is a  researcher at the Center for Wireless Communications, University of Oulu, Finland. He has been with the Faculty of Electrical and Computer Engineering, University of Tabriz, Tabriz, Iran from 2005 to 2018. His recent research interests focus are machine learning, signal processing, distributed control and optimization in 5G networks and beyond, URLLC, and Mean-Field Game theory. 
\end{IEEEbiography}

\begin{IEEEbiography}[{\includegraphics[width=1in,height=1.25in,clip,keepaspectratio]{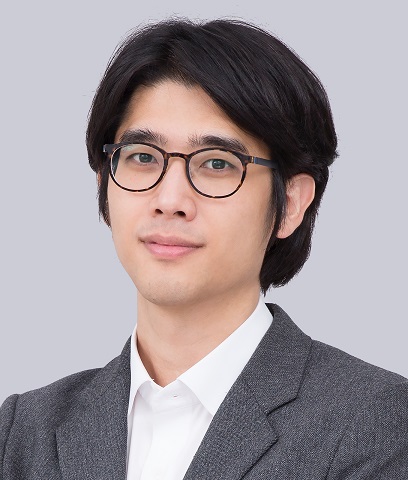}}]{Jihong Park}
	(S’09-M’16) is a Lecturer (assistant professor) at the School of IT, Deakin University, Australia. He received the B.S. and Ph.D. degrees from Yonsei University, Seoul, Korea, in 2009 and 2016, respectively. He was a Post-Doctoral Researcher with Aalborg University, Denmark, from 2016 to 2017; the University of Oulu, Finland, from 2018 to 2019. His recent research focus includes communication-efficient distributed machine learning, distributed control, and distributed ledger technology, as well as their applications for beyond 5G/6G communication systems. He served as a Conference/Workshop Program Committee Member for IEEE GLOBECOM, ICC, and WCNC, as well as NeurIPS, ICML, and IJCAI. He received the IEEE GLOBECOM Student Travel Grant in 2014, the IEEE Seoul Section Student Paper Contest Bronze Prize in 2014, and the 6th IDIS-ETNEWS (The Electronic Times) Paper Contest Award sponsored by the Ministry of Science, ICT, and Future Planning of Korea. Currently, he is an Associate Editor of Frontiers in Data Science for Communications, a Review Editor of Frontiers in Aerial and Space Networks, and a Guest Editor of MDPI Telecom SI on ``millimeter wave communiations and networking in 5G and beyond.”
\end{IEEEbiography}

\begin{IEEEbiography}[{\includegraphics[width=1in,height=1.25in,clip,keepaspectratio]{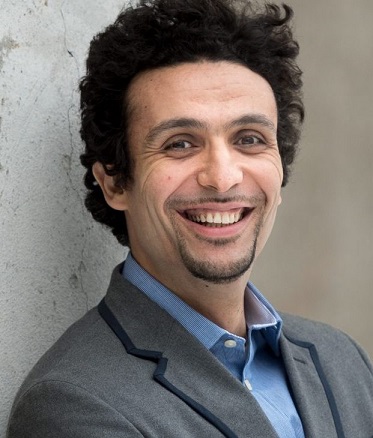}}]{Mehdi Bennis}
	is an Associate Professor at the Centre for Wireless Communications, University of Oulu, Finland, Academy of Finland Research Fellow and head of the intelligent connectivity and networks/systems group (ICON). His main research interests are in radio resource management, heterogeneous networks, game theory and distributed machine learning in 5G networks and beyond. He has published more than 200 research papers in international conferences, journals and book chapters. He has been the recipient of several prestigious awards including the 2015 Fred W. Ellersick Prize from the IEEE Communications Society, the 2016 Best Tutorial Prize from the IEEE Communications Society, the 2017 EURASIP Best paper Award for the Journal of Wireless Communications and Networks, the all-University of Oulu award for research and the 2019 IEEE ComSoc Radio Communications Committee Early Achievement Award. Dr Bennis is an editor of IEEE TCOM and Specialty Chief Editor for Data Science for Communications in the Frontiers in Communications and Networks journal. 
\end{IEEEbiography}

\end{document}